\documentclass{aa}  

\pdfoutput=1
\usepackage{longtable}
\usepackage{multirow}
\usepackage{natbib,afterpage}
\usepackage{color}
\usepackage{amssymb,times}
\usepackage[english]{babel}
\usepackage{subfigure}
\usepackage{ifpdf}
\usepackage{morefloats}
\usepackage[rightcaption]{sidecap}
\usepackage{lscape}
\usepackage{multicol}
\usepackage{booktabs}
\usepackage{slashbox, picture}
\usepackage{fixltx2e}
\usepackage{amssymb}
\usepackage{amsmath}

\newif\ifpdf \ifx\pdfoutput\undefined\pdffalse\else\pdftrue\fi
\ifpdf\pdfoutput=1
        \usepackage{graphicx}
        \usepackage{epstopdf}
        \DeclareGraphicsExtensions{.pdf,.png,.gif,.jpg}
        \else \usepackage[dvips]{color,graphicx} \fi

\def\Msun{\hbox{M$_{\odot}$}}               
\def\Lsun{\hbox{L$_{\odot}$}}               
\def\Rstar{\hbox{R$_{\star}$}}              
\def\Tstar{\hbox{T$_{\star}$}}              
\def\Mdot{\hbox{$\dot{M}$}}               
\def\arcsec{\hbox{$^{\prime\prime}$}}

\def\Al2O3{\hbox{Al$_2$O$_3$}}
\def\mic{\hbox{$\mu$m}}

\begin{document} 

\selectlanguage{english}
\newcommand{\red}{\textcolor[rgb]{1,0,0}}
\newcommand{\blue}{\textcolor[rgb]{0,0,1}}

\newlength{\fntxvi} \newlength{\fntxvii}
\newcommand{\chemical}[1]
{{\fontencoding{OMS}\fontfamily{cmsy}\selectfont
\fntxvi\the\fontdimen16\font
\fntxvii\the\fontdimen17\font
\fontdimen16\font=3pt\fontdimen17\font=3pt
$\mathrm{#1}$
\fontencoding{OMS}\fontfamily{cmys}\selectfont
\fontdimen16\font=\fntxvi \fontdimen17\font=\fntxvii}}

\title{Tracing the phase transition of Al-bearing species from molecules to dust in AGB winds} 
\subtitle{Constraining the  presence of gas-phase (Al$_2$O$_3$)$_n$ clusters}

   \author{L. Decin\inst{1}
   \and A.M.S.~Richards\inst{2}
   \and L.B.F.M.~Waters\inst{3,4}
   \and T.~Danilovich\inst{1}
   \and D.~Gobrecht\inst{5,1}
   \and T.~Khouri\inst{6}
   \and W.~Homan\inst{1}
   \and J.~Bakker\inst{7}
   \and M. Van de Sande\inst{1}
   \and J.A. Nuth\inst{8}
   \and E.~De Beck\inst{6}}

  \offprints{Leen.Decin@kuleuven.be}

  \institute{
  Instituut voor Sterrenkunde, Katholieke Universiteit Leuven, Celestijnenlaan 200D, 3001 Leuven, Belgium  
  \email{Leen.Decin@kuleuven.be}
  \and
  JBCA, Department Physics and Astronomy, University of Manchester, Manchester M13 9PL, UK
   \and
  SRON Netherlands Institute for Space Research, P.O. Box 800, 9700 AV Groningen, The Netherlands
   \and
  Anton Pannekoek Institute for Astronomy, University of Amsterdam, PO Box 94249, 1090 GE Amsterdam, The Netherlands
   \and
  Osservatorio Astronomico di Teramo, INAF, 64100 Teramo, Italy
   \and
  Department of Earth and Space Sciences, Chalmers University of Technology, Onsala Space Observatory, 439 92 Onsala, Sweden
   \and
  Radboud University, Institute for Molecules and Materials, FELIX Laboratory, Toernooiveld 7c, 6525 ED Nijmegen, The Netherlands
  \and
  NASA/GSFC, Mail Code: 690, Greenbelt, MD 20771, USA
  }


   \date{Received date; accepted date}

 
  \abstract
   {The condensation of inorganic dust grains in the winds of evolved stars is poorly understood. As of today, it is not yet known which (clusters of) molecular gas-phase species form the first dust grains in oxygen-rich (C/O$<$1) Asymptotic Giant Branch (AGB) winds. Aluminium oxides and iron-free silicates are often put forward as promising candidates for the first  dust seeds.}
   {We aim to constrain the dust formation histories in the winds of oxygen-rich AGB stars. }
   {We have obtained ALMA observations with a spatial resolution of $120\times150$\,mas tracing the dust formation region of a low mass-loss rate and a high mass-loss rate AGB star, respectively being R~Dor and IK~Tau. Emission line profiles of AlO, AlOH and AlCl are detected in the ALMA data and are used to derive a lower limit of atomic aluminium incorporated in molecules.    
   This constrains the aluminium budget that can condense into grains.} 
   {Radiative transfer models constrain the fractional abundances of AlO, AlOH, and AlCl in IK~Tau and R~Dor. We show that the gas-phase aluminium chemistry is completely different in both stars, with a remarkable difference in the AlO and AlOH abundance stratification. The amount of aluminium locked up in these 3 molecules is small, $\le$1.1$\times 10^{-7}$, for both stars, i.e.\ only $\le$2\% of the total aluminium budget. A fundamental result is that AlO and AlOH, being the direct precursors of alumina grains, are detected well beyond the onset of the dust condensation proving that the aluminium oxide condensation cycle is not fully efficient. The ALMA observations allow us to quantitatively assess the current generation of theoretical dynamical-chemical models for AGB winds. We discuss how the current proposed scenario of aluminium dust condensation for low mass-loss rate AGB stars at a distance of $\sim$1.5\,\Rstar, in particular for the stars R~Dor and W~Hya, poses a challenge if one wishes to explain both the dust spectral features in the spectral energy distribution (SED), in interferometric data, and in polarized light signal. In particular, the estimated grain temperature of \Al2O3\ is too high for the grains to retain their amorphous structure. We propose that large  gas-phase (\Al2O3)$_n$-clusters ($n>34$) can be the potential agents of the broad 11\,\mic\ feature in the SED and in the interferometric data and we explain how these large clusters can be formed. }
{The ALMA data provide us with an excellent diagnostic tool to study the gaseous precursors of the first grains in AGB winds. The observations enable us to constrain theoretical wind models and to refine our knowledge of the chemical sequence followed by aluminium species when going through the phase transition from gaseous to solid-state species. Aluminium-bearing molecules only lock up $\sim$2\% of aluminium in the inner wind of IK~Tau and R~Dor. If the rest of aluminium would form solid-state dust species, there remain challenges to explain different sets of observational data. We hypothesize that large gas-phase (\Al2O3)$_n$-clusters ($n>34$) are the carrier of the broad 11\,$\mu$m feature which is prominently visible in the SED of low mass-loss rate O-rich AGB stars.}

   
\titlerunning{Phase transition of Al-species in AGB winds}
  \maketitle

\section{Introduction} \label{Sec:Introduction}

It is well known from observations that low and intermediate mass stars (0.8\,\Msun$<$M$<$8\,\Msun) develop a low-velocity ($v\sim5-15$\,km/s) stellar wind at the end of their life during the Asymptotic Giant Branch (AGB) phase. At a mass-loss rate between $10^{-8}$ to $ 10^{-4}$\,\Msun/yr, material is stripped away from the stellar surface. Convection-induced pulsations lift material to greater heights where the temperature is low enough for gas to condense into dust grains. Radiation pressure on these grains is thought to be the main trigger for the onset of the stellar wind. While theoretical simulations based on this scenario predict stellar wind properties in broad agreement with observations of carbon-rich (C/O$>$1) AGB stars, more finetuning of the models is required for oxygen-rich (O-rich) AGB stars. Indeed, in these O-rich environments, the first dust seeds thought to form close to the stellar surface are metal oxides and pure Fe-free silicates such as Al$_2$O$_3$, SiO$_2$, Mg$_2$SiO$_4$, and MgSiO$_3$ \citep{Woitke2006A&A...460L...9W}. While their low near-infrared extinction efficiency prevents them from sublimating, it means that these glassy condensates make a negligible contribution to the radiative acceleration. The formation of both carbon and silicate grains \citep{Hofner2007A&A...465L..39H} or scattering of stellar radiation by micron-sized ($\sim$0.3\,\mic) Fe-free silicates \citep{Hofner2008A&A...491L...1H} have been proposed as possible alternatives to solve this acceleration deficit. After its postulation in 2008 by H{\"o}fner et al., the latter scenario got more and more support from observations, the first ones reported by \citet{Norris2012Natur.484..220N} using VLT/NACO data of three low mass-loss rate AGB stars (R~Dor, W~Hya, and R~Leo). The mid-infrared interferometric observations of the semi-regular AGB star RT~Vir with the MIDI instrument at VLTI also lend support to the presence of iron-free
silicates between 2 and 3 stellar radii \citep{Sacuto2013A&A...551A..72S}. Recent polarized light data obtained with the VLT/SPHERE instrument offer us another way to look at the inner wind chemical and morphological structure. Again, these data prove the presence of grains close to the star albeit putting stringent constraints on the grain size is not obvious \citep{Khouri2016A&A...591A..70K, Ohnaka2016A&A...589A..91O}. However, neither the NACO or the SPHERE data can pinpoint the exact chemical constitution of these micron-sized transparent grains.

Considering the abundance of molecular species and the temperatures at which they can exist in solid phase, possible candidates for the first nucleation seeds are SiO, SiO$_2$, MgO, Fe, Al$_2$O$_3$, TiO and TiO$_2$ \citep{Jeong2003A&A...407..191J}. However, taking into account the thermodynamic properties, MgO nucleation is thought to be completely negligible in O-rich AGB stars. Although Al$_2$O$_3$ dust is stable at very high temperatures, gaseous Al$_2$O$_3$ has a low abundance, calling into question its role as the first candidate.  For a long time, TiO and TiO$_2$ were considered to be the best candidates as primary condensates, since the nucleation rate of the more abundant SiO was thought to peak around 600\,K \citep[as based on the thermodynamic model of][using a single laboratory determination of the SiO vapour pressure]{Schick1960}, which is significantly lower than the typical temperatures at the inner edge of the dust shells. However, more recent laboratory experiments by \citet{Nuth2006ApJ...649.1178N}
and  \cite{Gail2013A&A...555A.119G} proved that the onset of SiO nucleation starts at much higher temperatures than was previously found, although the calculated SiO condensation temperatures are still $\sim$100\,K lower than the observed ones (being around 880\,K).

Another quest focusses on the potential chemical differentiation between grains formed in low mass-loss rate versus high mass-loss rate O-rich AGB stars. Observations indicate that low mass-loss rate stars form primarily dust that preserves the spectral properties of Al$_2$O$_3$, and stars with higher mass-loss rates form dust with properties of warm silicate oxides \citep{Karovicova2013arXiv1310.1924K}\footnote{Note that this differentiation does not mean that no silicates are present in low mass-loss rate stars: their presence might stay unnoticed in the spectral energy distribution (SED) due to, for instance, their glassy character.}. It is absolutely unknown if the silicates form via a heteromolecular homogeneous nucleation process consuming Mg, SiO, and H$_2$O  molecules \citep{Goumans2012MNRAS.420.3344G} or if the silicates form on top of the alumina grains (heterogeneous growth), hence gradually accelerating the wind. And be it now Fe-free silicates or aluminium-oxides that are the first solid-state species formed, another aspect that one might wonder about is if we could detect the (large) gas-phase \textit{clusters} which are the intermediate steps between the simple molecules and micron-sized grains.

A promising way to unravel the intriguing coupling between nucleation and wind generation is by studying at high spatial resolution the molecules contributing to the dust formation. 
This possibility is offered by ALMA (the Atacama Large Millimeter/sub-millimeter array). We were granted ALMA Cycle~2 observing time (project 2013.1.00166.S) to investigate the molecular content in the inner wind region of the low mass-loss rate semi-regular AGB star R~Dor (\Mdot$\sim 9 \times 10^{-8}$\,\Msun/yr) and the high mass-loss rate Mira star IK~Tau (\Mdot$\sim 4.5 \times 10^{-6}$\,\Msun/yr). We obtained the first ALMA spectral scan for O-rich AGB stars in band 7 at a spatial resolution of $\sim$0\farcs12$\times$0\farcs15. The full data scan will be presented in \citet{Decin2016scan}. In this paper, we aspire to elucidate the role of aluminium-bearing molecules, clusters, and grains in the stellar winds. The ALMA observations are presented in Sect.~\ref{Sec:ALMA_observations}. Since the aim is to correlate the gaseous aluminium content with the amount of aluminium locked up into grains, we briefly summarize the most important dust properties of the targets in Sect.~\ref{Sec:continuum}.  Important additional information on the dust formation history in O-rich AGB stars also stems from another low mass-loss rate O-rich AGB star, W~Hya. Where needed, we bring in additional constraints as derived from that target. Some properties of the three targets are introduced in Sect.~\ref{Sec:info_targets}. The results of the data analysis are presented in Sect.~\ref{Sec:Results} and are discussed in the context of molecule, cluster, and dust formation in Sect.~\ref{Sec:Discussion}. The conclusions are summarized in Sect.~\ref{Sec:conclusions}.


\section{Observations} \label{Sec:observations}

\subsection{Target selection}\label{Sec:info_targets}

To study the intricate coupling between gas and dust chemistry and the impact of the formation of some type of dust species on the wind dynamics, we have selected for our ALMA observations the two O-rich AGB stars which are the best representatives for the class of the low mass-loss rate and the high mass-loss rate O-rich AGB stars: R~Dor and IK~Tau. The reasons for their selection are their proximity (60\,pc and 260\,pc, respectively), their rich infrared and submillimeter spectra with signatures of many dust species and molecules, and the fact that both targets are very well studied with various instruments. Each target has a very different dust emission spectrum (see Sect.~\ref{Sec:continuum}), with the spectrum of IK~Tau being dominated by amorphous olivine dust (magnesium-iron silicates (Mg$_{2-x}$,Fe$_x$)SiO$_4$, with $0\le x \le 2$), while the infrared spectrum of R~Dor does not show the presence of olivine dust but species such as corundum (an aluminium oxide, \Al2O3), the aluminium-calcium bearing silicate gehlenite, the magnesium-aluminium member spinel and CO$_2$ gas have been detected \citep{Heras2005A&A...439..171H, KhouriPhD}. This supports the hypothesis that the dust condensation sequence in the low mass-loss rate object R~Dor has experienced a freeze-out after the formation of aluminium oxides, with only a small amount of silicon condensing into dust grains. In IK~Tau the formation of silicon-bearing dust species seems to be much more efficient. This can happen through either a one-step process (growth of micron-sized Fe-free silicates close the central star) or a two-steps process (growth of aluminium oxides around a few stellar radii (\Rstar), after which the grains where coated with silicates around $\sim$10 stellar radii) might take place. 

R~Dor is the closest AGB star to our Sun, with a gas mass-loss rate value of $\sim$0.9$-$2$\times10^{-7}$\,\Msun/yr and an expansion velocity of only $\sim$5.7\,km/s \citep{Schoier2004A&A...422..651S, KhouriPhD, Maercker2016A&A...591A..44M, VandeSande2017}. It is a semi-regular variable with two pulsation modes having periods of 332\,days and about 175\,days, the latter one with much smaller amplitude \citep{Bedding1998MNRAS.301.1073B}. Its angular diameter is $\sim$60\,mas \citep{Bedding1997MNRAS.286..957B, Norris2012Natur.484..220N}.
From an analysis of Herschel and sub-millimeter data,  \citet{KhouriPhD} and \citet{VandeSande2017} derived that some 90\% of atomic silicon is locked up in gaseous SiO, the remainder in silicates. VLTI/NACO and SPHERE/ZIMPOL data support the presence of a close halo of large transparent grains ($\sim$0.3\,\mic) at $\sim$1.5$-$2\,\Rstar\ \citep{Norris2012Natur.484..220N, Khouri2016A&A...591A..70K}.

IK~Tau is a high mass-loss rate Mira-type AGB star with a gas mass-loss rate of $\sim$4.5$\times10^{-6}$\,\Msun/yr, an expansion velocity of 17.7\,km/s \citep{Decin2010A&A...516A..69D, Maercker2016A&A...591A..44M}, and a pulsation period of $\sim$470\,days \citep{Wing1973ApJ...184..873W}. The anguldar diameter is estimated around 60\,mas \citep{Decin2010A&A...516A..69D}. Currently, a dozen different molecules and some of their isotopologues have been discovered in IK~Tau, including CO, HCN, SiO, SiS, SO, SO$_2$, NaCl, NS, NO, HCO$^+$, H$_2$CO, PO, PN, H$_2$S \citep{Milam2007ApJ...668L.131M, Kim2010A&A...516A..68K, Decin2010A&A...516A..69D, Decin2010A&A...521L...4D, DeBeck2013A&A...558A.132D, DeBeck2015ASPC..497...73D, Prieto2016arXiv160901904V}. Some molecules including SiO and SiS are depleted in the circumstellar envelope, possibly due to condensation onto dust grains. Other molecules such as HCN demonstrate the impact of pulsation-driven shocks in the inner wind zone which forces the chemistry into a  non-equilibrium state.

W~Hya resembles in many aspects the low mass-loss rate AGB star R~Dor. Thanks to is proximity \citep[$\sim$78\,pc,][]{Knapp2003A&A...403..993K} and its brightness it is also well studied with various observational techniques from the visible to the radio. This semi-regular variable has a light curve showing a clear periodicity with period of 389 days \citep{Uttenthaler2011A&A...531A..88U}. The molecular abundance stratification in its circumstellar envelope was studied in detail by \citet{Khouri2014A&A...561A...5K, Khouri2014A&A...570A..67K}. Based on an extensive grid of CO, SiO, H$_2$O and HCN lines (and some of their isotopologues) the thermodynamic structure of the wind was derived and they deduced that about one-third of the silicon atoms is locked up in dust particles. The SED of W~Hya is characterized by the same dust emission peaks as R~Dor (see Sect.~\ref{Sec:continuum}) and also for this star large scattering grains in a halo close the star are detected via polarized light signal \citep{Norris2012Natur.484..220N, Ohnaka2016A&A...589A..91O}. 
W~Hya was also observed by ALMA in band~7 (project 2015.1.01466.S, P.I.\ A.\ Takigawa) with a spatial resolution of $\sim$40\,mas at 3 selected frequency bands (330.40--331.34\,GHz, 342.40-343.34\,GHz, 344.03-345.91\,GHz) encompassing the AlO N=9-8 and AlOH J=11-10 transition. AlO is detected in their data \citep{Takigawa2016LPICo1921.6543T}.

\subsection{ALMA observations} \label{Sec:ALMA_observations}

\subsubsection{ALMA observing strategy} \label{Sec:ALMA_strategy}

Spectral scans of IK~Tau and R~Dor were obtained covering the
frequencies between 335--362\,GHz, using four separate observations
each, during August-September 2015. The observations and data
reduction used the ALMA pipeline for calibration, followed by
identifying and imaging the line-free continuum, which was then used
for self-calibration.  The solutions were applied to all data and
spectral cubes were made adjusted to constant $v_{\mathrm{LSR}}$. The
process is described in more detail in \citet{Decin2016scan}; here, we concentrate on
the parameters affecting the Al compound cubes. The astrometric
accuracy is $\sim$17\,mas and the flux scale is accurate to
$\sim$7\%. The relative accuracy (e.g., aligning different lines in the
same source) is limited only by the signal-to-noise, S/N, and would be $<$1\,mas for S/N of
200, or 33\,mas (about 1 pixel) for a faint 3$\sigma$ detection of a
compact source.  The rms noise ($\sigma_{\rm{rms}}$) varies across the band depending on
intrinsic atmospheric transmission and the weather at the time of
observations. The final image parameters are summarised in
Table~\ref{Table:obs}. Owing to the different elevations as seen from ALMA, we achieved a higher resolution and used a smaller pixel size (30 mas versus 40 mas) for R Dor versus IK Tau. For each line, we made cubes at higher and lower spectral resolution
(see info in Table~\ref{Table:obs}), by using different weighting and spectral
averaging. In addition, for AlO, we further smoothed the wide-channel
data set with a 500\,mas Gaussian kernel. The other lines detected and the sub-mm continuum properties will be discussed in future papers.

\begin{table}
\caption{ALMA observation parameters for the Al-compounds in IK~Tau and R~Dor.}
\label{Table:obs}
\setlength{\tabcolsep}{1mm}
\begin{tabular}{lclll}
\hline \hline
 Species  & Chan\tablefootmark{a} & Vel\tablefootmark{b}        &Beam                       & $\sigma_{\rm{rms}}$  \\
         & (MHz)&(km s$^{-1})$&(mas$\times$mas, P.A.)\tablefootmark{c}     & (mJy)\\
\hline
 & \multicolumn{4}{c}{IK~Tau}\\
 \cmidrule(r){2-5} 
Continuum&15.2$\times$10$^3$&             &180$\times$160,  27$^{\circ}$&0.047 \\
AlCl     &1.95  & 1.7         &160$\times$130,  43$^{\circ}$&1.6   \\
AlCl     &5.9   & 5           &200$\times$160,  42$^{\circ}$&1.4   \\
AlOH     &1.95  & 1.7         &150$\times$140, $-$22$^{\circ}$&5.0   \\
AlOH     &5.9   & 5           &190$\times$170, $-$1$^{\circ}$&2.4   \\
AlO      &1.95  & 1.7         &160$\times$130, $-$15$^{\circ}$&5.0   \\
AlO      &5.9   & 5           &190$\times$170, $-$20$^{\circ}$&2.6   \\
\hline
 & \multicolumn{4}{c}{R~Dor}\\
 \cmidrule(r){2-5} 
Continuum&11.6$\times$10$^3$&             &150$\times$140,--5$^{\circ}$ &0.047 \\
AlCl     &0.97  & 0.9         &150$\times$130,  20$^{\circ}$&2.7   \\
AlCl     &2.5   & 2.2         &180$\times$180,  10$^{\circ}$&1.4   \\
AlOH     &0.97  & 0.9         &170$\times$120, $-$48$^{\circ}$&3.8   \\
AlOH     &2.5   & 2.2         &200$\times$160, $-$48$^{\circ}$&2.4   \\
AlO      &0.97  & 0.9         &160$\times$130, $-$26$^{\circ}$&4.0   \\
AlO      &2.5   & 2.2         &190$\times$160, $-$30$^{\circ}$&1.8   \\
\hline
\end{tabular}
\tablefoot{
\tablefoottext{a}{`Chan' is the channel width for the line cubes, or the total line-free bandwidth
(spread between 335--362\,GHz) for the continuum.}\\
\tablefoottext{b}{`Vel' is the spectral resolution of the image cube.}\\
\tablefoottext{c}{P.A.\ is the position angle measured from North to East.}}
\end{table}

\subsubsection{Al-bearing species detected with ALMA} \label{Sec:Al_detected_ALMA}

Chemical models show that the most likely carriers of aluminium in the wind of oxygen-rich evolved stars are Al, AlH, AlO, AlOH, AlO$_2$, AlCl, Al$_2$, and Al$_2$O \citep[see Fig.~\ref{Fig:chem_Al_radius} in Sect.~\ref{Sec:disc_Al_gas}, a result by][]{Gobrecht2016A&A...585A...6G}. The main isomers of AlO$_2$ and Al$_2$O are linear and their rotational lines are not observable; nor is the homonuclear Al$_2$ easily observable \citep{Cai1991JChPh..95...73C, Koput2004JChPh.121..130K, Kaminski2016A&A...592A..42K}.  AlH does not have a rotational line transition in the targeted ALMA frequency range. The other three aluminium-bearing molecules are detected in IK~Tau and R~Dor with ALMA: AlO N=9$-$8 (rest frequency, $\nu_{\rm{rest}}$, at 345.457\,GHz, lower state energy E$_{\rm{low}}$\,=\,66\,K), AlOH J=11$-$10 ($\nu_{\rm{rest}}$\,=\,346.156\,GHz, E$_{\rm{low}}$\,=\,83\,K), and AlCl J=24$-$23 ($\nu_{\rm{rest}}$\,=\,349.444\,GHz, E$_{\rm{low}}$\,=\,193\,K). 

The transitions of these three molecules are split up by hyperfine components due to the nuclear spin, $I=5/2$, of aluminium. AlOH and AlCl are closed-shell species with a $^1\Sigma^+$ ground state.
AlO is a radical with a $^2\Sigma$ ground state. The rotational levels of this molecule are therefore split by both fine and hyperfine interactions. The net result is that each rotational transition consists of 10$-$12 favourable, closely spaced, hyperfine components \citep{Tenenbaum2009ApJ...694L..59T, Tenenbaum2010ApJ...712L..93T}.

\begin{figure}[!htp]
\vspace{0pt}{\includegraphics[angle=0,width=.48\textwidth]{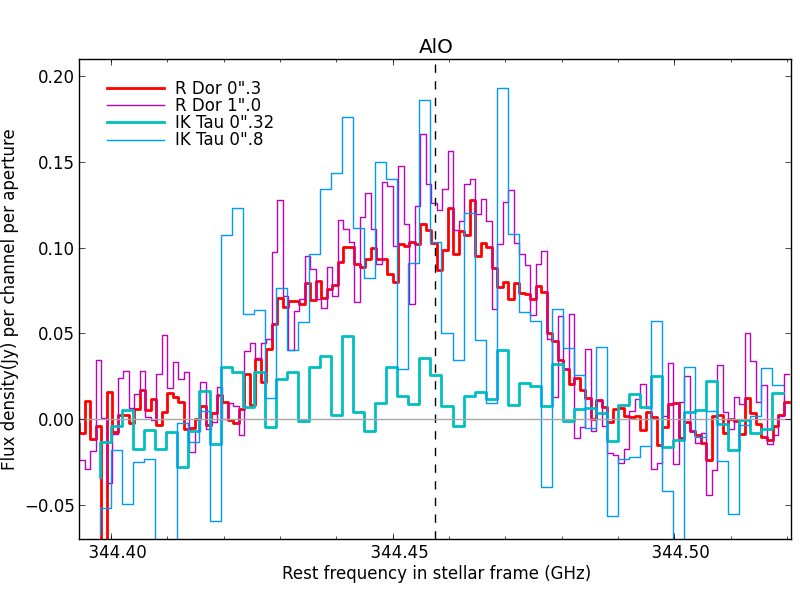}}
 \vspace{0pt}{\includegraphics[angle=0,width=.48\textwidth]{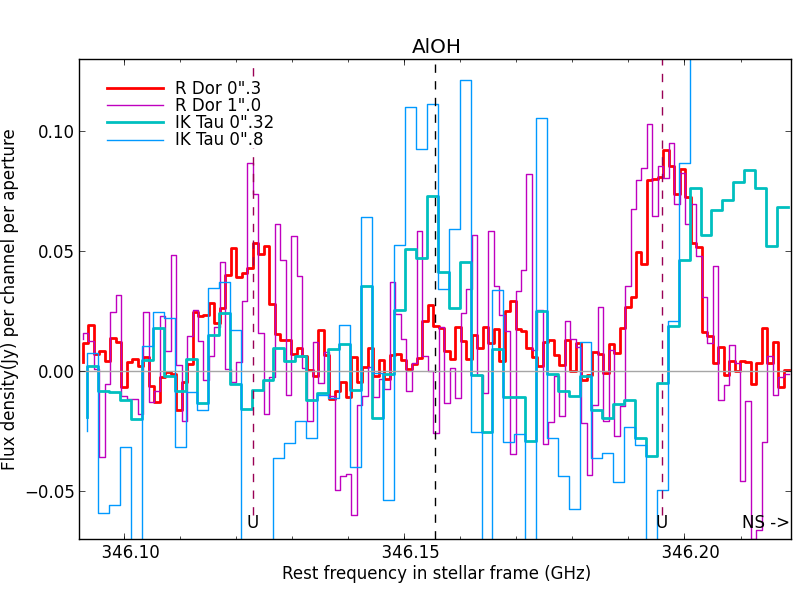}}
 \vspace{0pt}{\includegraphics[angle=0,width=.48\textwidth]{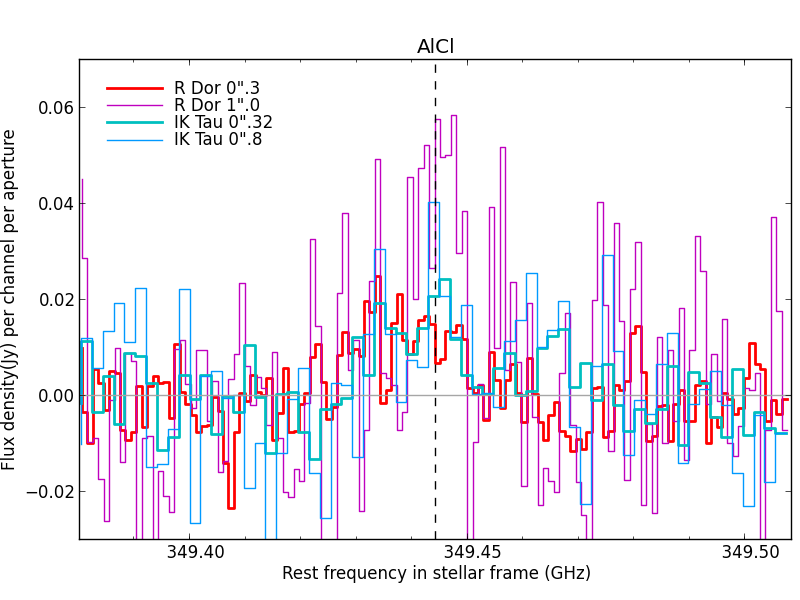}}
 \caption{ALMA AlO, AlOH, and AlCl data extracted for an aperture size of 320\,mas (300\,mas) and 800\,mas (1000\,mas) for IK~Tau (R~Dor). The frequency range is shifted to rest frequency using a local standard of rest velocity, $v_{\rm{LSR}}$, of 35\,km/s for IK~Tau and of 7\,km/s for R~Dor. Close to the AlOH emission, NS $^2\Pi_{1/2}$ J=15/2--13/2 transition is detected in IK~Tau (at rest frequency 346.2201\,GHz) and two unidentified lines ('U') in R~Dor at a rest frequencies of 346.123\,GHz and 346.196\,GHz.}
 \label{Fig:ALMA_data}
 \end{figure}

Figure~\ref{Fig:ALMA_data} shows the flux densities extracted for an aperture size of 320\,mas (300\,mas) and 800\,mas (1000\,mas) for IK~Tau (R~Dor). The channel maps for the individual transitions are shown in Fig.~\ref{Fig:channel_IKTau_AlO}--\ref{Fig:channel_IKTau_AlCl} for IK~Tau and in Fig.~\ref{Fig:channel_RDor_AlO}--\ref{Fig:channel_RDor_AlCl} for R~Dor.  Table~\ref{Table:Al_ALMA} gives the measured properties of each molecular transition, including the velocity width, the shift of the centroid w.r.t.\ the LSR velocity of the star, the angular width, and the peak flux and integrated flux within the specified velocity interval. The angular width has been obtained from the zeroth moment (total intensity) by measuring the mean flux in annuli of 60 (80) mas, centred on the position of R Dor (IK Tau), assumed to be the continuum peak. The angular size was taken as the distance out to 2\,$\sigma_{\rm{rms}}$.
The measurements were also used to compute the azimuthally averaged flux densities, in order to compare with 1D models solving the non-LTE radiative transfer equation for the specific molecule (Sect.~\ref{Sec:Results}) and with models solving the chemical network (so-called `forward' chemistry models) by \citet{Gobrecht2016A&A...585A...6G} in Sect.~\ref{Sec:Discussion}. The emission of some molecular transitions shows some irregular morphologies (as described in the next subsections) but there is no obvious preferred direction, so this is a reasonable approximation although not allowing for clumpiness.

\begin{table*}
\setlength{\tabcolsep}{1mm}
\caption{Spectral properties of AlO, AlOH, and AlCl as observed with ALMA in IK~Tau (upper part of the table) and R~Dor (lower part). Values for the velocity width, velocity asymmetry, angular width, and peak and integrated flux are only given in the case of a clear detection.}
\label{Table:Al_ALMA}
\begin{tabular}{lrrrrrrrrr}
\hline \hline
\multicolumn{1}{c}{(1)} & \multicolumn{1}{c}{(2)} & \multicolumn{1}{c}{(3)} & \multicolumn{1}{c}{(4)} & \multicolumn{1}{c}{(5)} & \multicolumn{1}{c}{(6)} & \multicolumn{1}{c}{(7)} & \multicolumn{1}{c}{(8)} & \multicolumn{1}{c}{(9)} & \multicolumn{1}{c}{(10)} \\
 \multicolumn{1}{c}{Molecule} &  \multicolumn{1}{c}{Velocity} & \multicolumn{1}{c}{Velocity\tablefootmark{a}}  & \multicolumn{1}{c}{Peak} & \multicolumn{1}{c}{Integrated} & \multicolumn{1}{c}{Velocity} & \multicolumn{1}{c}{Velocity\tablefootmark{a}}  & \multicolumn{1}{c}{Peak} & \multicolumn{1}{c}{Integrated} &\multicolumn{1}{c}{Angular} \\
 &  \multicolumn{1}{c}{width} & \multicolumn{1}{c}{asymmetry}  &  \multicolumn{1}{c}{flux} & \multicolumn{1}{c}{flux} & \multicolumn{1}{c}{width} & \multicolumn{1}{c}{asymmetry} & \multicolumn{1}{c}{flux} & \multicolumn{1}{c}{flux}   & \multicolumn{1}{c}{width} \\
 & [km s$^{-1}$] & [km s$^{-1}$] &  [Jy] & [Jy km s$^{-1}$] & [km s$^{-1}$] & [km s$^{-1}$] &  [Jy] & [Jy km s$^{-1}$] & [mas] \\
 \hline
 & \multicolumn{9}{c}{IK~Tau}\\
 \cmidrule(r){2-10} 
 & \multicolumn{4}{c}{Aperture 320\,mas} & \multicolumn{4}{c}{Aperture 800\,mas} \\
  \cmidrule(r){2-5}
  \cmidrule(l){6-9}
AlO N=9$-$8 & 3.40 &  $-$0.32 &     0.036 &   0.247 & 32.30 &  14.98 &     0.186 &   3.211 & 800 \\
AlOH J=11$-$10  & 11.84 &   1.06  &   0.073 &   0.466 & 11.84 &   1.06 &     0.121 &   0.794 & 640 \\
AlCl J=24$-$23 &  6.70 &  $-$2.92 &   0.024 &   0.228 &  3.35 &  $-$2.92 &      0.040 &   0.255 & 160 \\
\hline
 & \multicolumn{9}{c}{R~Dor}\\
 \cmidrule(r){2-10} 
 & \multicolumn{4}{c}{Aperture 300\,mas} & \multicolumn{4}{c}{Aperture 1000\,mas} \\
  \cmidrule(r){2-5}
  \cmidrule(l){6-9}
AlO N=9$-$8 &  51.00\tablefootmark{b} &   3.67 &     0.128 &   4.032 & 47.60 &  10.47 &     0.166 &   4.570 & 840\\
AlOH J=11$-$10 &  3.38 &  $-$0.36 &     0.027 &   0.119 &  $-$  &  $-$   &  $-$  &     $-$  & 240 \\
AlCl J=24$-$23 &   3.35 &   2.65 &      0.021 &   0.113 &  9.22 &  $-$1.54 &      0.058 &   0.326 & 240 \\
\hline
\end{tabular}
\tablefoot{
\tablefoottext{a}{The shift of the centroid w.r.t.\ the LSR velocity of the star is obtained by dividing the value of the `velocity asymmetry' by two.}\\
\tablefoottext{b}{Large velocity width is caused by the hyperfine structure; see Sect.~\ref{Sec:Results}.}
}
\end{table*}

 \subsubsection{IK Tau}\label{Sec:ALMA_IKTau}
 
\begin{figure*}
\includegraphics[width=0.98\textwidth]{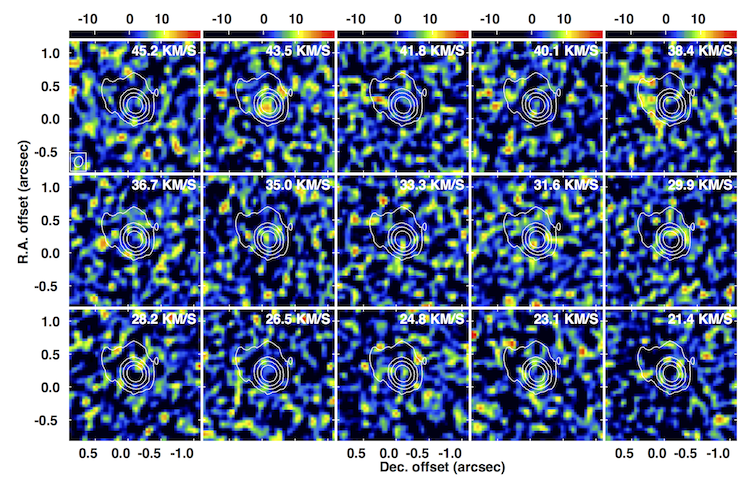}
\caption{Channel maps (colour scale in mJy) of the AlO N=9$-$8 transition in IK~Tau. White contours are the continuum at ($-1$, 1, 4, 16, 64, 256)$\times$0.3\,mJy. The ordinate and co-ordinate axis give the offset of the right ascension and declination, respectively, in units of arcseconds. The ALMA beam size is shown in the bottom left corner of the upper panel.}
\label{Fig:channel_IKTau_AlO}
\end{figure*}

\begin{figure*}
\includegraphics[width=0.98\textwidth]{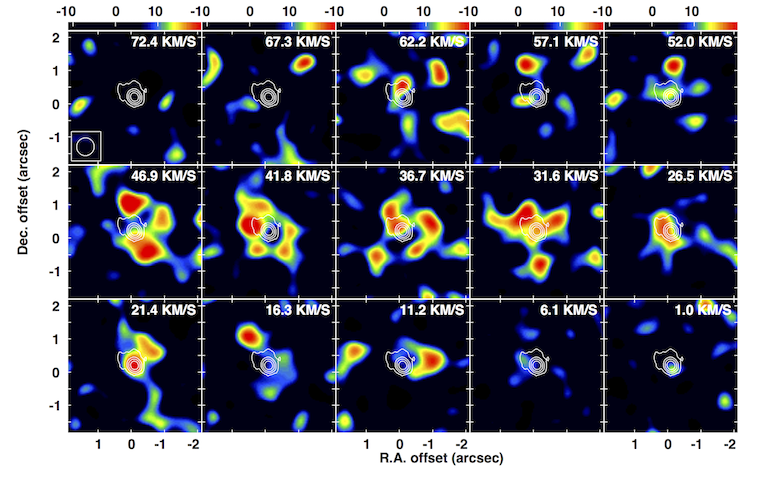}
\caption{Same as for Fig.~\ref{Fig:channel_IKTau_AlO}, but the AlO emission is averaged over 5 channels and smoothed with 0\farcs5 Gaussian kernel.}
\label{Fig:channel_IKTau_AlO_smooth}
\end{figure*}

 \paragraph{AlO:} Looking at Fig.~\ref{Fig:channel_IKTau_AlO}, AlO remains at first sight barely detected in IK~Tau. This is reflected by the line profile for the 320\,mas extraction aperture (Fig.~\ref{Fig:ALMA_data}). However, enlarging the extraction aperture to 800\,mas, a clear signature of AlO emission is seen with a peak flux of 0.19\,Jy and integrated line flux of 3.2\,Jy\,km\,s$^{-1}$. 
 The SMA observations by \citet{Debeck2017A&A...598A..53D} also show a weak AlO emission feature with an integrated line flux of 2.33\,Jy\,km\,s$^{-1}$ for an aperture of $1\times1$\arcsec.
 This points toward faint extended AlO emission that is probably genuinely clumpy. To verify this idea, we have averaged the AlO emission over 5 channels and smoothed the emission with a 0\farcs5 Gaussian kernel; see Fig.~\ref{Fig:channel_IKTau_AlO_smooth}.  Albeit the signal-to-noise ratio is still low and we caution that the smoothing is on a scale comparable to the maximum recoverable scale (MRS), this figure indeed hints at extended clumpy emission.
 
\begin{figure*}
\includegraphics[width=0.98\textwidth]{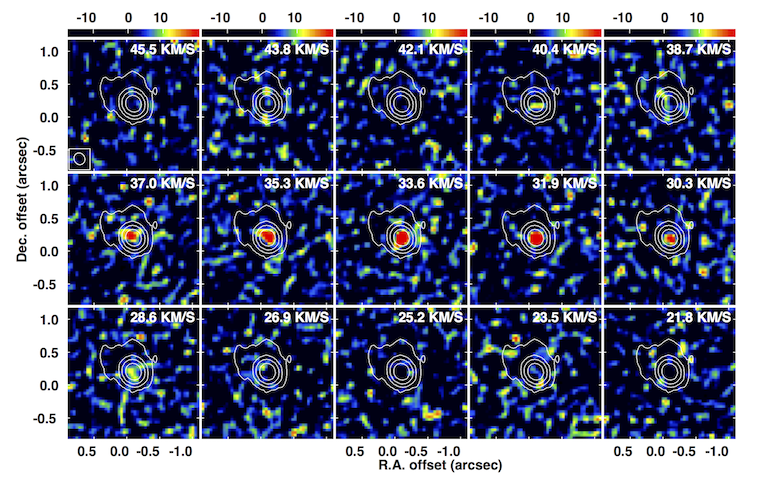}
\caption{Same as for Fig.~\ref{Fig:channel_IKTau_AlO}, but for AlOH J=11$-$10.}
\label{Fig:channel_IKTau_AlOH}
\end{figure*}

 \paragraph{AlOH:} The channel maps of IK~Tau show a clear detection of AlOH with a peak strength around 0.07\,$-$\,0.12\,Jy (depending on the extraction aperture; Fig.~\ref{Fig:channel_IKTau_AlOH}). The emission is centred around the peak of the continuum emission and has an angular width of $\sim$240\,mas.
 
\begin{figure*}
\includegraphics[width=0.98\textwidth]{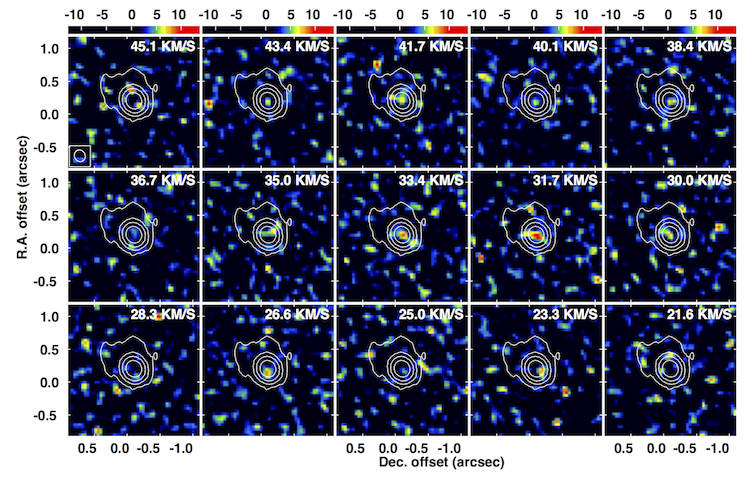}
\caption{Same as for Fig.~\ref{Fig:channel_IKTau_AlO}, but for AlCl J=24$-$23.}
\label{Fig:channel_IKTau_AlCl}
\end{figure*}

 \paragraph{AlCl:} Around the $v_{\rm{LSR}}$ of $\sim$35\,km/s almost no AlCl emission is seen, but the emission increases in strength at blue-shifted velocities (Fig.~\ref{Fig:channel_IKTau_AlCl}). This is reflected in the asymmetric AlCl line profile in Fig.~\ref{Fig:ALMA_data}.

\subsubsection{R~Dor}

\begin{figure*}[htp]
\sidecaption
\includegraphics[width=120mm]{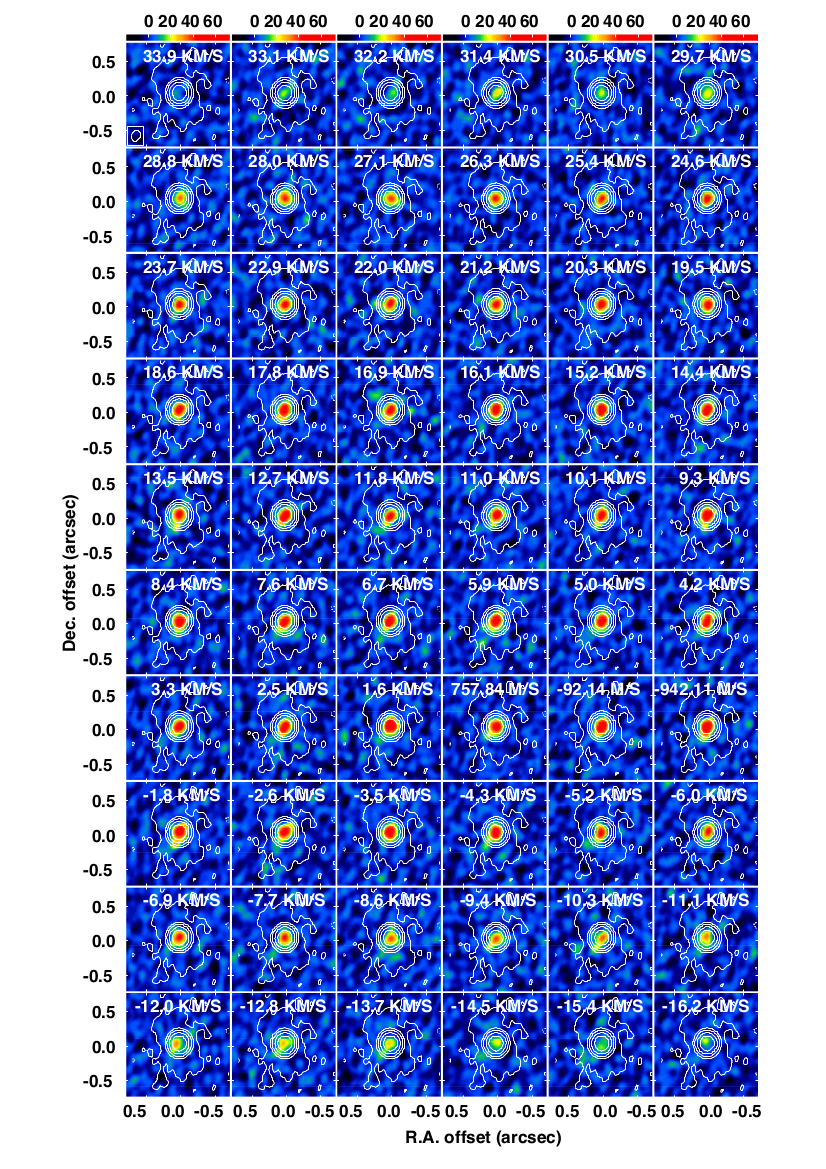}
\caption{Channel maps (colour scale in mJy) of the AlO N=9-8 transition in R~Dor. White contours are the continuum at ($-1$, 1, 4, 16, 64, 256)$\times$0.15\,mJy. The ordinate and co-ordinate axis give the offset of the right ascension and declination, respectively, in units of arcseconds. The ALMA beam size is shown in the bottom left corner of the upper panel.}
\label{Fig:channel_RDor_AlO}
\end{figure*}

\begin{figure*}[htp]
\sidecaption
\includegraphics[width=120mm]{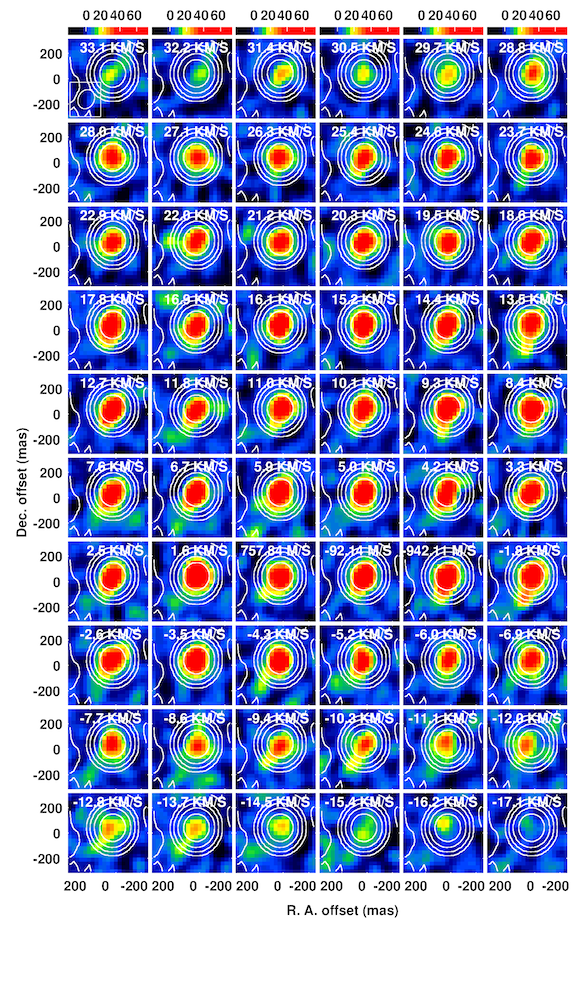}
\caption{Same as for Fig.~\ref{Fig:channel_RDor_AlO}, but zooming into the AlO emission in the inner 0\farcs2 region.}
\label{Fig:channel_RDor_AlO_ZOOM}
\end{figure*}

\paragraph{AlO:} The bright AlO N=9-8 is easily picked up by the ALMA instrument, with a peak flux of $\sim$0.13\,--\,0.17\,Jy for the two different extraction apertures. The emission is smooth and concentrated around the peak of the continuum emission (Fig.~\ref{Fig:channel_RDor_AlO}). At an offset of $\sim$280\,mas to the south-east, enhanced AlO emission is seen in many channel maps (Fig.~\ref{Fig:channel_RDor_AlO_ZOOM}). At the same offset location other molecules \citep[including AlOH, SO, SiO, CO, and SO$_2$; see][]{Decin2016scan} show excess emission as well with the main peak having a  blue-shifted frequency offset of 15\,$-$\,20\,MHz.

\begin{figure*}[htp]
\includegraphics[width=0.98\textwidth]{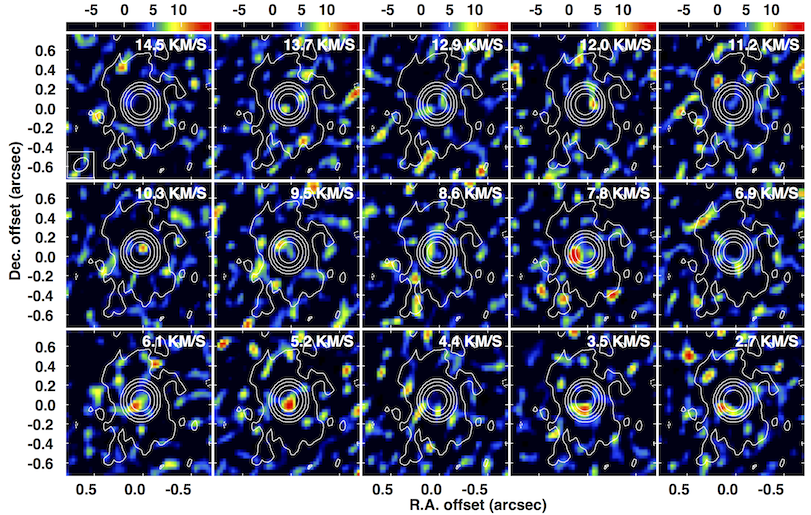}
\caption{Same as for Fig.~\ref{Fig:channel_RDor_AlO}, but for AlOH J=11$-$10.}
\label{Fig:channel_RDor_AlOH}
\end{figure*}

\begin{figure*}[htp]
\sidecaption
\includegraphics[width=120mm]{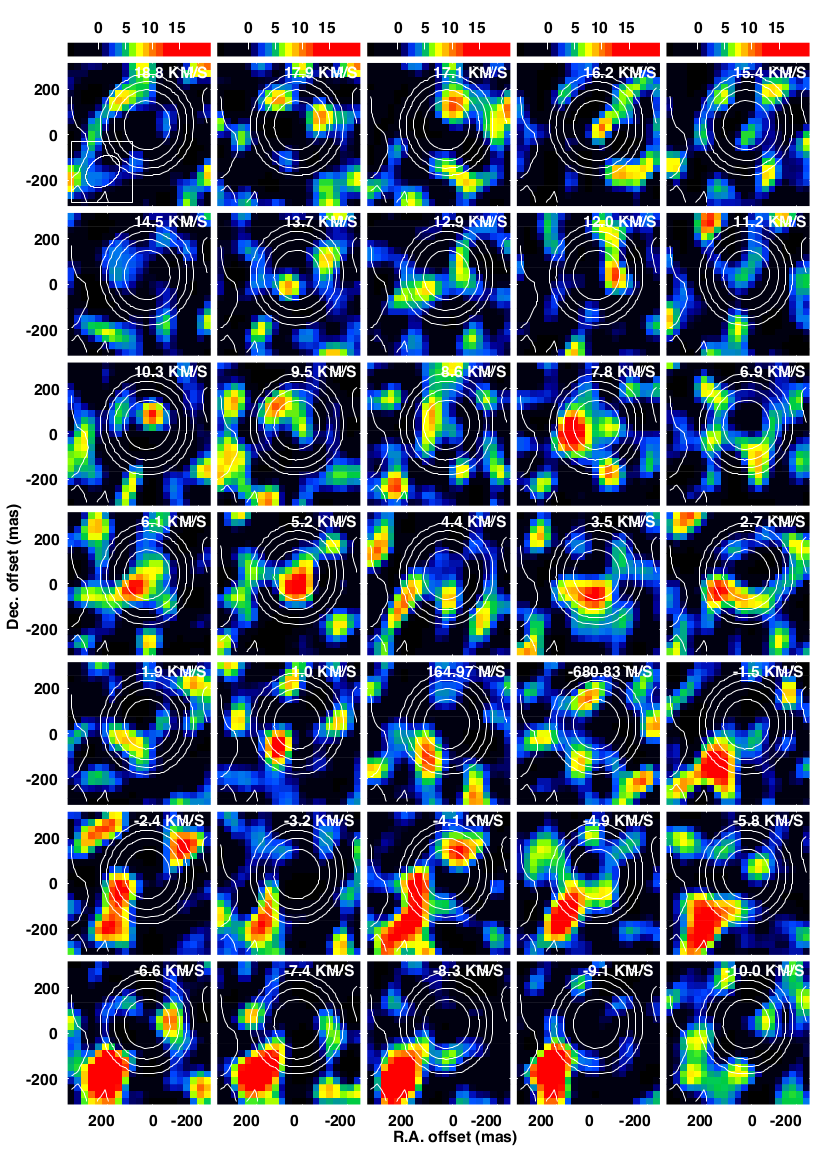}
\caption{Same as for Fig.~\ref{Fig:channel_RDor_AlOH}, but zooming into the AlOH emission in the inner 0\farcs2 region.}
\label{Fig:channel_RDor_AlOH_zoom}
\end{figure*}

\paragraph{AlOH:} The AlOH emission in R~Dor is very weak and just above 3\,$\sigma_{\rm{rms}}$ for the 300\,mas extraction aperture (Fig.~\ref{Fig:ALMA_data}). The main emission around $v_{\rm{LSR}}$ is slightly off-center (as compared to the peak of the continuum emission; Fig.~\ref{Fig:channel_RDor_AlOH}). As noted already for AlO, a brighter clump of emission is visible around 280\,mas to the south-east (Fig.~\ref{Fig:channel_RDor_AlOH_zoom}).

\begin{figure*}[htp]
\includegraphics[width=0.98\textwidth]{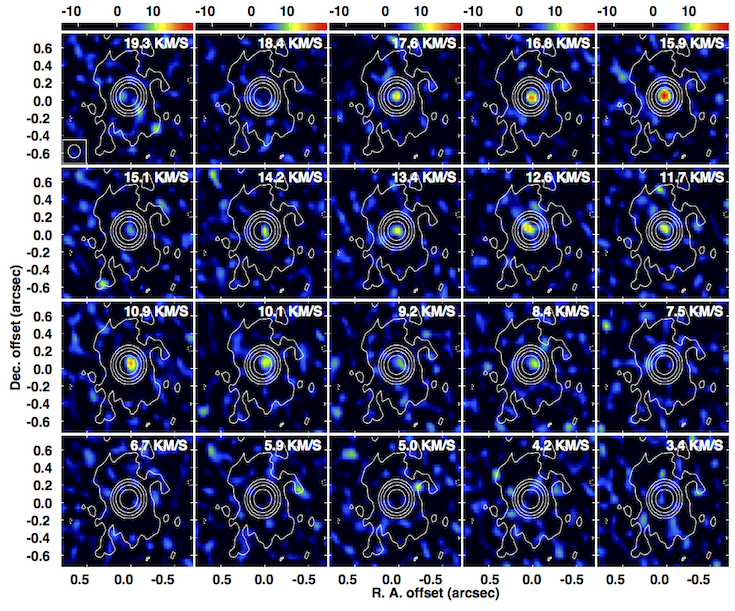}
\caption{Same as for Fig.~\ref{Fig:channel_RDor_AlO}, but for AlCl J=24$-$23.}
\label{Fig:channel_RDor_AlCl}
\end{figure*}

\paragraph{AlCl:} Almost no AlCl emission is visible around $v_{\rm{LSR}}$ for the 300\,mas aperture (Fig.~\ref{Fig:channel_RDor_AlCl}), but brighter emission becomes apparent at larger velocities with a peak flux around 0.02\,$-$\,0.06\,Jy and an integrated flux of 0.1\,$-$\,0.3\,Jy\,km\,s$^{-1}$ (depending on the extraction aperture).

The different velocity offsets for AlCl (on the red-shifted versus blue-shifted sides of the systemic velocity) for R~Dor versus IK~Tau cautions us to consider a potential blend or misidentification of this species.

%
%
 
\begin{figure*}[htp]
\begin{minipage}[t]{.32\textwidth}
        \centerline{\resizebox{\textwidth}{!}{\includegraphics[angle=0]{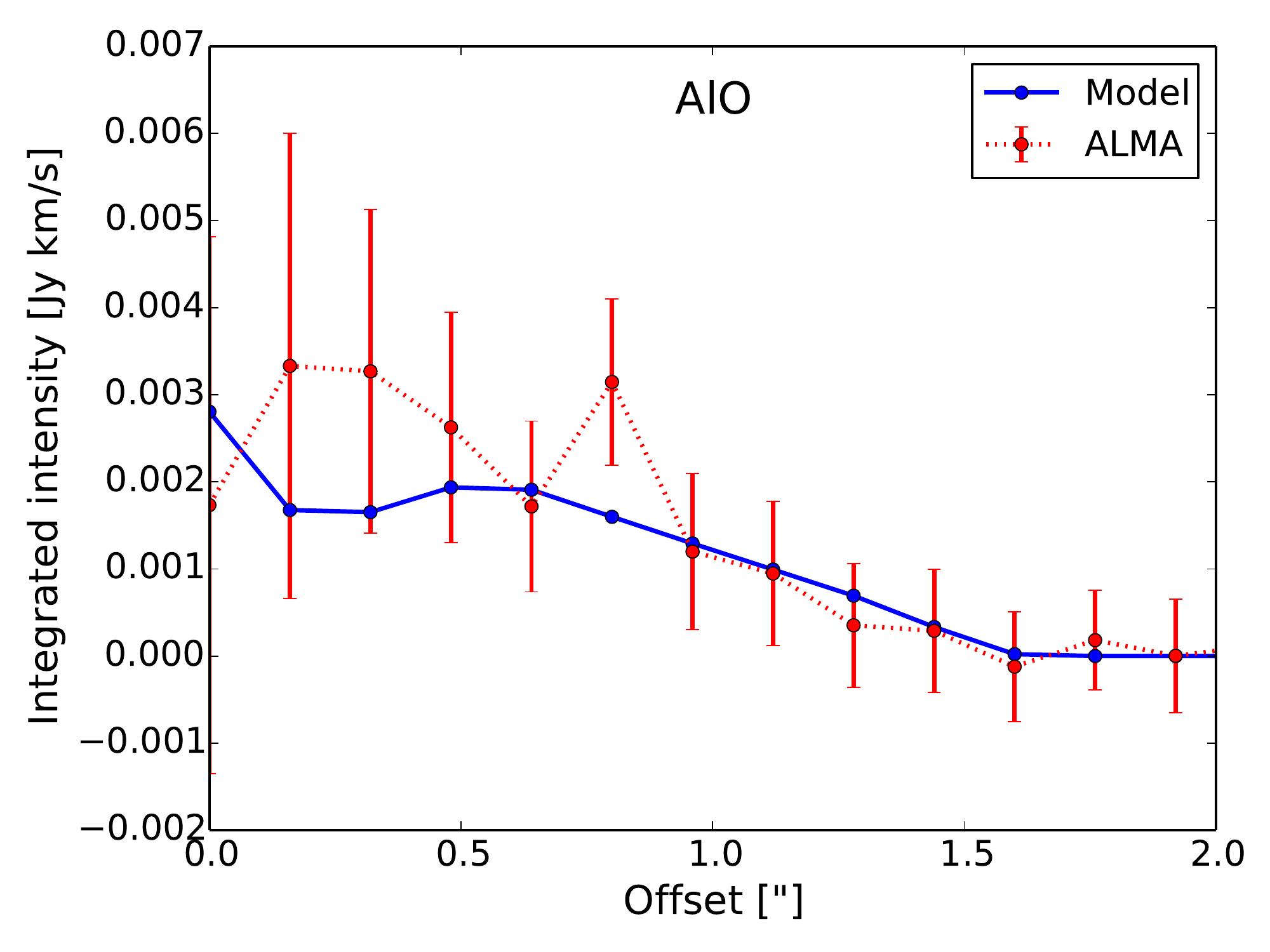}}}
    \end{minipage}
    \hfill
\begin{minipage}[t]{.32\textwidth}
        \centerline{\resizebox{\textwidth}{!}{\includegraphics[angle=0]{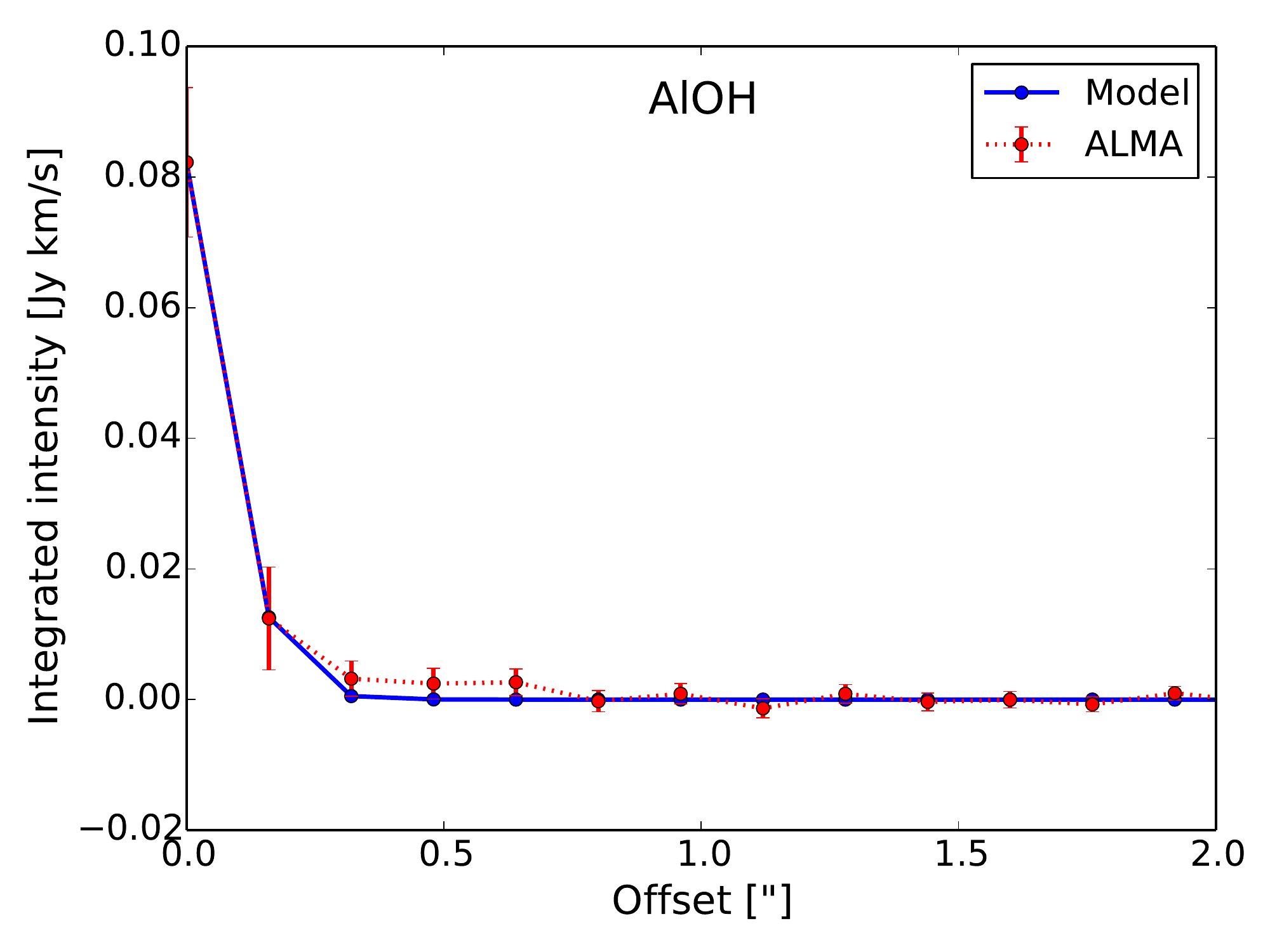}}}
    \end{minipage}
    \hfill
\begin{minipage}[t]{.32\textwidth}
        \centerline{\resizebox{\textwidth}{!}{\includegraphics[angle=0]{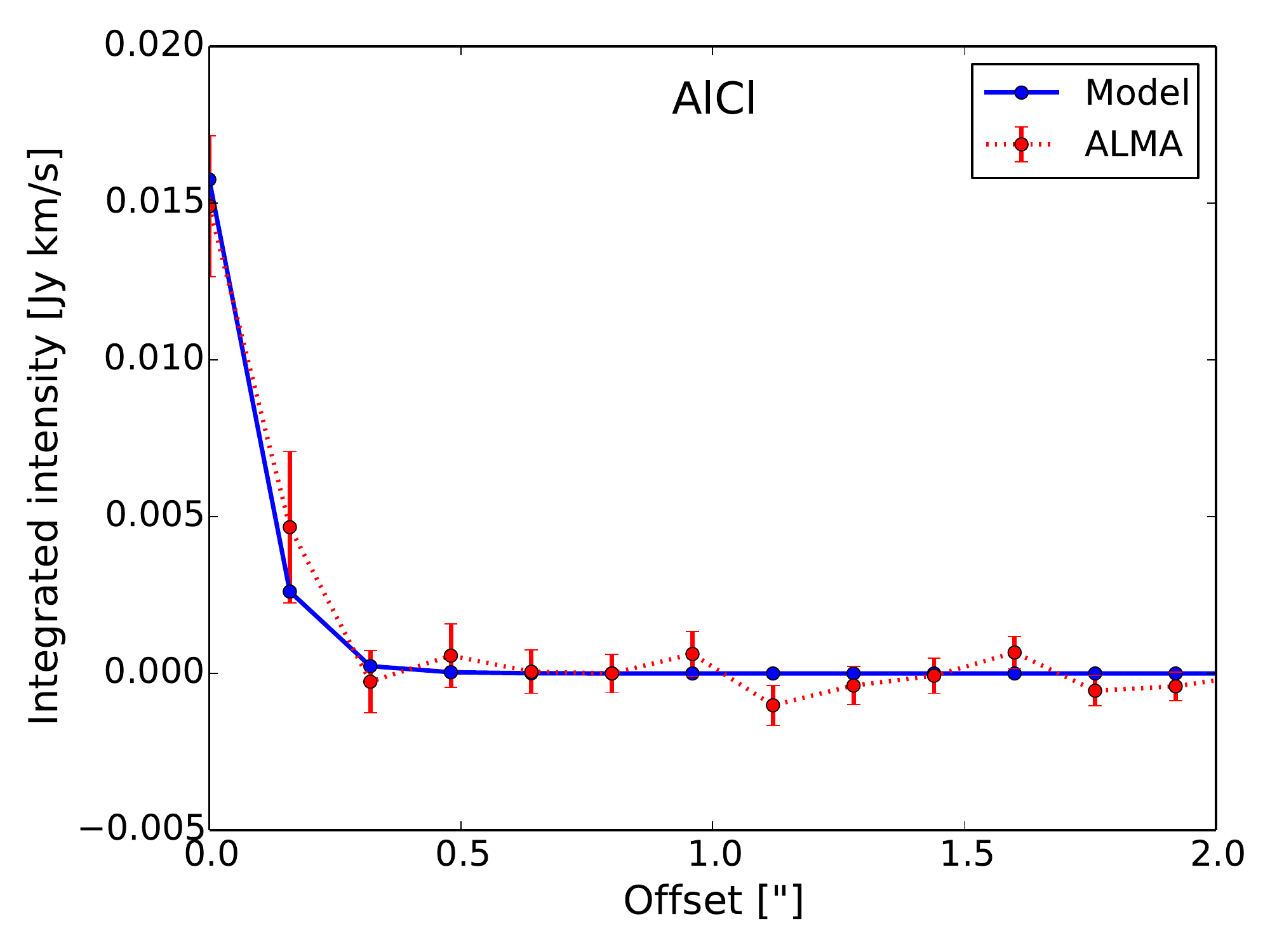}}}
    \end{minipage}
\begin{minipage}[t]{.32\textwidth}
        \centerline{\resizebox{\textwidth}{!}{\includegraphics[angle=0]{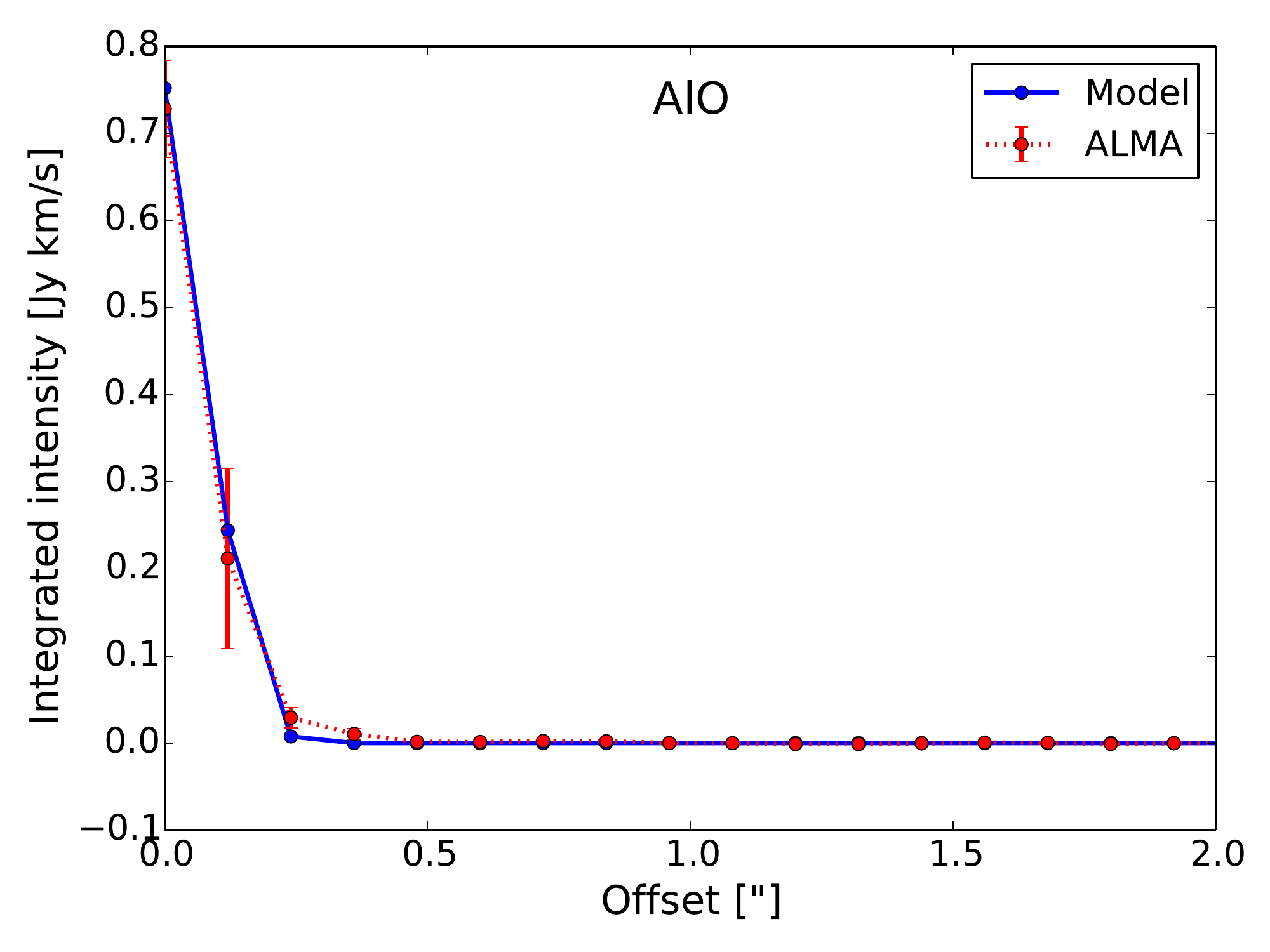}}}
    \end{minipage}
    \hfill
\begin{minipage}[t]{.32\textwidth}
        \centerline{\resizebox{\textwidth}{!}{\includegraphics[angle=0]{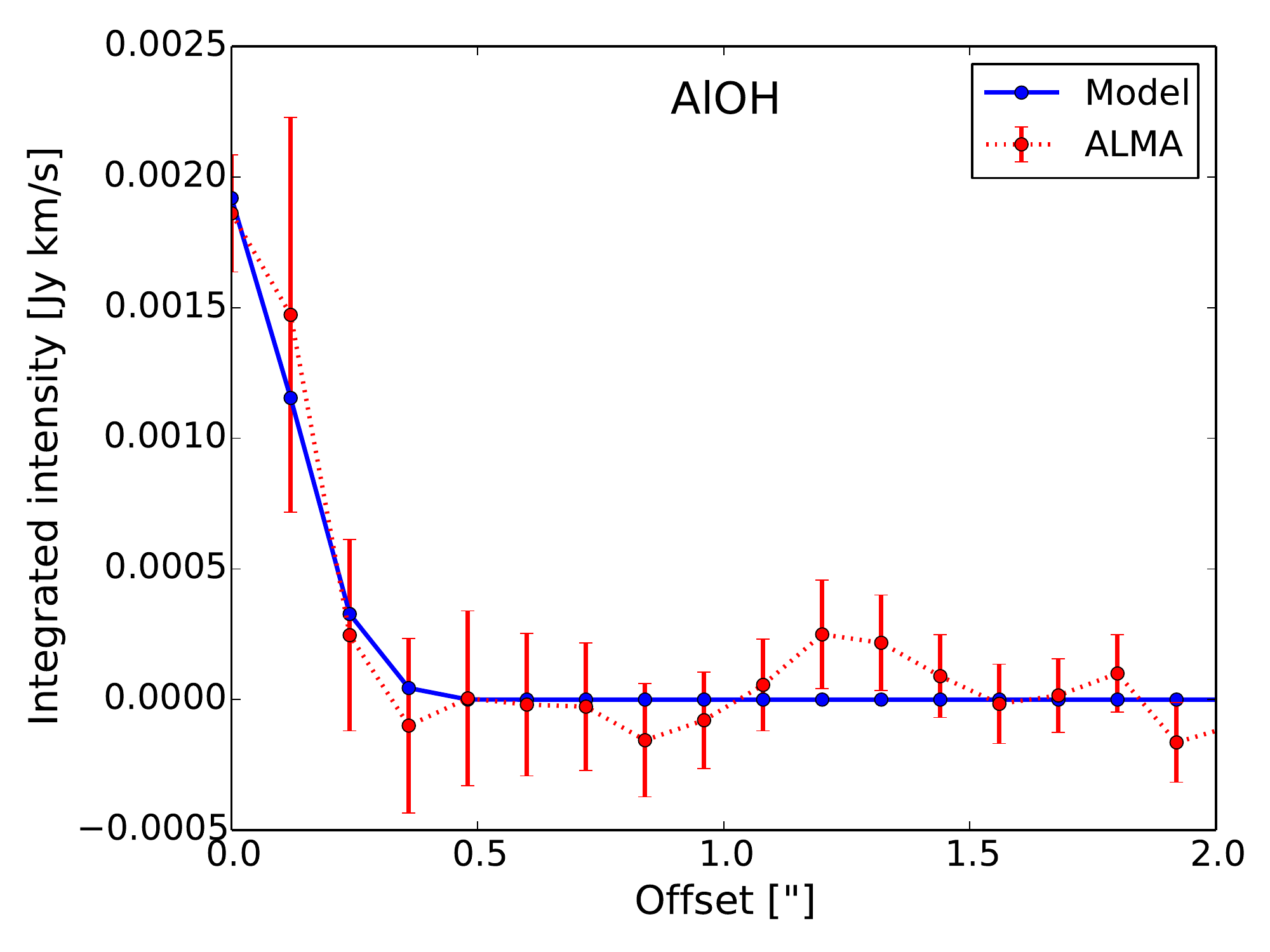}}}
    \end{minipage}
    \hfill
\begin{minipage}[t]{.32\textwidth}
        \centerline{\resizebox{\textwidth}{!}{\includegraphics[angle=0]{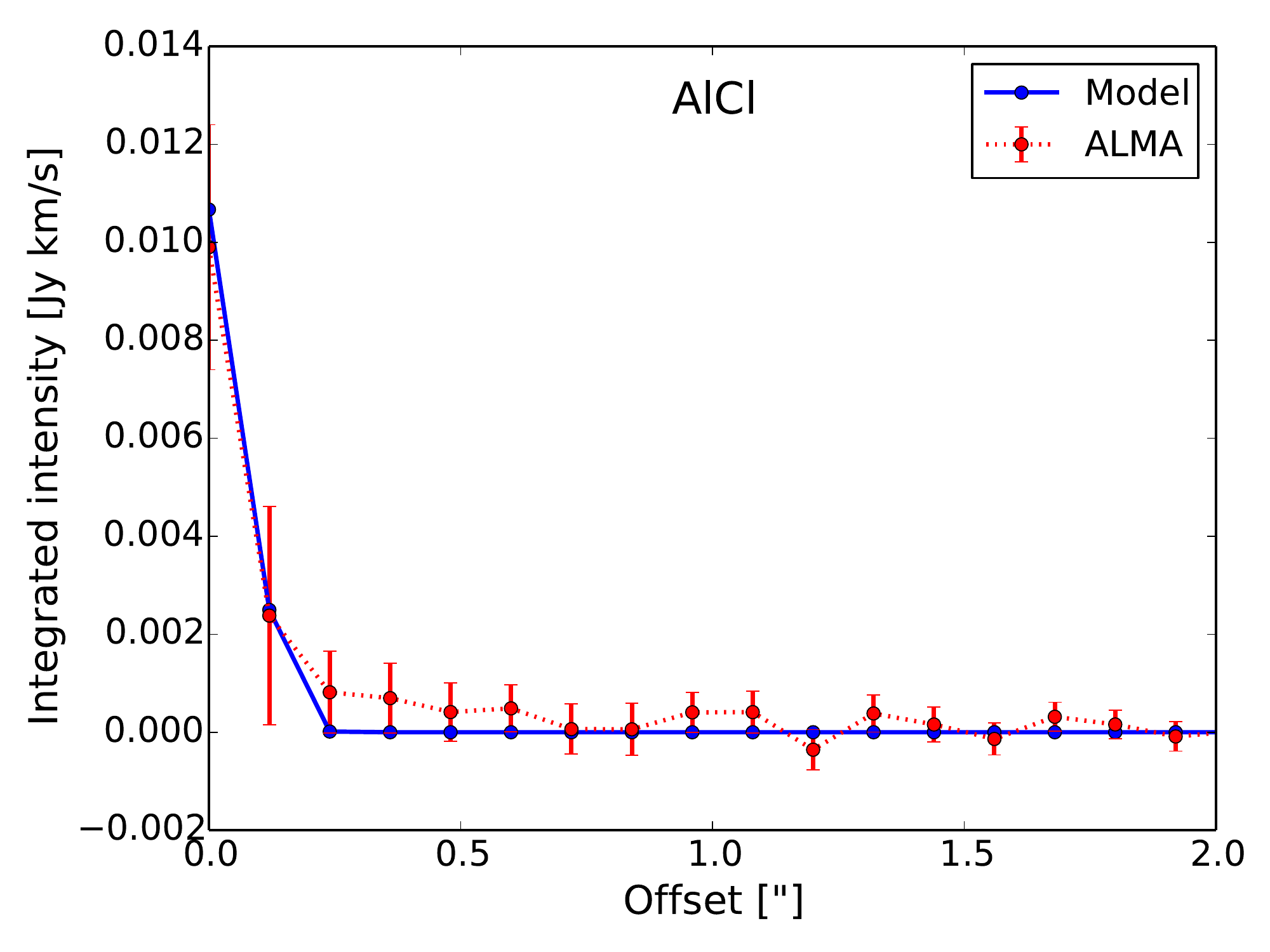}}}
    \end{minipage}
 \caption{Azimuthally averaged flux density in function of the angular distance from the central star for IK~Tau (top) and R~Dor (bottom), respectively. The ALMA data are shown in red, with the vertical bars denoting the uncertainty. The radiative transfer model predictions are displayed in blue (see Sect.~\ref{Sec:Results}).}
 \label{Fig:Al_ang_profile}
\end{figure*}

 \subsection{Infrared continuum emission} \label{Sec:continuum}
 
R~Dor, IK~Tau, and W~Hya have been studied using different spectroscopic, interferometric, polarimetric, and imaging instruments. With the aim to discuss the aluminium fraction locked up into grains, we shortly summarize in this section the results published in the literature that are relevant for this study.

\begin{figure}[htp]
 \begin{center}
  \includegraphics[angle=0,width=0.47\textwidth]{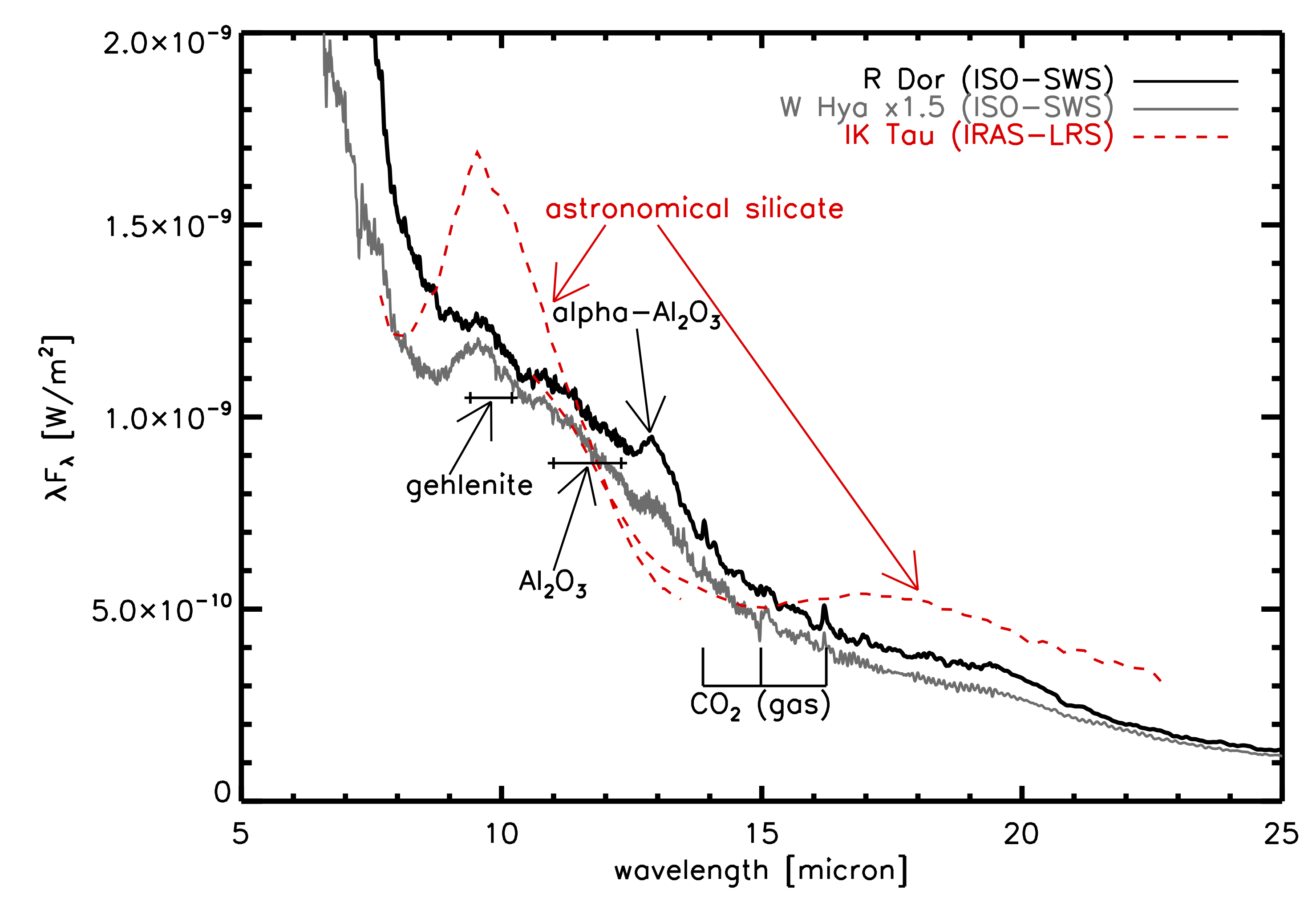}
 \end{center}
\caption{Infrared spectrum of R~Dor (black, full line), W~Hya (grey, full line, scaled with a factor 1.5) and IK~Tau (red, dashed line). Spectral features arising from dust species such as crystalline and amorphous Al$_2$O$_3$, gehlenite (Ca$_2$Al$_2$SiO$_7$), and so-called astronomical silicates \citep[amorphous olivines with different relative magnesium and iron fractions;][]{deVries2010A&A...516A..86D} are indicated.}
\label{Fig:SED_RDor_IKTau_WHya}
\end{figure}

 Both R~Dor and W~Hya have been observed with the Short Wavelength Spectrometer (SWS) on board of the Infrared Space Observatory (ISO) using the AOT01 observing mode. R~Dor was observed at so-called speed 1, while for W~Hya both speed 1 and speed 3 were used. The spectral resolution for the speed 3 data is around 700--1000, and is a factor of $\sim$2 worse for the speed 1 data. Interferences in the instrument cause however fringes in the ISO band~3 data  (12-29\,$\mu$m), particularly in the case of the higher resolution observations.
 Unfortunately, no ISO-SWS data are available for IK~Tau, the only infrared spectrum available being  IRAS-LRS data (see Fig.~\ref{Fig:SED_RDor_IKTau_WHya}).

SPHERE/ZIMPOL data were recently obtained for R~Dor and W~Hya \citep{Khouri2016A&A...591A..70K, Ohnaka2016A&A...589A..91O} confirming the earlier result by \citet{Norris2012Natur.484..220N} that a halo of large transparent grains is present at $\sim$1.5--3\,\Rstar. Estimated grain sizes are around 0.3-0.5\,\mic\ and Fe-free silicates or \Al2O3\ are put forward as the possible dust species that produce the scattered light.

\citet{Zhao2011A&A...530A.120Z} monitored W~Hya with MIDI/VLTI (8--13\,\mic, spectral resolution of 30) during nearly three pulsation cycles. From the spectrally dispersed visibility data, the authors concluded that the Fe-rich silicate emission seen around 10\,\mic\ in the ISO-SWS data arises from a region fairly far away from the star, at radii larger than 30 times the photospheric radius. Using image reconstruction techniques, \citet{Zhao2015PASP..127..732Z} were able to spatially resolve the dust formation region close to the star and deduced that the data signal the presence of a dust layer of presumably amorphous aluminium oxide (\Al2O3) at $\sim$2\,\Rstar. The observations suggest that the formation of amorphous \Al2O3\ occurs mainly around or after visual minimum \citep{Zhao2012A&A...545A..56Z}.

VLTI/MIDI data are also available for R~Dor. However the fringe-track data do not contain any interferometric signal. Probably the beams were not overlapped correctly, as the data were taken for test purposes. No MIDI data are available for IK~Tau.

To derive the dust chemical composition and shed light on the mass-loss mechanism, the SED of W~Hya together with infrared images and optical scattered light fractions were analysed by \citet{Khouri2015A&A...577A.114K}. The only way to explain all observed signatures, and in particular the amorphous Al$_2$O$_3$ emission in the SED, was by implementing in their radiative transfer model a so-called gravitationally bound dust shell (GBDS) close to the star at around 1.5 to 3 stellar radii. This GBDS is a shell of dust particles for which the ratio of the radiation pressure on the grains to the gravitational attraction (often called the $\Gamma$-factor) is still smaller than 1 so that the dust particles can reside close to the star without being pushed outward. This allows the material to potentially grow in size, and only once their size is larger than 0.1-0.3\,\mic, $\Gamma$ can become larger than 1. This idea of a GBDS was recently shown to fit well with the theoretical dynamical wind models of \citet{Hofner2016arXiv160509730H}, who  allowed for the formation of composite grains with an Al$_2$O$_3$ core and a silicate mantle at slightly larger distances than the Al$_2$O$_3$ GBDS shell so to give the wind a kick start. The location of the GBDS coincides with the region where large transparent grains are detected in polarized light images \citep{Norris2012Natur.484..220N, Ohnaka2016A&A...589A..91O}, albeit this does not imply by necessity that the species responsible for the \Al2O3\ emission in the SED are the same ones that cause the stellar light to be scattered. Thus,  it is still possible that the scattering agents are Fe-free silicates, since these can remain fairly cold, and so not produce signature in the thermal infrared.

\begin{figure*}[htp]
\centering\includegraphics[angle=270, width=0.8\textwidth]{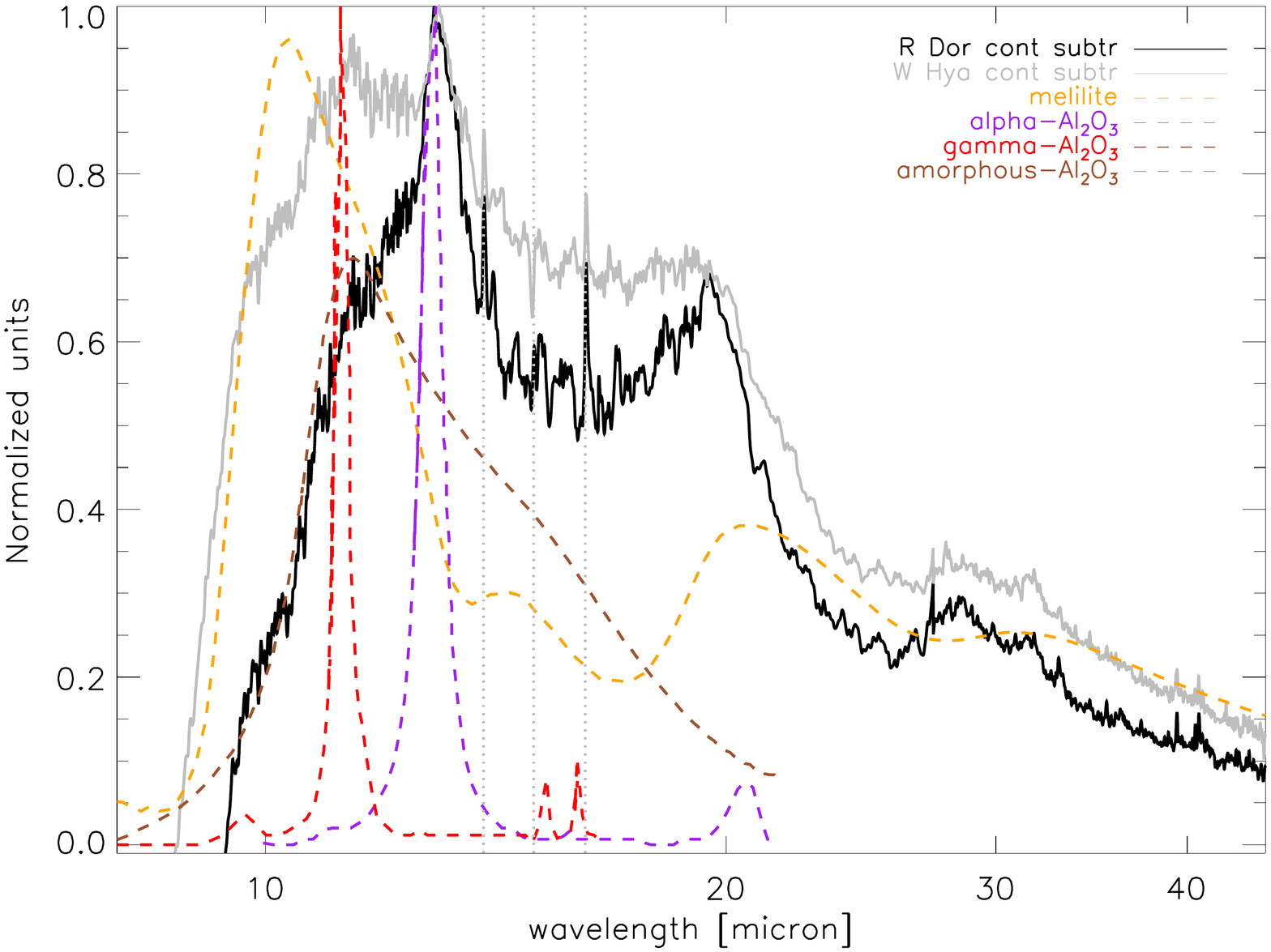}
\caption{Continuum subtracted and normalized SED of R~Dor and W~Hya (in full black and gray lines, respectively). Characteristic (scaled) emission patterns of different dust species are shown in the coloured dashed lines. The gray vertical dotted lines indicate CO$_2$ molecular emission \citep{Justtanont1998A&A...330L..17J, Ryde1999A&A...341..579R}.}
\label{Fig:SED_RDor_WHya}
\end{figure*}

The poor quality of the infrared continuum data of IK Tau led us to decide not to analyse the dust content in this object, but to focus on the dust composition of R~Dor and W~Hya. 
We combined the available IR data for the latter two objects and subtracted the stellar continuum (represented as a black-body with temperature of 3000\,K for R~Dor and of 2500\,K for W~Hya) in order to visualise better their dust spectral features; see Fig.~\ref{Fig:SED_RDor_WHya}. The emission bands of some dust aluminium bearing dust species are plotted, as well as some CO$_2$ and H$_2$O molecular emission features.

\section{Results: Abundances of aluminium-bearing molecules} \label{Sec:Results}

To establish the abundance and distribution of AlO, AlOH, and AlCl, we have modelled the ALMA data using a non-LTE radiative transfer model of the circumstellar envelope (CSE). 
We  use a code based on the ‘Accelerated Lambda Iteration’ (ALI) method \citep{Maercker2008A&A...479..779M}, which allows us to retrieve the global mean molecular density by assuming a 1D geometry.

Collisional excitation rates have not been published for any of these three molecules. For that reason, we have used the values for the HCN-H$_2$ system \citep{Green1974ApJ...191..653G} as substitute for AlOH \citep[cf.][]{Tenenbaum2010ApJ...712L..93T}, and of SiO-H$_2$ \citep{Dayou2006A&A...459..297D} for AlO and AlCl \citep[cf.][]{Kaminski2016A&A...592A..42K}, scaling for the difference in molecular weight. The collisional rates are used in the form they appear in the LAMDA database\footnote{http://home.strw.leidenuniv.nl/$\sim$moldata/} \citep{Schoier2005A&A...432..369S}. To check for the dependency on collisional excitation rates, we also ran models using the CS-H$_2$ cross sections \citep{Turner1992ApJ...399..114T} and found no significant difference in results.
The first 30 rotational levels in the ground vibrational state were considered for AlOH and the first 40 ones for AlO and AlCl. Our treatment of the radiative transfer does not deal with the (hyper)fine structure components, and is therefore limited to the prediction of rotational lines $N\rightarrow N^\prime$ or $J\rightarrow J^\prime$ with $\Delta\,N$ or $\Delta\,J$ equal to 1. Therefore, all levels treated are described by one quantum number, $N$ or $J$. The final line profile is computed by splitting the predicted total intensity over the different (hyper)fine components according to their relative quantum-mechanical line strength, $S$, as given in CDMS\footnote{https://www.astro.uni-koeln.de/cdms/catalog} \citep{Muller2001A&A...370L..49M, Muller2005JMoSt.742..215M}. This approach is only valid under LTE conditions.

For an angular diameter of 20\,mas and a distance of 260\,pc, the stellar radius of IK~Tau is $\sim$3.8$\times$10$^{13}$\,cm, while the angular diameter of 57\,mas and the distance of 60\,pc translates into a stellar radius of $\sim$2.5$\times$10$^{13}$\,cm for R~Dor. The gas kinetic temperature is assumed to follow a power law
\begin{equation}
T_{\rm{gas}} = T_{\rm{\star}}\left(\frac{\Rstar}{r}\right)^{-\epsilon}\,,
\end{equation}
and the gas velocity profile is parametrized using a $\beta$-type law
\begin{equation}
v_{\rm{gas}} = v_0 + (v_\infty - v_0)\left(1 - \frac{r_0}{r}\right)^\beta\,,
\end{equation}
with $r$ the distance to the star, $v_\infty$ the terminal velocity, and $v_0$ the velocity at radius $r_0$ at which the wind is launched and which is set to the local sound speed. The gas velocity within the dust condensation radius, $r<r_0$, is assumed to be 2\,km/s. The turbulent velocity for both stars is assumed to be 1\,km/s. The gas density, $\rho(r)$, is calculated from the equation of mass conservation
\begin{equation}
\Mdot = 4 \pi r^2 v(r) \rho(r)\,.
\end{equation}
The main input parameters for R~Dor and IK~Tau are listed in Table~\ref{Table:input_RT}, following \citet{Maercker2016A&A...591A..44M}. These properties are simplified estimates, for example the velocity structure inside the dust formation radius is complex, but are adequate for 1D modelling in the regions where these Al compounds have been observed.
For all molecular species, the fractional abundance structure with respect to H$_2$ was varied until a fit with the ALMA azimuthally averaged flux density was achieved (see Fig.~\ref{Fig:Al_ang_profile}). In some cases, the fractional abundance declines according to a Gaussian profile centred on the star, and hence is described by following equation
\begin{equation}
f(r) = f_0 \exp\left(-\left(\frac{r}{r_e}\right)^2\right)\,,
\end{equation}
with $f_0$ the initial abundance and $r_e$ the $e$-folding radius.

\begin{table}[htp]
\begin{center}
\caption{Input parameters for the non-LTE radiative transfer modeling}
\label{Table:input_RT}
\begin{tabular}{lrr}
\hline
 & R Dor & IK Tau \\
 \hline
 \Tstar [K] & 2400 & 2100 \\
 $\epsilon$ & 0.65 & 0.6 \\
 $v_\infty$ [km/s] & 5.7 & 17.5 \\
 $\beta$ & 1.5 & 1.2\\
 $r_0$ [cm] & 5.3$\times10^{13}$ & 2.38$\times10^{14}$ \\
 \Mdot [\Msun/yr] & 1.6$\times10^{-7}$ & 5$\times10^{-6}$ \\
 L [\Lsun] & 6500 & 7700 \\
 \hline
\end{tabular}
\end{center}
\end{table}

\begin{figure*}[htp]
\begin{minipage}[t]{.48\textwidth}
        \centerline{\resizebox{\textwidth}{!}{\includegraphics[angle=0]{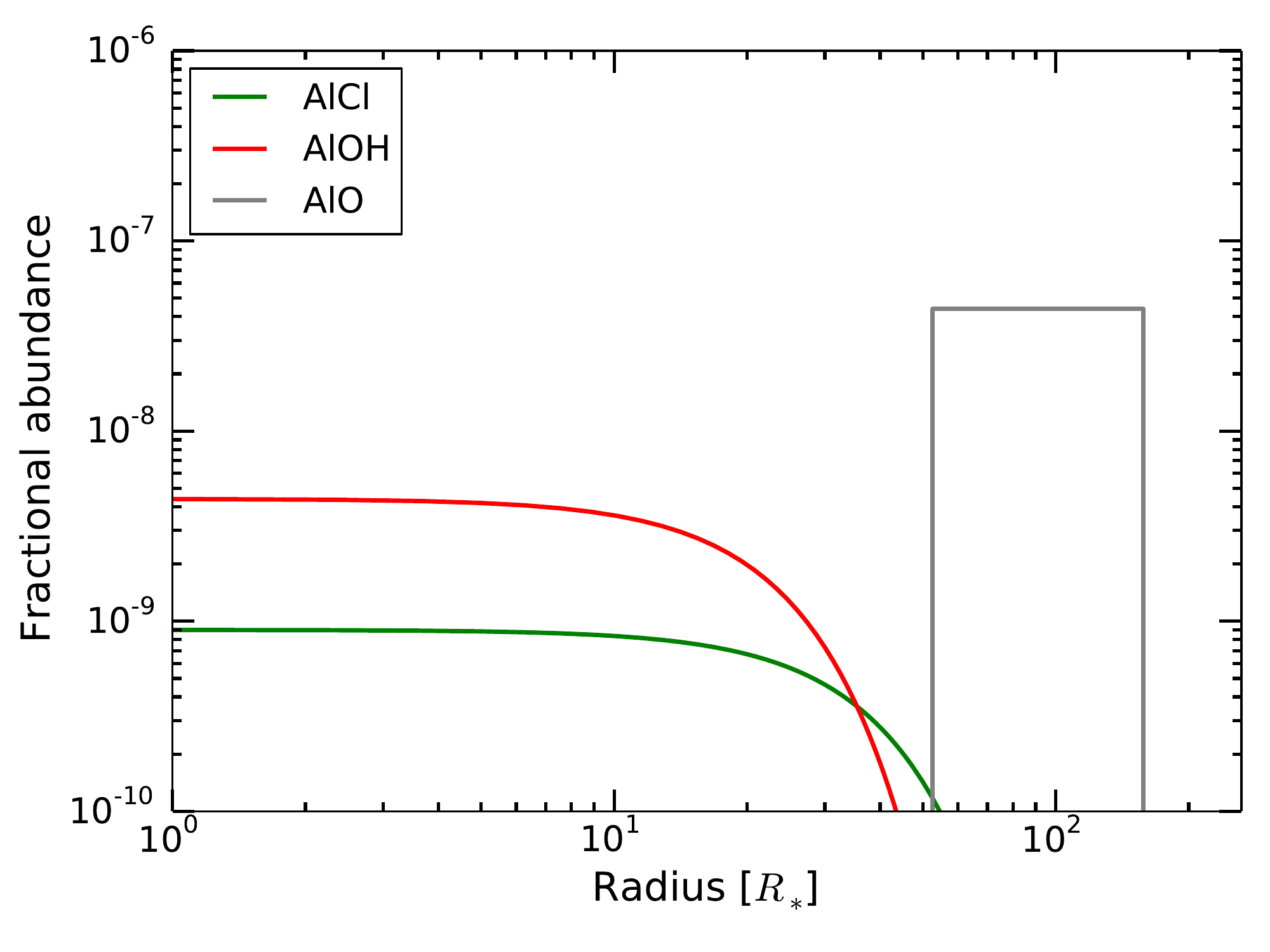}}}
    \end{minipage}
    \hfill
\begin{minipage}[t]{.48\textwidth}
        \centerline{\resizebox{\textwidth}{!}{\includegraphics[angle=0]{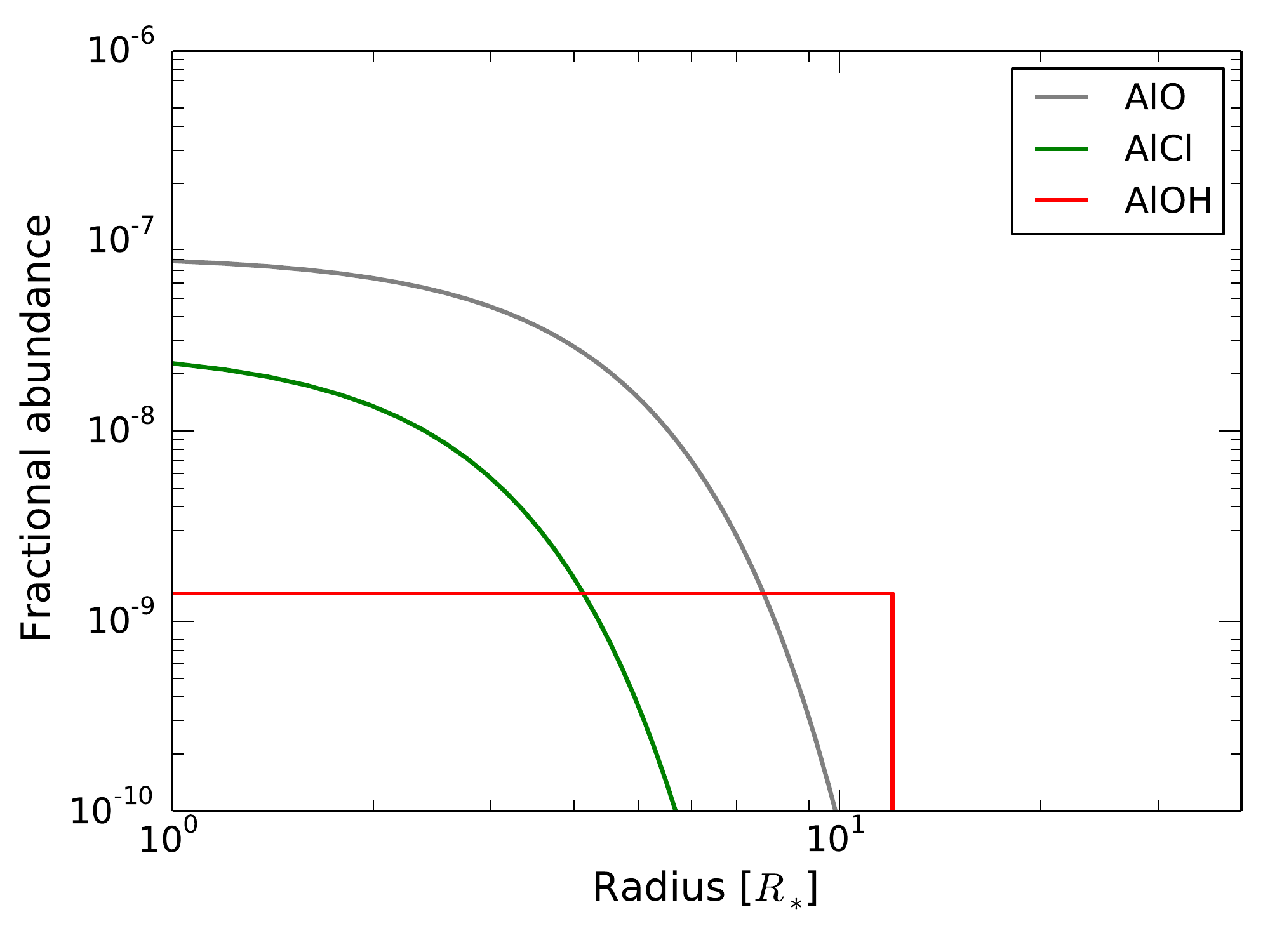}}}
    \end{minipage}
 \caption{Retrieved fractional abundance structure [AlO/H$_2$], [AlOH/H$_2$], and [AlCl/H$_2$] for IK~Tau (left) and R~Dor (right).}
 \label{Fig:retrieved}
\end{figure*}

\begin{figure*}[htp]
\begin{minipage}[t]{.48\textwidth}
        \centerline{\resizebox{\textwidth}{!}{\includegraphics[angle=0]{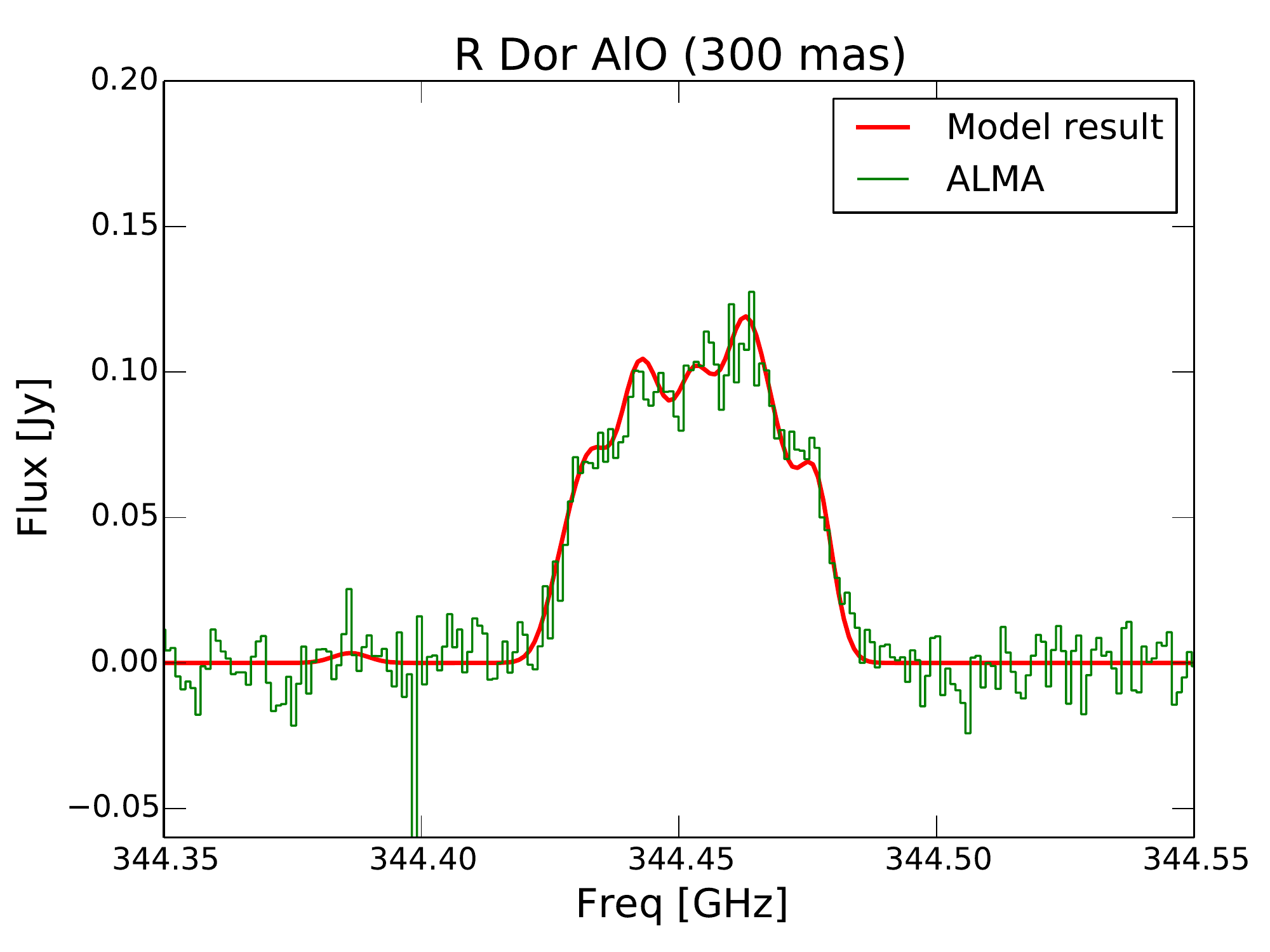}}}
    \end{minipage}
    \hfill
\begin{minipage}[t]{.48\textwidth}
        \centerline{\resizebox{\textwidth}{!}{\includegraphics[angle=0]{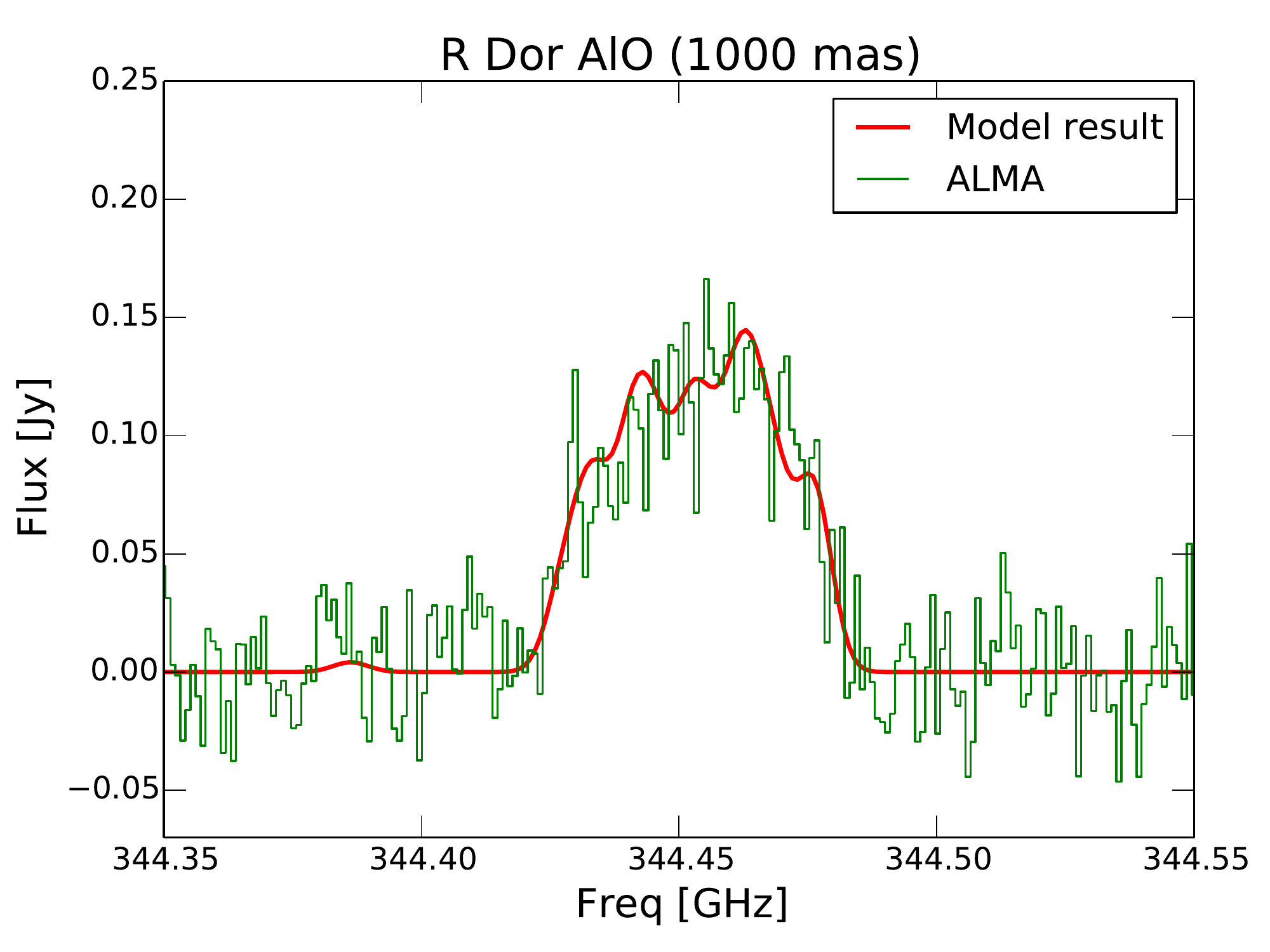}}}
    \end{minipage}
 \caption{Comparison between the extracted ALMA AlO spectra of R~Dor (green) and the predicted line profile (red) for an AlO abundance structure as given in Fig.~\ref{Fig:retrieved} for the 300\,mas (left) and 1000\,mas (right) extraction aperture.}
 \label{Fig:RDor_AlO_spectra_model}
\end{figure*}

The retrieved fractional abundance structures are shown in Fig.~\ref{Fig:retrieved}; the fit to the ALMA data in Fig.~\ref{Fig:Al_ang_profile}. Clearly both stars show a very different Al-chemistry. For IK~Tau, AlOH could be modelled assuming a Gaussian distribution with a peak central abundance of $4.4\times10^{-9}$ and an $e$-folding radius, $r_e$, of 22\,\Rstar. As could be deduced already from Fig.~\ref{Fig:ALMA_data} and Fig.~\ref{Fig:channel_IKTau_AlO_smooth}, the AlO emission might be clumpy not showing a centrally peaked abundance structure. The data could be modelled assuming a constant abundance of 4.4$\times$10$^{-8}$ from 50--160\,\Rstar. If the AlCl identification is correct (see Sect.~\ref{Sec:ALMA_observations}), then the data are compatible with a Gaussian distribution with peak central abundance of $9\times10^{-10}$ and $r_e$ of 40\,\Rstar.
The AlOH emission in R~Dor can be approximated with a constant slab with abundance of 1.4$\times$10$^{-9}$ from the stellar surface out to 12\,\Rstar. The AlO and AlCl distribution in R~Dor follow a Gaussian distribution with peak central abundance of $8.4\times10^{-8}$ and $2.5\times10^{-8}$ and $e$-folding radius of 3.8\,\Rstar\ and 2.4\,\Rstar, respectively. The good correspondence between the observed and predicted emission of AlO in R~Dor is shown in Fig.~\ref{Fig:RDor_AlO_spectra_model}. 

While we can trace the fractional abundance above $1\times10^{-9}$ within $\sim$150\,\Rstar\ for IK Tau, the ALMA Al-data of R~Dor only cover the first 25\,\Rstar.
The factor $\sim$6 difference in spatial extent is in line with the factor $\sim$50 in mass-loss rate (and hence sensitivity to some particle density assuming a density law which is proportional to $r^{-2}$).

Assuming a solar aluminium content of Al/H = 3$\times$10$^{-6}$, our results imply that for both stars only a small fraction of aluminium ($\la$2\%) is locked up in AlO, AlOH, and AlCl.


\section{Discussion} \label{Sec:Discussion}

The analysis of the ALMA data leads to a direct determination of the mean fractional abundances of AlO, AlOH, and AlCl in the inner winds of R~Dor and IK~Tau. These results can be compared quantitatively to the theoretical model predictions solving the chemical network for IK~Tau by \citet{Gobrecht2016A&A...585A...6G}, and hence bring about the first qualitative assessment of this kind of dynamical-chemical models (Sect.~\ref{Sec:disc_Al_gas}). The amount of aluminium locked up in gaseous species turns out to be low, leaving ample of room for aluminium to be incorporated into dust grains. However, as we will discuss in Sect.~\ref{Sec:disc_Al_dust}, the estimated grain temperatures for aluminium oxide (\Al2O3) at $\sim$2\,\Rstar\ is high suggesting that the grains are annealed and hence have a crystalline lattice structure.  This outcome  implies a tension in the interpretation of the 11\,\mic\ spectral feature, present in the SED and interferometric data. We hypothesize that large gas-phase aluminium oxide clusters are the cause of the 11\,\mic\ feature, and explain how these large clusters can be formed (Sect.~\ref{Sec:disc_Al_clusters}).

\subsection{The gas-phase aluminium budget} \label{Sec:disc_Al_gas}

\begin{figure}[htp]
\begin{minipage}[t]{.4\textwidth}
        \centerline{\resizebox{\textwidth}{!}{\includegraphics{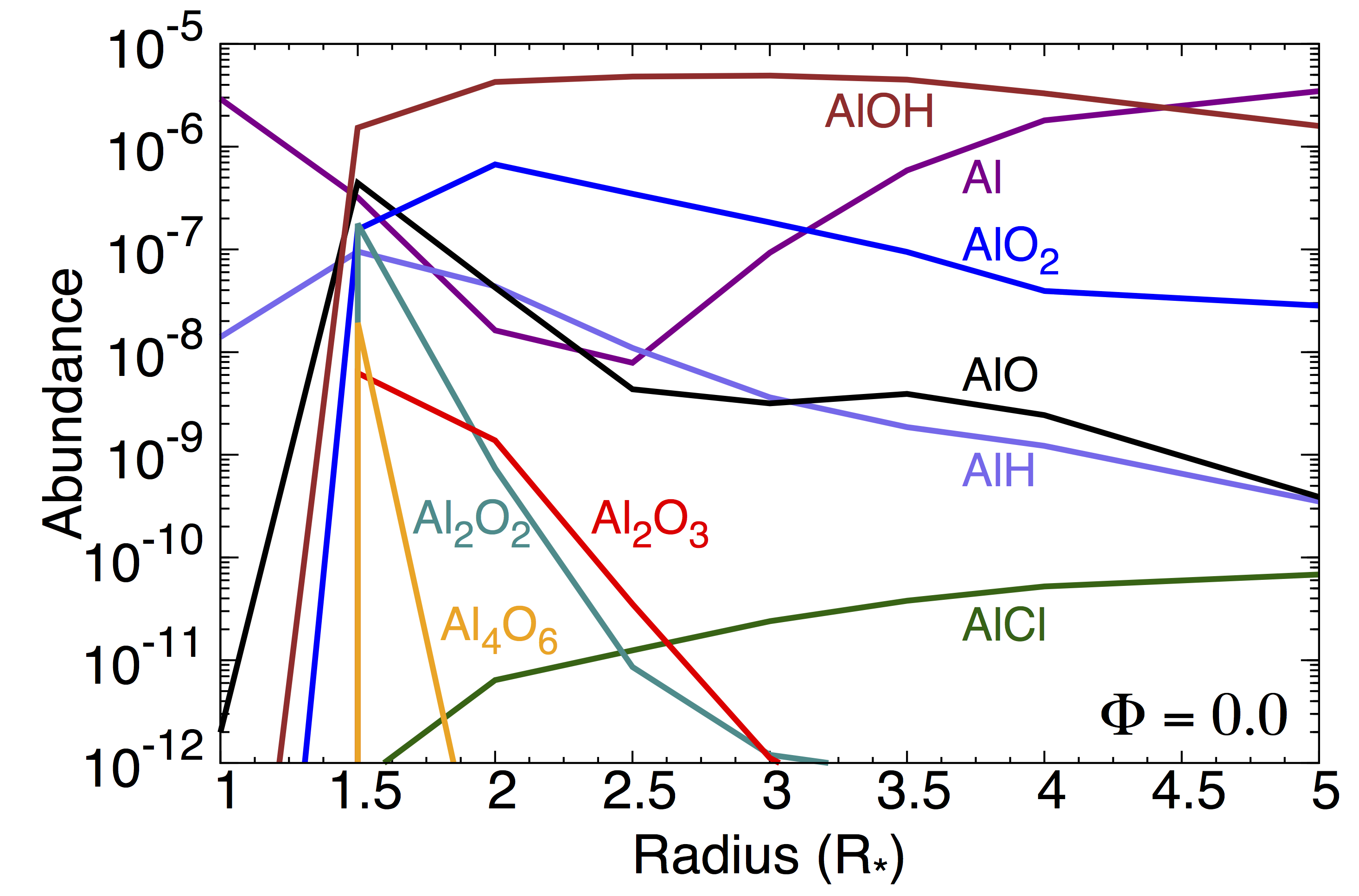}}}
   \end{minipage}
    \hfill
\begin{minipage}[t]{.4\textwidth}
        \centerline{\resizebox{\textwidth}{!}{\includegraphics{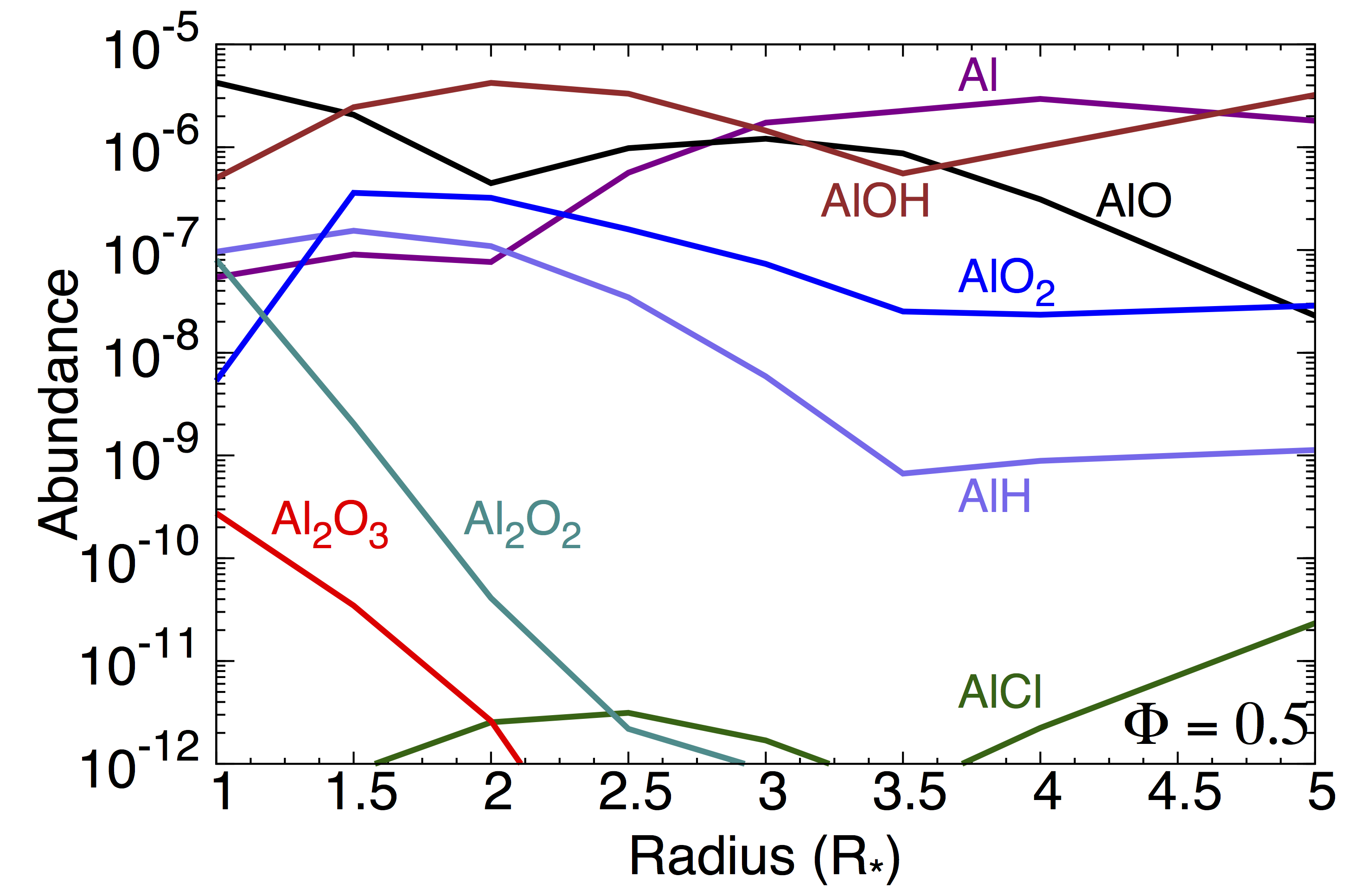}}}
    \end{minipage}
        \caption{Predicted abundances of aluminium-bearing molecules with respect to the total gas number density as a function of radius. Chemical models are based on the pulsation-induced shock dynamics presented by \citet{Gobrecht2016A&A...585A...6G}. Model simulations for a star resembling IK~Tau at phase $\phi$=0.0 (top) and $\phi=0.5$ (bottom).}
\label{Fig:chem_Al_radius}
\end{figure}

\begin{figure}[htp]
\begin{minipage}[t]{.4\textwidth}
        \centerline{\resizebox{\textwidth}{!}{\includegraphics{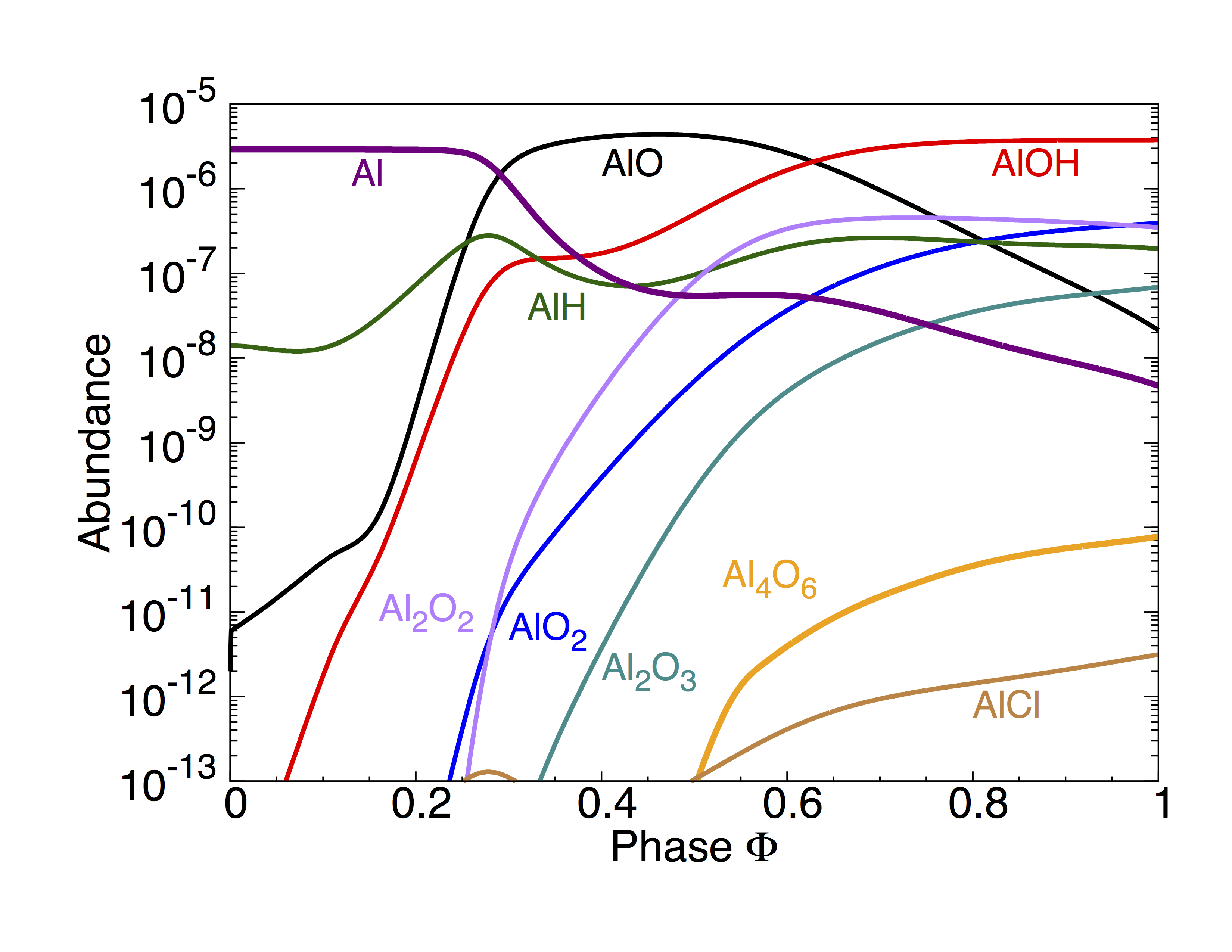}}}
   \end{minipage}
    \hfill
\begin{minipage}[t]{.4\textwidth}
        \centerline{\resizebox{\textwidth}{!}{\includegraphics{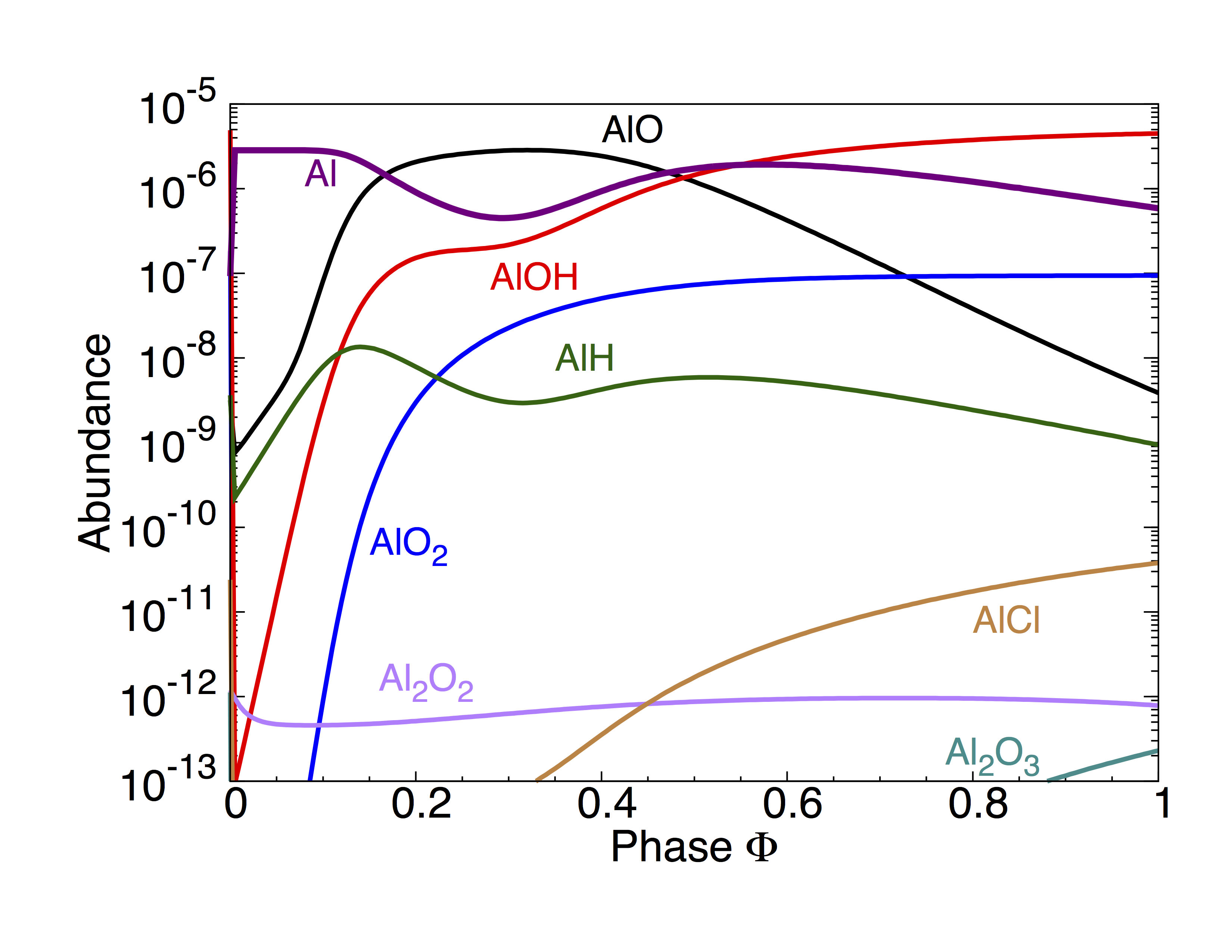}}}
    \end{minipage}
    \hfill
\begin{minipage}[t]{.4\textwidth}
        \centerline{\resizebox{\textwidth}{!}{\includegraphics{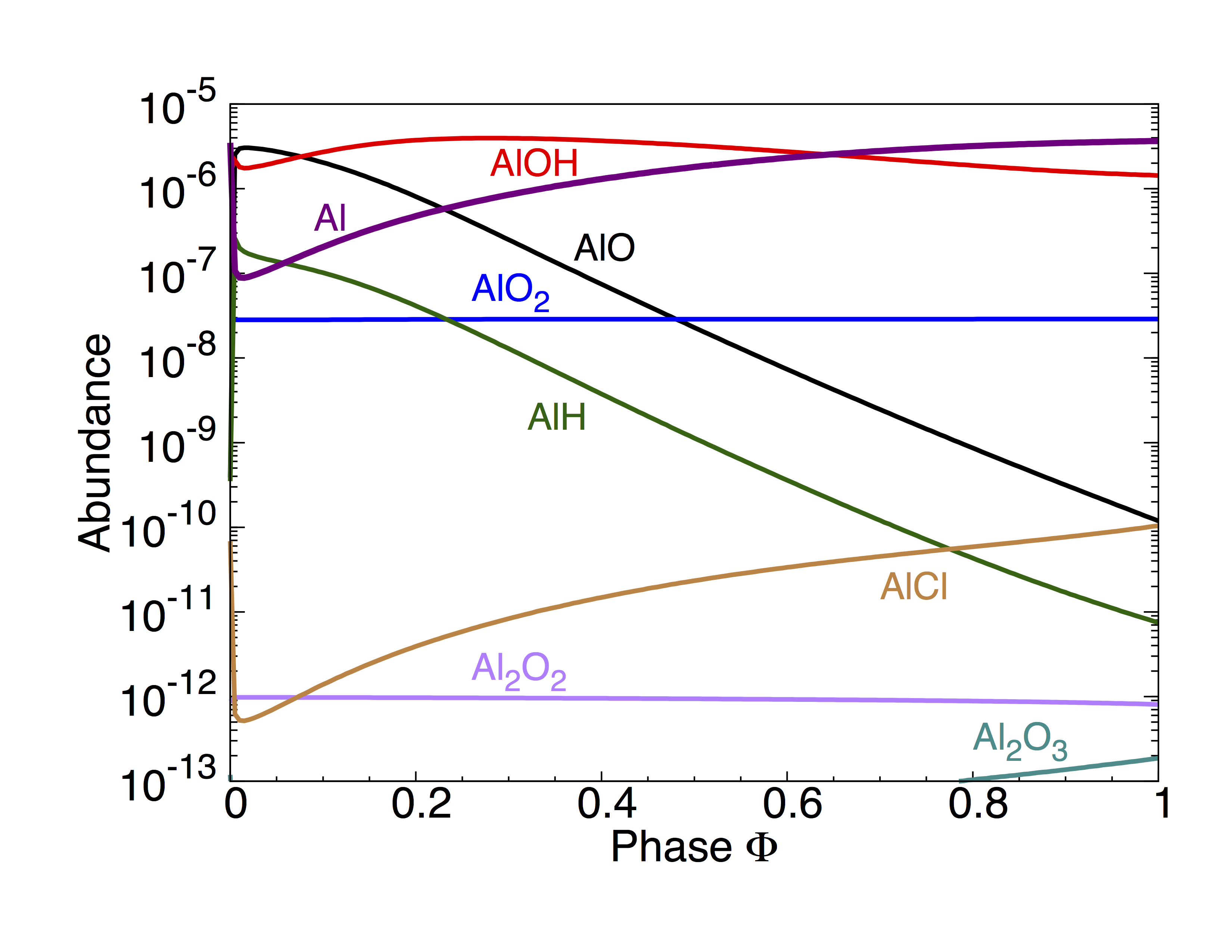}}}
    \end{minipage}
        \caption{Predicted abundances of aluminium-bearing molecules with respect to the total gas number density as a function of phase $\phi$ for three specific radii (top: 1\,\Rstar, middle: 3\,\Rstar, bottom: 5\,\Rstar). Chemical models are based on the pulsation-induced shock dynamics presented by \citet{Gobrecht2016A&A...585A...6G} for a star resembling IK~Tau.}
\label{Fig:chem_Al_phase}
\end{figure}

The deduced AlO, AlOH, and AlCl abundances of IK~Tau can be compared directly to the chemical predictions by \citet{Gobrecht2016A&A...585A...6G}. They modelled the synthesis of molecules and dust in the inner wind of IK~Tau ($r<5$\,\Rstar) by considering the effect of periodic shocks induced by the stellar pulsation. The non-equilibrium chemistry in the shocked gas-layers was followed between 1\,\Rstar\ and 10\,\Rstar\ by solving a chemical kinetic network including 100 species and 424 reactions. The condensation of dust is described by a Brownian formalism using the simplifying assumption that the gaseous dimers of forsterite (Mg$_2$SiO$_4$), enstatite (MgSiO$_3$) and aluminia (\Al2O3) are the initial dust seeds and considering coalescence and coagulation of these dimers to form the dust grains. 

The aluminium species in the network of \citet{Gobrecht2016A&A...585A...6G} include Al, AlH, AlO, AlOH, AlO$_2$, AlCl, Al$_2$, and Al$_2$O. The aluminium dust precursors are described by Al$_2$O$_2$, Al$_2$O$_3$, and Al$_4$O$_6$. The predicted abundances of the aluminium species as function of radius and phase $\phi$\footnote{The pulsation phase is defined as the decimal part of $(t-T_0)/P$, with $T_0$ the epoch of the start of the pulsation cycle and $P$ the period.} are shown in Fig.~\ref{Fig:chem_Al_radius} and Fig.~\ref{Fig:chem_Al_phase}, respectively. We note that Al$_2$ and Al$_2$O have very low abundances and therefore  do not appear in the plots. 
The plots in Fig.~\ref{Fig:chem_Al_radius} can be interpreted as a snapshot of the inner wind just after the passage of a shock (at phase $\phi=0.0$) and at half of the pulsation cycle (phase $\phi=0.5$), assuming shock fronts every 0.5\,\Rstar. A full pulsation cycle corresponds to phase 1.0. Phase 0.0 corresponds also to a snapshot after a full pulsation cycle. However, at this point, the gas densities have been rescaled from the previous radial position and just have been shocked again at the new radial position \citep[see description in][]{Gobrecht2016A&A...585A...6G}.

For a period of 470~days and with $T_0$ at JD$\sim$2453025 \citep{Matsumoto2008PASJ...60.1039M}, the ALMA data of IK~Tau are obtained around $\phi = 0.98$.
The methodology applied by \citet{Gobrecht2016A&A...585A...6G} implies that a \textit{relative} comparison between the retrieved and predicted molecular abundance stratifications is best to be done using  the first panel of Fig.~\ref{Fig:chem_Al_radius}, while the absolute scaling can be seen in Fig.~\ref{Fig:chem_Al_phase}. Taking the predicted abundances at 5\,\Rstar\ (bottom panel in Fig.~\ref{Fig:chem_Al_phase}) at phase $\phi=0.98$ as representative values for the comparison with the profiles retrieved from the ALMA data (left panel in Fig.~\ref{Fig:retrieved}), we can see that both the observations and chemical models predict a higher abundance for AlOH than for AlO. However, the retrieved AlOH abundance at 5\,\Rstar\ is 2 orders of magnitudes lower than predicted. The ALMA data indeed show that the fractional AlO abundance ($r<50$\,\Rstar) is very low ($<10^{-10}$) in accord with the predictions, but remarkably we see that an active Al-chemistry is taking place further out in the wind yielding an increase in AlO emission. We derive for IK~Tau a mean AlO fractional abundance of $\sim$5$\times$10$^{-8}$ between 50--150\,\Rstar, with the emission mainly arising from clumps at that radial distance. The increase in AlO abundance coincides with a strong decrease in AlOH abundance suggesting that AlOH is converted to AlO. Another possibility is that desorption or sputtering release AlO from grains back into its gaseous form. If no line-blend is blurring the AlCl identification, then the ALMA data show that the AlCl abundance in IK~Tau is a factor $\sim$40 higher than predicted.

The stellar angular diameter of IK~Tau is $\theta_D\sim20$\,mas \citep{Decin2010A&A...516A..69D}. The inner radius of dust detected by \citet{Dehaes2007MNRAS.377..931D} and inferred by \citet{Richards2012A&A...546A..16R} is at 6--10\,$R_{\star}$, but these observations were not sensitive to the potential presence of transparent grains closer to the star. The ALMA beam of $\sim$0\farcs12$\times$0\farcs15 is hence too large to confirm the predicted strong radial variation in molecular abundance within 5\,\Rstar\ (as visible in Fig.~\ref{Fig:chem_Al_radius}). The angular diameter of R~Dor is much larger, $\theta_D\sim57$\,mas \citep{Bedding1997MNRAS.286..957B, Norris2012Natur.484..220N, Khouri2016A&A...591A..70K}. Although a direct comparison with Fig.~\ref{Fig:chem_Al_radius}--\ref{Fig:chem_Al_phase} should be done with care --- since the shock velocities are expected to be smaller for R~Dor --- our ALMA data do not corroborate  any strong radial variation in AlO and AlOH abundance within the first 10 stellar radii. In addition, the AlCl abundance decreases with radial distance in contrast to the chemical network predictions. 

The resolution of our  ALMA observations does allow us to trace the molecular abundances at  the smallest radii at which dust has been detected  (being around 6–-10\,$R_{\star}$ for IK~Tau and around 2\,$R_{\star}$ for R~Dor). AlO and AlOH are directly associated with the nucleation of alumina and their retrieved abundances are indeed much lower than in the pure gas-phase models of \citet{Gobrecht2016A&A...585A...6G}. But clearly, not all available AlO and AlOH molecules participate in that dust formation process. In addition, the ALMA results on R~Dor prove that a second phase of dust formation, proposed to occur slightly beyond the GBDS, is (almost) not consuming AlO. This supports the scenario suggested by \cite{Hofner2016arXiv160509730H} that the growth of silicates (potentially on top of already existing alumina grains) is a trigger for the wind to get launched.

It would be of interest to monitor the changes in abundance structure in function of the pulsation phase and confront them with the predictions shown in Fig.~\ref{Fig:chem_Al_phase}. This kind of data could then be used to benchmark and improve current forward chemistry models that implement gas dynamics. We note that \citet{Kaminski2016A&A...592A..42K} have reported some variability in the AlO emission of the low mass-loss rate Mira star $o$~Cet, but currently no link with the pulsation cycle is apparent. 

Although the current comparison points toward improvements to be made in the chemical model predictions, a conclusion from the ALMA data in combination with these models is that the fraction of aluminium that is locked up in simple aluminium-bearing molecules seems small. I.e., the combined fractional abundance of Al-molecules which are thought to form easily in the inner wind region is only $\sim$5$\times$10$^{-8}$. Even more, since the formation of the more complex \Al2O3\ and its dimer (\Al2O3)$_2$ is described by  termolecular recombination and hence requires high densities, these higher order molecules will also only consume a tiny fraction of the available aluminium (see Fig.~\ref{Fig:chem_Al_phase}, where one can see that the alumina dimers only form at 1\,\Rstar\ and at late phases); i.e.\ these larger molecules too can be neglected as major sink of aluminium.

The most difficult gaseous aluminium form to constrain is that of atomic aluminium. The atomic aluminium abundance is predicted to be high when the parcel of material is close to the star ($r<2.5$\,\Rstar) for phases $\phi>$0.2 (see Fig.~\ref{Fig:chem_Al_phase}). Exactly at the shock front, phase $\phi=0$, the high temperature ensures a high Al abundance, but  as a consequence of the declining temperature for later phases the atomic Al abundance decreases rapdily. Beyond 3\,\Rstar, one can note a transformation of molecular in atomic Al due to the decreasing gas densities with distance from the star. We have performed an in-depth study of the high spectral resolution optical Mercator/Hermes \citep{Raskin2011A&A...526A..69R} data of IK~Tau and W~Hya (no observations are available for R~Dor) to check for the presence of atomic Al in the outer atmosphere/inner wind of both stars. Molecular line veiling complicates the analysis of the data. We could not identify either absorption or emission atomic Al lines. This does not imply that no atomic Al is present, but at least the quantity is too low for it to be detected amongst the molecular lines. An analogous conclusion was drawn by \citet{Kaminski2016A&A...592A..42K} for the low mass-loss rate AGB star $o$~Cet, for which reason they concluded that Al is locked in molecules and dust.

This ensemble of considerations supports the conclusion that only a minor fraction of Al resides in Al-bearing molecules and that  the atomic Al abundance is difficult to constrain. An important outcome from these ALMA data is that the aluminium condensation process, considered to be the first condensation cycle to start in AGB stars \citep{Tielens1990fmpn.coll..186T},  is for sure not fully efficient since the gaseous precursors of the aluminium grains are detected in and beyond the dust formation region.

\subsection{Aluminium depletion into grains} \label{Sec:disc_Al_dust}

Presolar grains prove that aluminium oxide grains are formed in oxygen-rich AGB stars.  The circumstellar \Al2O3-grains are relatively abundant in meteorites \citep{Nittler1997ApJ...483..475N}, albeit perhaps not as abundant as one would expect if all aluminium would condense in these grains \citep{Alexander1997AIPC..402..567A}. This result from meteorite studies now gets some support by our analysis of the ALMA data, confirming that a fraction of the aluminium is still present in its gaseous form after its passage through the dust formation region.
At least for a few silicate grains, the meteoretic studies show that the aluminium oxides are at the core of the silicate grain suggesting  that \Al2O3\ is the first solid to form in the CSE \citep{Nittler2008ApJ...682.1450N, Vollmer2006LPI....37.1284V}. Both crystalline and amorphous \Al2O3\ enclosures have been detected \citep{Stroud2004Sci...305.1455S}. The variation in the crystal structure and Ti content of these presolar grains demonstrate that \Al2O3\ can condense in the absence of TiO$_2$ seed clusters but that Ti may be important in determining the lattice structure.

\begin{figure}[htp]
\includegraphics[width=0.48\textwidth]{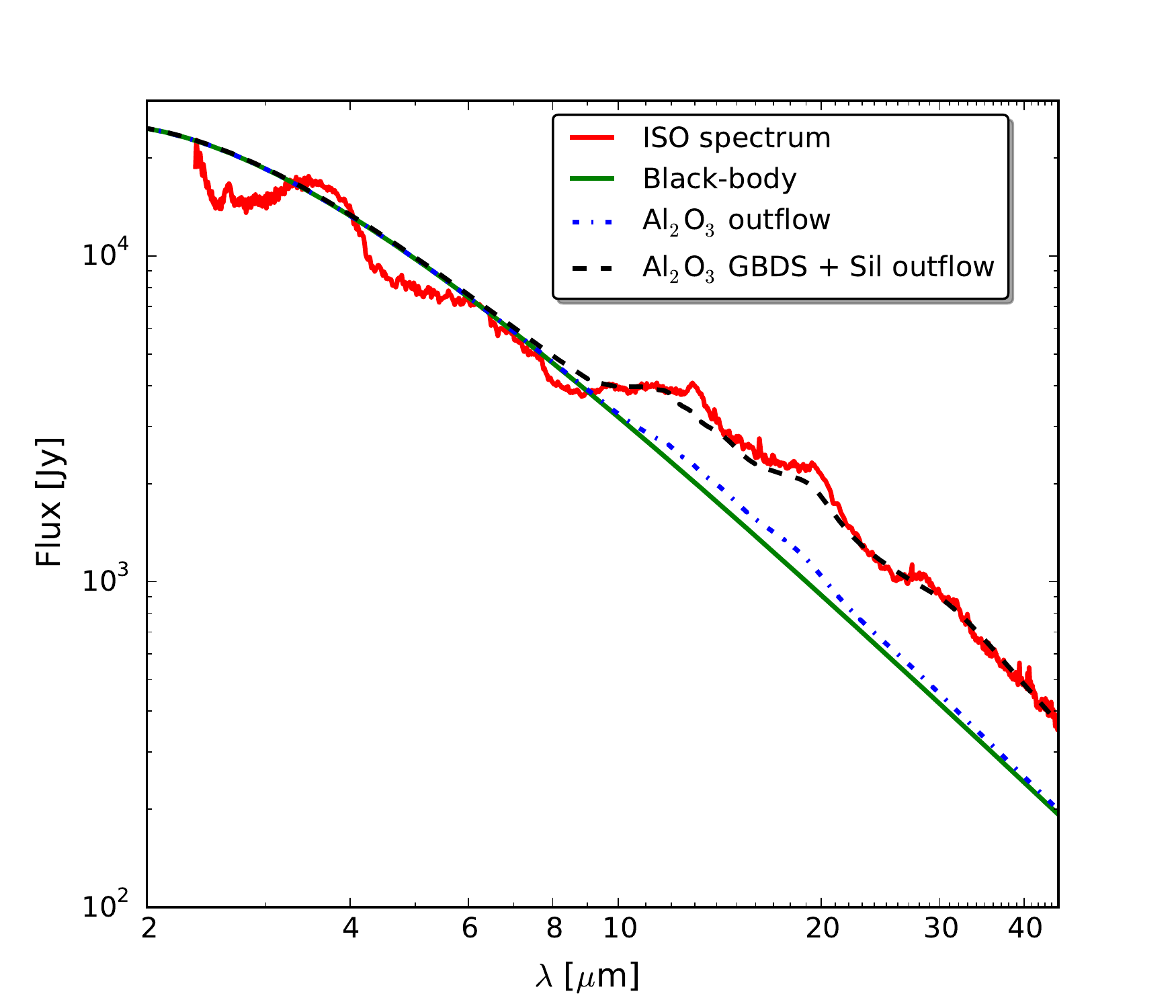}
\caption{Infrared SED of R~Dor. ISO data are shown in red; the blackbody represents the stellar emission at a temperature of 3000\,K, the dotted line is model  assuming a 100\% condensation of aluminium into amorphous \Al2O3\ that resides in the stellar outflow, while the dashed line represents a model fit for amorphous \Al2O3\ being present in the so-called GBDS and in the stellar outflow. 
}
\label{Fig:RDor_GBDS}
\end{figure}

The SED analysis of R~Dor and W~Hya turned out to be challenging in terms of the aluminium content \citep{Khouri2015A&A...577A.114K, KhouriPhD}. Based on constraints from
gas-phase models and solar abundances, Khouri et al.\ were unable to model the SEDs of both stars in the spectral region around 11\,\mic\ when considering aluminium condensates only to be present in the stellar outflow (see dotted line in Fig.~\ref{Fig:RDor_GBDS}). One can fit the broad 11\,\mic\ feature in the SED by invoking the presence of a gravitational bound dust shell (GBDS; see Sect.~\ref{Sec:continuum} for a more detailed explanation on this terminology) of amorphous \Al2O3\ grains at $\sim$2\,\Rstar\ (see dashed line in Fig.~\ref{Fig:RDor_GBDS}). The GBDS can contain both small and large \Al2O3\ grains. The model for R~Dor (W~Hya) requires an Al$_2$O$_3$ dust mass in the GBDS of $\sim 5.7 \times 10^{-10}$~M$_\odot$ ($\sim 1.5 \times 10^{-9}$~M$_\odot$), which would correspond to a gas mass of $7.5 \times 10^{-6}$~M$_\odot$ ($2 \times 10^{-5}$~M$_\odot$) if one would assume full aluminium condensation.
If this is compliant (or not) with the total aluminium content is difficult to assess since the region where the GBDS resides is known to be highly variable in density due to the stellar pulsations, with variations up to one order of magnitude \citep{Hofner2016arXiv160509730H}.

Polarized light data confirm also the presence of large grains close to W~Hya and R~Dor at 1.5--3\,\Rstar\ \citep{Norris2012Natur.484..220N, Khouri2016A&A...591A..70K, Ohnaka2016A&A...589A..91O}. From the scattered light data, one can not determine the composition of the grains, and both aluminium oxide grains and cold Fe-free silicates are suggested to be good candidates due to their glassy character (Sect.~\ref{Sec:continuum}).

The GBDS results \citep{KhouriPhD, Khouri2015A&A...577A.114K}, the scattered light data \citep{Norris2012Natur.484..220N, Khouri2016A&A...591A..70K, Ohnaka2016A&A...589A..91O}, the interferometric results \citep{Zhao2015PASP..127..732Z}, and the theoretical models \citep{Hofner2016arXiv160509730H}  build up evidence for a shell of aluminium oxide grains at 1.5 -- 3\,\Rstar. The kinetic temperature in that region is around 1500--2000\,K. Optical constants are measured for amorphous \Al2O3\ grains, but only for wavelengths beyond 0.2\,\mic\ \citep{Koike1995Icar..114..203K} or beyond 7.8\,\mic\ \citep{Begemann1997ApJ...476..199B}. Depending on the replica used for the short-wavelength range and under the assumption of thermodynamic equilibrium,  the estimated temperature of the amorphous aluminina grains at that spatial region would be around 1200--1600\,K \citep{Khouri2015A&A...577A.114K}.  However, amorphous alumina can only exist below 1000\,K \citep{Levin1998PSSAR.166..197L, Levin2005}. At temperatures around 1500\,K, crystalline $\alpha$-alumina can form, while below 1300\,K crystalline $\gamma$-alumina can exist. With the grain temperature in the 1.5--3\,\Rstar\ region being higher than the annealing temperature, this would imply a structural transformation of \Al2O3\ to its crystalline form. The thermal infrared spectrum of R~Dor and W~Hya is not compatible with the presence of \Al2O3\ grains with a crystalline $\gamma$-lattice structure (see Fig.~\ref{Fig:SED_RDor_WHya}). As shown in Fig.~\ref{Fig:SED_RDor_WHya} crystalline $\alpha$-\Al2O3\ can indeed explain the 13\,\mic\ feature in the SED. A crystallinity degree of 5\% can explain the 13\,\mic\  feature in W~Hya, while for R~Dor the estimated crystallinity degree is around 12\%. These crystalline grains might be the scattering agents of the polarized light emission.

However, this questions how to explain the broad 11\,\mic\ feature in the SED of R~Dor and W~Hya and, in particular, the MIDI data of W~Hya which clearly show that emission arises very close to the star. A potential answer to that question is that the 11\,\mic\ feature does \textit{not} signal the presence of solid-state amorphous aluminium oxide, but --- to the contrary --- points towards the existence of \textit{large neutral gas-phase aluminium oxide clusters} (see Sect.~\ref{Sec:disc_Al_clusters}). Individual (Al$_2$O$_3$)$_n$ clusters may exist above 1500\,K. We note that our analysis does not exclude the existence of amorphous aluminium oxides in the wind, but only at larger distances from the star where temperatures are cooler.

\subsection{Constraining the presence of aluminium oxide clusters} \label{Sec:disc_Al_clusters}

\begin{figure}[htp]
\centering\includegraphics[width=0.4\textwidth]{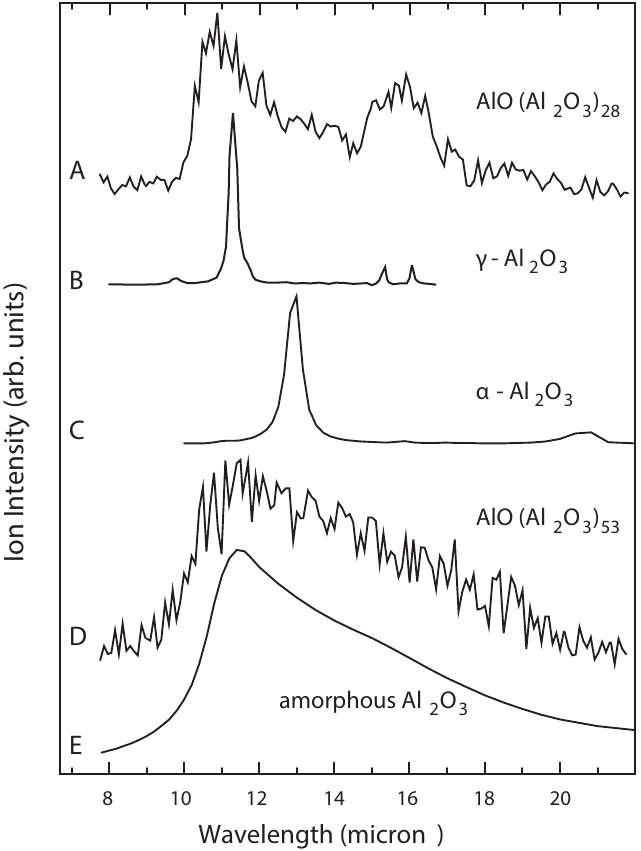}
\caption{Comparison of gas-phase vibrational spectra for AlO$\cdot$(\Al2O3)$_n$ with $n=28$ (A) and for $n=53$ (D) to the infrared absorption spectra of bulk aluminium oxide lattices (B = crystalline $\gamma$-\Al2O3; C = crystalline $\alpha$-\Al2O3; E = amorphous \Al2O3). Result obtained by \citet{vanHeijnsbergen2003PCCP....5.2515V}.}
\label{Fig:VanHeijnsbergen}
\end{figure}

If amorphous alumina dust grains are not the carrier of the MIDI and ISO spectral signature at 11\,$\mu$m, then the interesting possibility exists that large aluminium oxide clusters are present, and detected, just above the stellar surface. Indeed, an outcome of the experimental IR-REMPI studies by \citet{vanHeijnsbergen2003PCCP....5.2515V}  is that large AlO$\cdot$(\Al2O3)$_n$ clusters ($n>34$) exhibit a spectral signature that is very similar to that of amorphous \Al2O3, while smaller AlO$\cdot$(\Al2O3)$_n$ clusters (around $n=15$) show the best match with that of crystalline $\gamma$-\Al2O3\ (see Fig.~\ref{Fig:VanHeijnsbergen}). Due to the non-linear regime in which the IR-REMPI results are obtained, one can however not estimate the amount of clusters present in the beam. This renders the impossibility of quantifying the amount of large aluminium oxide clusters that would be needed in order to explain the 11\,\mic\ feature.

If the 11\,\mic\ feature marks the presence of large ($n > 34$) aluminium oxide clusters, then one should wonder if smaller clusters are present as well and if they can be detected. The experiments by \citet{vanHeijnsbergen2003PCCP....5.2515V} and \citet{Demyk2004A&A...420..547D} show that clusters with $n=15-40$ show two vibrational bands, one around 11.8\,\mic\ and another one around 15\,\mic. Comparing this to the emission features of solid-state aluminium oxide lattices identifies these gas phase clusters to have a structure similar to crystalline $\gamma$-\Al2O3\ instead of $\alpha$-\Al2O3 (corundum), which is the thermodynamically most stable form of the bulk (see Fig.~\ref{Fig:VanHeijnsbergen}). The infrared observations of W~Hya and R~Dor do not display an obvious emission component around 15\,\mic. This does not imply by necessity  that these $n=15-40$ sized clusters are not present, but at least they are not abundant enough to leave any spectral imprint.

%
Due to variety of processes prevalent in the vast interstellar space such as collisions, condensation, coagulation, coalescence the size distribution of interstellar dust grains is dominated by small particles, i.e.\ $n_d(a)\,da \propto a^{-7/2}\,da$ \citep{Mathis1977ApJ...217..425M}. Also in our own Earth's atmosphere, fresh aerosols which are created in situ from the gas-phase by nucleation display a number distribution heavily skewed towards the very small particles \citep{Seinfeld1998PhT....51j..88S}. In analogy, one might expect a significantly larger fraction of very small aluminium oxide clusters as compared to large clusters in the stellar wind. Current capacities in quantum chemical computations give us the possibility to test this proposition on the prevailing presence of very small clusters; see next section.

\subsubsection{Emission of very small (\Al2O3)$_n$ ($n \le 4$) aluminium oxide clusters}\label{Sec:disc_Al_clusters_small}

A way to test if very small \Al2O3\ polymers are present is via the detection of their vibrational frequencies. These bands arise owing to the presence of internal bending and stretching modes in the cluster. As the energy and geometry, the vibrational spectrum differs from isomer to isomer,  and thus can be seen as a spectral fingerprint. 

\begin{figure*}[htp]
\begin{minipage}[t]{.48\textwidth}
        \centerline{\resizebox{\textwidth}{!}{\includegraphics[angle=270]{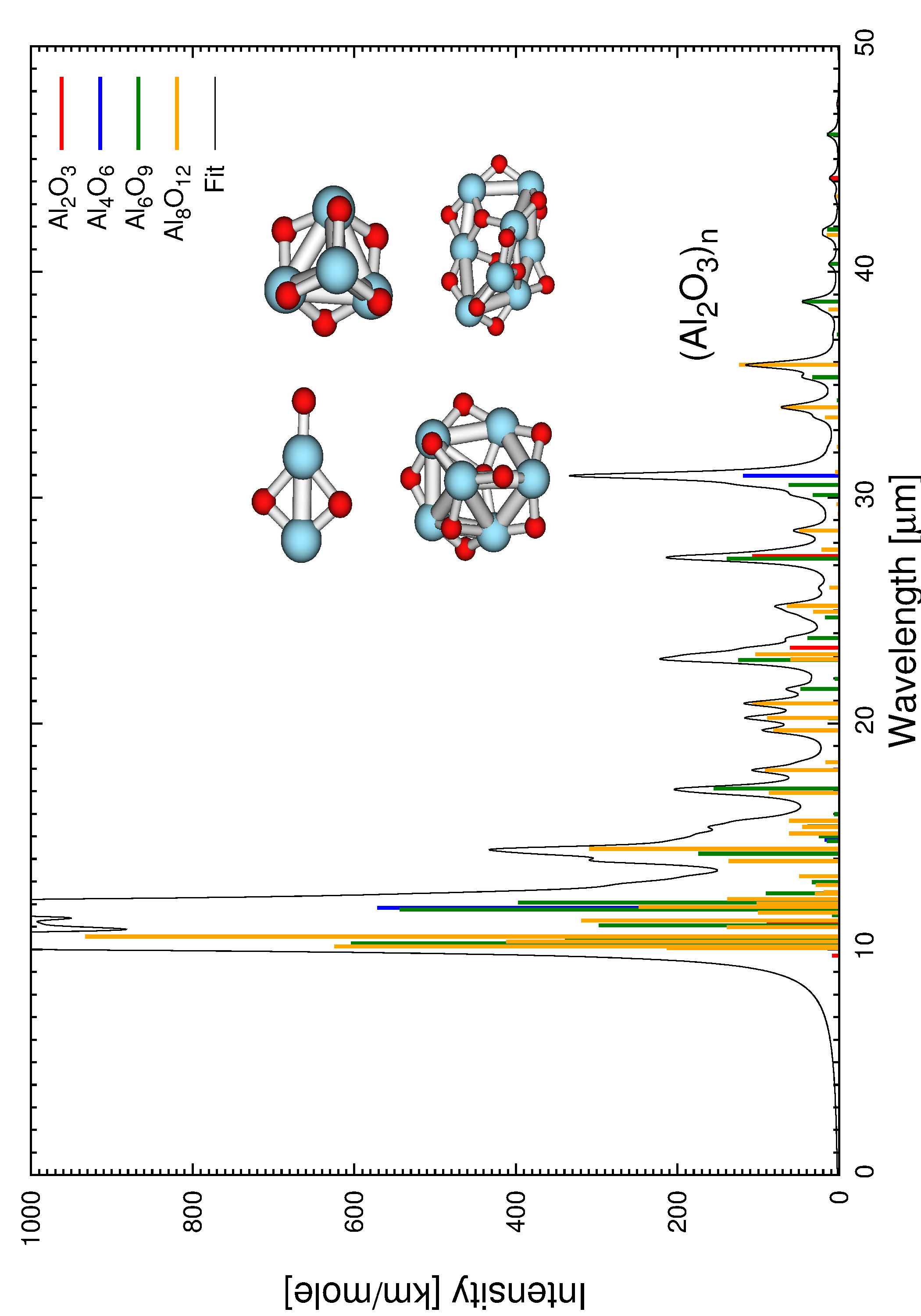}}}
    \end{minipage}
    \hfill
\begin{minipage}[t]{.48\textwidth}
        \centerline{\resizebox{\textwidth}{!}{\includegraphics[angle=270]{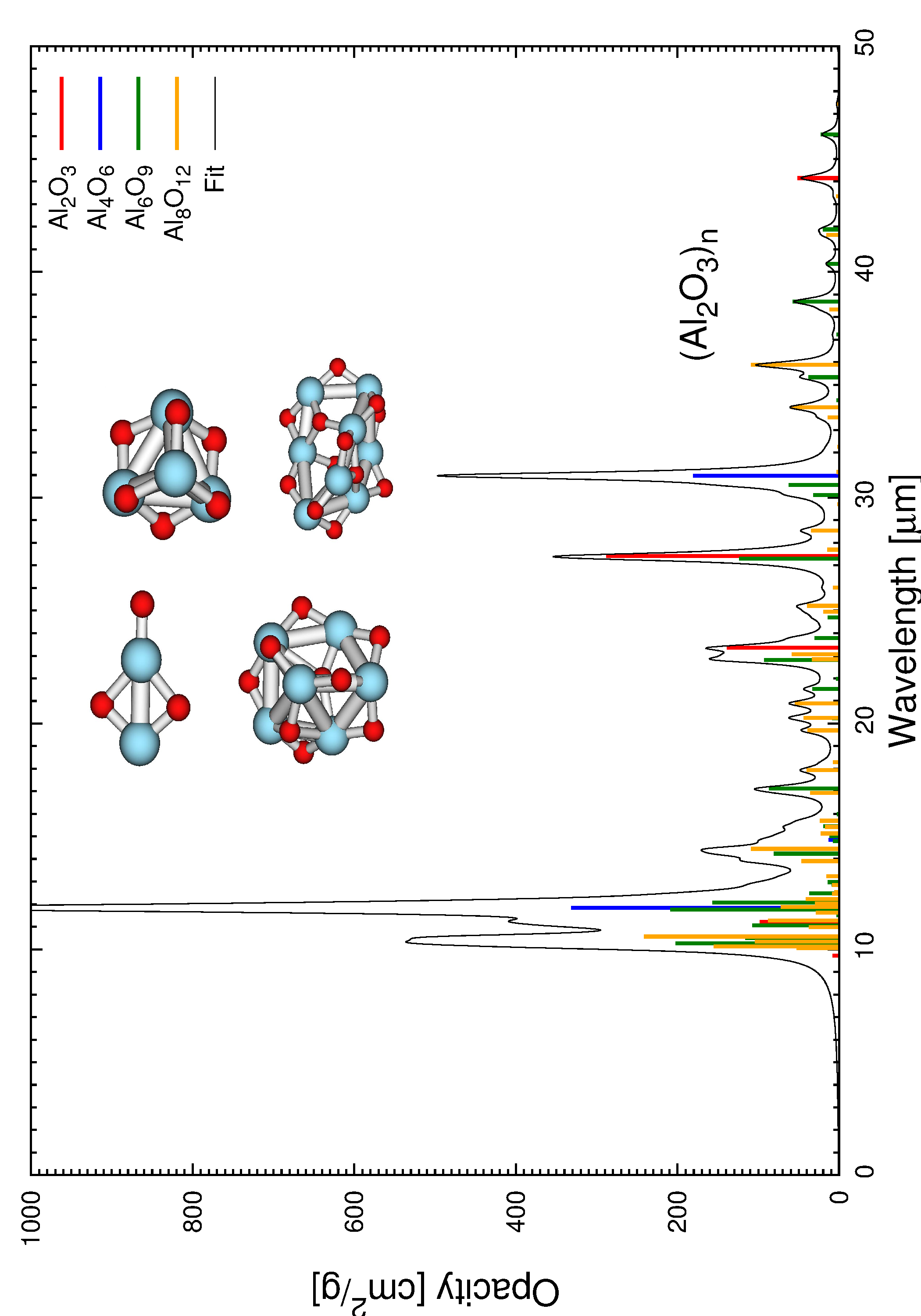}}}
    \end{minipage}
 \caption{Vibrational spectrum of the (\Al2O3)$_n$, $n=1-4$, ground state clusters and their corresponding structures. The structures (for $n=1-4$) are depicted with increasing $n$ from left to right and from top to bottom. The black line represents a Lorentzian distribution with HWHM parameter $\gamma=0.2$ for all peaks. The left panel shows the intensity in units of km/mole; the right panel the opacity in units of cm$^2$/g. }
 \label{Fig:Gobrecht}
\end{figure*}

Because there is no way to currently predict the most likely cluster composition in the outflow and because these clusters would probably range over a wide range in both stoichiometry as well as atomic sizes, we have taken the simple approach of calculating the spectral properties of a range of stoichiometric alumina clusters to estimate where observable lines might exist.
For $n=1 - 4$ we have determined the most stable structures of the neutral aluminium oxide clusters using density functional theory (DFT); see Fig.~\ref{Fig:Gobrecht}. The computations were accomplished by the GAUSSIAN~09 package \citep{Frisch09} using the hybrid B3LYP \citep[Becke, three parameter, Lee-Yang-Parr;][]{Lee1988PhRvB..37..785L, Miehlich1989CPL...157..200M, Becke1993JChPh..98.5648B} exchange-correlation functional with a 6-311+g$^*$ basis set. The binding energy and stability of electronic systems including clusters are pressure and temperature-dependent, respectively. Variations in temperature and density imply that the energy levels of the clusters may shift and that the order of the most stable clusters may change. We have calculated how the stability\footnote{ calculated from the Gibbs free energy. The entropy is given by the partition function (with contributions from electronic, translational, vibrational and vibrational motion) of which the  translational part depends on the pressure.}  of the considered small alumina clusters is affected in various ways for conditions occurring in the CSE (with pressure $p=0.001-50$\,Pa, and temperature $T=500-3000$\,K). 
For \Al2O3, the first excited structural isomer is a linear chain and has a potential energy 0.1\,eV above the kite-shaped triplet ground state. Above 1000\,K the linear form is slightly more stable, below 1000\,K the kite-shaped triplet structure is favoured. 
Considering (\Al2O3)$_2$, the energy difference of the ground state to the first excited state is  relatively large (0.48\,eV) and therefore structure 2A \citep[see Fig.~1 in][]{Li2012} is also the most stable in circumstellar conditions. 
For $n=3$, we find five stable structures in an energy interval of only 0.13\,eV for the alumina trimer. In circumstellar conditions, however,  the first excited state 3B \citep{Li2012} is the most stable configuration. 
For a temperature of 1500\,K, the ground and first excited state of (\Al2O3)$_4$ are almost identical in potential energy. Below 1500\,K isomer 4A is favoured, above 1500\,K structure 4B \citep{Li2012} which is only stable for pressures above 0.1\,Pa.
Overall, the Gibbs Free energy of formation for alumina clusters depends significantly on T, but not much on density or pressure.

\begin{figure}[htp]
\includegraphics[height=0.48\textwidth,angle=270]{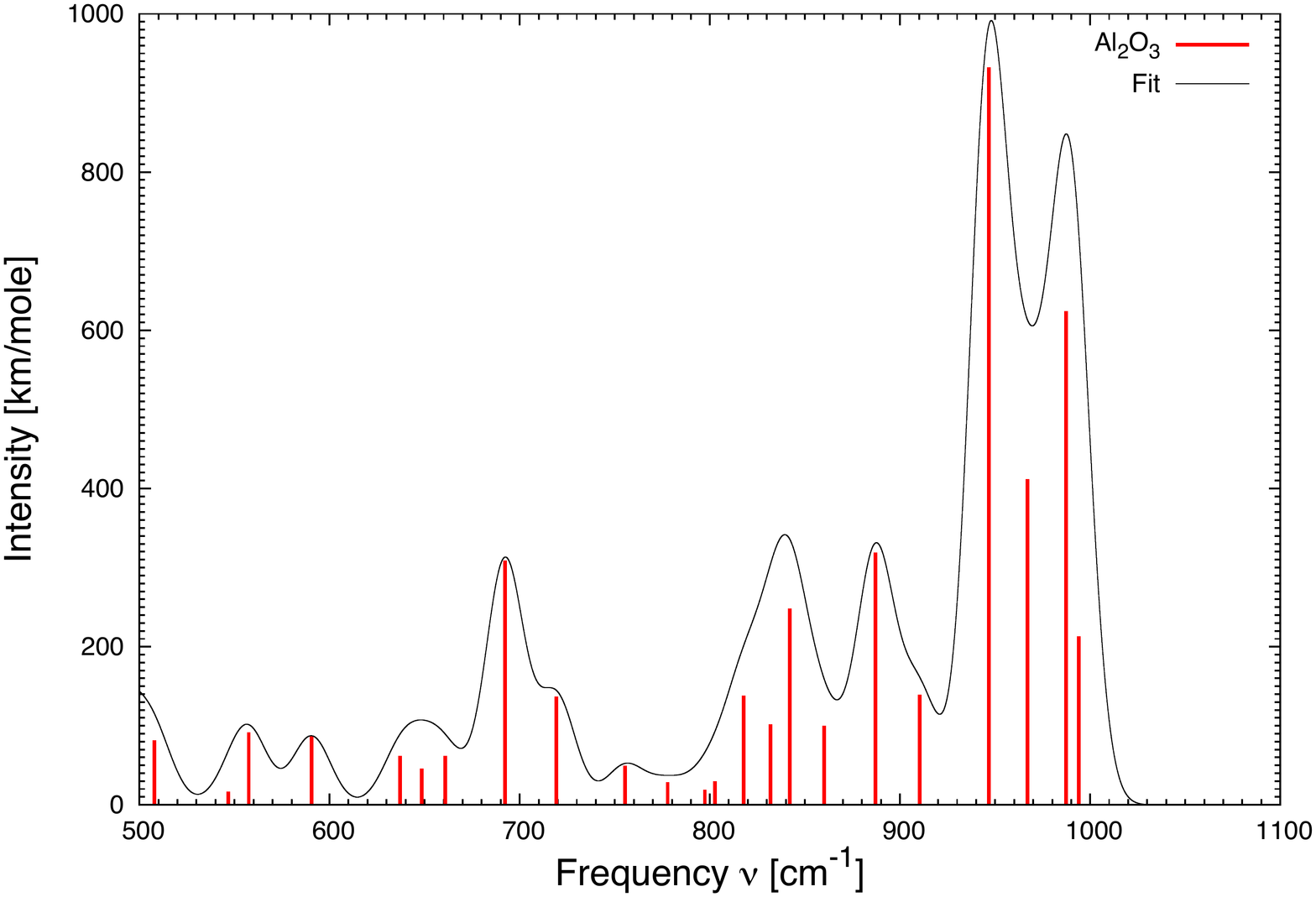}
\caption{Intensity of the (\Al2O3)$_4$ ground state cluster versus frequency convoluted with a Lorentzian function with FWHM of 24\,cm$^{-1}$. To be compared with Fig.\ 2d in \citet{Sierka2007}.}
\label{Fig:Gobrecht_Sierka}
\end{figure}

For each of the optimized structures, the vibrational spectrum of the ground state clusters was calculated (see Fig.~\ref{Fig:Gobrecht}). The predicted spectra compare well with the experimental results by \citet{Demyk2004A&A...420..547D} and the calculated linear infrared absorption spectra by \citet{Sierka2007} for small cluster sizes (see Fig.~\ref{Fig:Gobrecht_Sierka}), with the strongest peaks in the 10-11\,\mic\ region and an ensemble of smaller peaks around 14\,\mic. 
Implementing the stable clusters in thermo-chemistry calculations for densities down to $1\times10^{-4}$ dyne\,cm$^{-2}$ \citep{Hofner2016arXiv160509730H} and temperatures up to 1800\,K  and allowing the small clusters to grow via monomer addition indicates that (\Al2O3)$_3$ seems to be the best candidate for possible detection in CSEs (see Table~\ref{Table_David}).

\begin{table}
\caption{Overview of the aluminia nucleation by monomer (\Al2O3) addition at different temperatures $T$ (in K) and pressures $p$ (in dyne\,cm$^{-2}$). An energetically feasible nucleation is marked with $\surd$. Suppressed nucleation with (large) energy barriers is marked with x. If the nucleation pathway is partially favourable, the largest preferential cluster is given. }
\label{Table_David}
\begin{center}
\begin{tabular}{l|cccc}
\hline
\backslashbox[0pt][l]{}{}
  & \makebox(0,0)[lb]{\put(-1.25\tabcolsep,0){\llap{$T$}}}%
1000 & 1200 & 1500 & 1800 \\\hline
\makebox(0,0){\put(0,2.25\normalbaselineskip){\rlap{$p$}}}%
1 & $\surd$ & $\surd$ & $\surd$  & (\Al2O3)$_3$ \\
0.1 & $\surd$ & $\surd$ & $\surd$  & (\Al2O3)$_3$ \\
0.01 & $\surd$ & $\surd$ & $\surd$  & x \\
$10^{-3}$ & $\surd$ & $\surd$  & (\Al2O3)$_3$ & x \\
$10^{-4}$& $\surd$ & $\surd$  & (\Al2O3)$_3$ & x \\
\hline
\end{tabular}
\end{center}
\end{table}

\begin{figure*}[htp]
\centering \includegraphics[height=0.9\textheight]{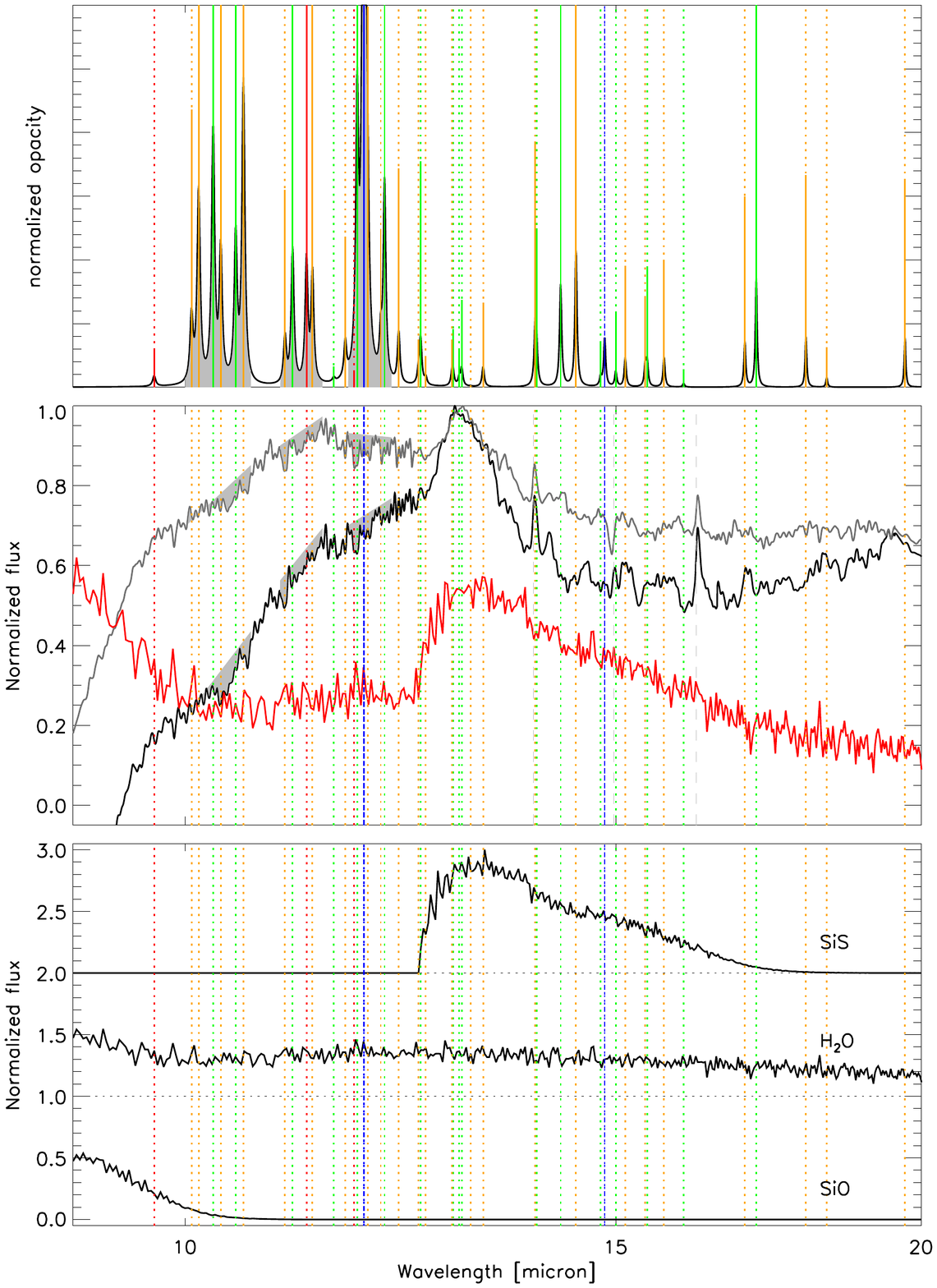}
\vspace*{-2cm}
\caption{Upper panel: normalized opacity of the (\Al2O3)$_n$ clusters convoluted with a Lorentzian profile with a FWHM of 0.03\,\mic. The full coloured lines represent the opacities normalized for each cluster entity separately (with the same colour coding as in Fig.~\ref{Fig:Gobrecht}), while the dotted coloured lines indicate the wavelength position of all ground-state vibrational frequencies of the stable (\Al2O3)$_n$ ($n=1-4$) clusters.
Middle panel: Comparison between the (stellar continuum subtracted and normalized) SED of R~Dor (full black line) and W~Hya (full gray line) with the predicted vibrational frequencies of the stable (\Al2O3)$_n$ ($n=1-4$) clusters (vertical coloured lines). Grey areas indicate the region with some of the largest (\Al2O3)$_n$ opacities. Dashed gray lines indicate the CO$_2$ lines as discussed by \citet{Justtanont2004A&A...417..625J}. The full red line is the (stellar continuum subtracted and normalized) emission for a slab model containing H$_2$O, SiO, and SiS. Column density for all species were taken to be $1\times10^{20}$\,cm$^{-2}$ at 0.5\Rstar\ above the stellar surface for a temperature of 2000\,K. The individual contributions of H$_2$O, SiO, and SiS are shown in the bottom panel. Note that the column densities of the molecules were not chosen to fit the ISO/SWS data, but serve as illustrative examples on the molecular line veiling.}
\label{Fig:molecules_clusters}
\end{figure*}

We have tested if the spectral features present in the infrared SED of R~Dor and W~Hya could be explained by the vibrational frequencies of these small aluminium oxide clusters (see Fig.~\ref{Fig:molecules_clusters}). However, the low spectral resolution (around 300) of the ISO/SWS speed 1 data for both stars makes it impossible to draw any firm conclusion. Albeit a positive correlation can be noticed between various absorption peaks in the ISO data and the predicted cluster spectra, we caution against over-interpretation of these results. Specifically, molecular line veiling by other molecules which might be present in the 'extra-molecular' layer created by pulsation-induced density enhancements result in an ensemble of emission and absorption features that complicates any potential identification. High spectral resolution data in the 10-20\,\mic\ regime (as might be obtained in the future using the E-ELT/METIS instrument) is paramount to answer the question on the potential presence and vibrational excitation of these small aluminium oxide clusters.

The more pronounced evidence of large ($n>34$) (\Al2O3)$_n$-clusters (Fig.~\ref{Fig:VanHeijnsbergen} and Fig.~\ref{Fig:SED_RDor_WHya}) warrants a discussion on the possibility if a physico-chemical situation might occur resulting in a skewed cluster-size distribution toward large cluster sizes. 

\subsubsection{Skewed cluster-size distribution} \label{Sec:skewed}

First of all, one might wonder if the \textit{grain}-size distribution in the inner wind region of AGB stars is similar to the one found in the ISM, $n_d(a)\,da \propto a^{-7/2}\,da$ \citep{Mathis1977ApJ...217..425M}. \citet{Dominik1989A&A...223..227D} derived a particle size distribution for carbon-rich stars with spectral index of $-5$ using Classical Nucleation Theory (CNT) with heteromolecular homogeneous grain formation as developed by \citet{Gail1988A&A...206..153G}. 
A different approach was pursued by \cite{Gobrecht2016A&A...585A...6G} who included pulsation-induced shocks in the inner wind of an oxygen-rich AGB star and assumed coalescence and coagulation of (molecular) dimers to form dust grains. Using a Brownian formalism to compute the grain size distribution, they showed that coagulation is only efficient for dust growth for late phase ($\phi > 0.8$), resulting in very small aluminium oxide grains in the size range of 7--20\,nm after one oscillation. Assuming that the aluminium oxide grains survive the next passage of the periodic shocks and that growth occurs through surface deposition gives rise to a grain size distribution at 2 stellar radii after 12 pulsation cycles with two components, one peaking at $\sim$30\,nm, the other around 300\,nm. This result is already a first indication that size distributions in CSEs might differ from the ISM characteristics mentioned above.

Two direct pathways for clusters to grow is via adding monomers and by collisions and sticking together. Since the number of seeds is low, the chance of colliding with monomers will be higher than of collision with larger polymers, so the first pathway seems to be preferred. The region around 2\,\Rstar\ coincides with the GBDS, i.e.\ a higher density region where species can stay longer without being pushed outward. This might  create a favourable condition for clusters to grow to large sizes. These  large aluminium oxides can be very stable at high temperatures due to their ability of undergoing thermo-ionic emission rather than dissociation \citep{Demyk2004A&A...420..547D}. I.e.\ the clusters preferentially emit an electron, a unique property which is only observed for very strongly bound species. This ability might also greatly enhance the growth efficiencies of the clusters, since the transient complex that is formed during the growth reaction process might also stabilize via the emission of an electron.

Growth by adding monomers is hence a potential process for the creation of larger clusters, with a preference for specific  cluster sizes initiating coagulation. Energy considerations in this highly turbulent region just above the stellar atmosphere should be taken into account to answer the question on the favourable cluster size for coagulation. No calculations including the whole spectrum from small to large aluminium oxide clusters are currently available to answer this question. Generally, the alumina clusters (including the very small ones) show a ionic character with alternating Al-O bondings. Moreover, some of the low-lying small clusters exhibit a trigonal symmetry (like bulk-corundum). One thus expects that the coagulation might start at rather small cluster sizes ($n=3,4$).
Our calculations including only the $n=1-4$ (\Al2O3)$_n$ clusters show indeed a preference for the $n=3$ cluster size. A similar behaviour is noted for (SiO)$_n$ for which the matrix isolation experiment by\citet{Stranz1980PhDT........80S} shows that SiO monomers, dimers and trimers can be seen, but then the material seems to transition to the bulk phase. If coagulation were to commence from small cluster sizes, then we would not expect to see (much) large clusters, but we assume that these are in fact present, as indicated by the observed 11\,$\mu$m spectral feature.

Large clusters could, however, be formed by alternative pathways. One possibility is the formation of large clusters by surface processes; i.e.\ a heterogeneous nucleation process during which large clusters are formed on the surface of existing grains and subsequently ejected. Nozzle experiments show that a substantial quantity of large metal clusters can be formed and ejected that way \citep{Knauer1987}. The NACO and SPHERE data prove that grains (of still unknown composition) are ubiquitous close to the star. These grains might serve as surface for cluster growth to occur.
Vaporization of solid material might offer another explication for a skewed cluster-size distribution toward large sizes. The \Al2O3\ grain temperatures in the region around 2\,\Rstar\
are high (Sect.~\ref{Sec:disc_Al_dust}) and might lead to the crystallization of the bulk material. If energetically favourable, the material might also partly get vaporized producing large clusters. This kind of behaviour is noted during formation kinetics experiments and computer simulations for the nucleation and growth rates of metal vapors \citep{Yamada1987}. The recent SPHERE data of \cite{Khouri2016A&A...591A..70K} show that the solid-state material is grouped into irregular, clumpy structures. Backwarming occurring between the dust clumps and the star  might locally increase the temperature \citep{Fleischer1995A&A...297..543F}, favouring vaporization to take place.


\section{Conclusions}\label{Sec:conclusions}

In this paper we have studied the aluminium content in a low and a high mass-loss rate oxygen-rich AGB star, R~Dor and IK~Tau respectively. The aim was to reveal the role of aluminium species in the initiation of the stellar wind. We therefore have observed both stars with ALMA at high spatial resolution ($\sim$150\,mas) to trace the dust formation region via aluminium bearing molecules (AlO, AlOH, and AlCl). The azimuthally averaged flux densities in function of the angular distance from the central star were fitted using a non-LTE radiative transfer code to retrieve the fractional abundances of the molecules.
The gas-phase aluminium chemistry is largely different for both stars, in particular the AlO and AlOH fractional abundances. In IK~Tau AlOH is clearly detected, while traces of clumpy structures containing AlO are only apparent beyond 50\,\Rstar. In contrast, AlO is $\sim$70 times more abundant than AlOH in R~Dor. These three molecules consume only $\la$2\% of the total aluminium budget. Hence, there is ample opportunity to form aluminium-bearing dust species. AlO and AlOH are direct precursors of these aluminium-bearing grains and both molecules are detected well beyond the dust condensation radius. This implies that the aluminium dust condensation cycle is not fully efficient.

To assess the aluminium dust content in oxygen-rich AGB stars, we have also included the well-studied low mass-loss rate star W~Hya. We discuss how the spectral features present in the spectral energy distribution (SED), polarized light signal, and interferometric data point toward the presence of a halo of large (and small) transparent grains close the star. A candidate species for these grains is amorphous \Al2O3. However, at a distance of only 1.5\,\Rstar\ the estimated grain temperature is $>1000$\,K, i.e.\ too high for the grains to retain their amorphous lattice structure. The grains should anneal and form a crystalline compound. Indeed, the typical 13\,\mic\ features signalling the presence of crystalline \Al2O3 ($\alpha$-\Al2O3) is seen in the SEDs of W~Hya and R~Dor. One is then left with the question how to explain the broad 11\,\mic\ feature in the SED and interferometric data. We propose that large gas-phase (\Al2O3)$_n$-clusters ($n>34$) are the species that produce the signatures seen in the SED and interferometric data. We explain how these large clusters can be formed. Using density functional theory (DFT) we calculate the stability and emissivity of very small clusters ($n\la4$). We show how future E-ELT/METIS observations will be crucial to answer the question if very small clusters ($n\la4$) are present in the inner winds of these oxygen-rich AGB stars.

\begin{acknowledgements}
LD acknowledges support from the ERC consolidator grant 646758 AEROSOL and the FWO Research Project grant G024112N. We acknowledge the CINECA award under the ISCRA initiative for the availability of high performance computing resources and support.
This paper makes use of the following ALMA data: ADS/JAO.ALMA2013.0.00166.S. ALMA is a partnership of ESO (representing 
its member states), NSF (USA) and NINS (Japan), together with NRC 
(Canada) and NSC and ASIAA (Taiwan), in cooperation with the Republic of 
Chile. The Joint ALMA Observatory is operated by ESO, AUI/NRAO and NAOJ. We thank Willem Jan de Wit for his contribution to the reduction of the VLTI/MIDI data of R~Dor. We acknowledge with thanks the variable star observations from the AAVSO International Database contributed by observers worldwide. Based on observations obtained with the HERMES spectrograph, which is supported by the Research Foundation - Flanders (FWO), Belgium, the Research Council of KU Leuven, Belgium, the Fonds National de la Recherche Scientifique (F.R.S.-FNRS), Belgium, the Royal Observatory of Belgium, the Observatoire de Genève, Switzerland and the Thüringer Landessternwarte Tautenburg, Germany. 
\end{acknowledgements}

\bibliographystyle{aa}
\bibliography{ALMA_AlO}

\begin{thebibliography}{87}
\expandafter\ifx\csname natexlab\endcsname\relax\def\natexlab#1{#1}\fi

\bibitem[{{Alexander}(1997)}]{Alexander1997AIPC..402..567A}
{Alexander}, C.~M.~O. 1997, in American Institute of Physics Conference Series,
  Vol. 402, American Institute of Physics Conference Series, ed. E.~K. {Zinner}
  \& T.~J. {Bernatowicz}, 567

\bibitem[{{Becke}(1993)}]{Becke1993JChPh..98.5648B}
{Becke}, A.~D. 1993, \jcp, 98, 5648

\bibitem[{{Bedding} {et~al.}(1998){Bedding}, {Zijlstra}, {Jones}, \&
  {Foster}}]{Bedding1998MNRAS.301.1073B}
{Bedding}, T.~R., {Zijlstra}, A.~A., {Jones}, A., \& {Foster}, G. 1998, \mnras,
  301, 1073

\bibitem[{{Bedding} {et~al.}(1997){Bedding}, {Zijlstra}, {von der Luhe},
  {Robertson}, {Marson}, {Barton}, \& {Carter}}]{Bedding1997MNRAS.286..957B}
{Bedding}, T.~R., {Zijlstra}, A.~A., {von der Luhe}, O., {et~al.} 1997, \mnras,
  286, 957

\bibitem[{{Begemann} {et~al.}(1997){Begemann}, {Dorschner}, {Henning},
  {Mutschke}, {G{\"u}rtler}, {K{\"o}mpe}, \&
  {Nass}}]{Begemann1997ApJ...476..199B}
{Begemann}, B., {Dorschner}, J., {Henning}, T., {et~al.} 1997, \apj, 476, 199

\bibitem[{{Cai} {et~al.}(1991){Cai}, {Carter}, {Miller}, \&
  {Bondybey}}]{Cai1991JChPh..95...73C}
{Cai}, M., {Carter}, C.~C., {Miller}, T.~A., \& {Bondybey}, V.~E. 1991, \jcp,
  95, 73

\bibitem[{{Dayou} \& {Balan{\c c}a}(2006)}]{Dayou2006A&A...459..297D}
{Dayou}, F. \& {Balan{\c c}a}, C. 2006, \aap, 459, 297

\bibitem[{{De Beck} {et~al.}(2017){De Beck}, {Decin}, {Ramstedt}, {Olofsson},
  {Menten}, {Patel}, \& {Vlemmings}}]{Debeck2017A&A...598A..53D}
{De Beck}, E., {Decin}, L., {Ramstedt}, S., {et~al.} 2017, \aap, 598, A53

\bibitem[{{De Beck} {et~al.}(2015){De Beck}, {Kami{\'n}ski}, {Menten}, {Patel},
  {Young}, \& {Gottlieb}}]{DeBeck2015ASPC..497...73D}
{De Beck}, E., {Kami{\'n}ski}, T., {Menten}, K.~M., {et~al.} 2015, in
  Astronomical Society of the Pacific Conference Series, Vol. 497, Why Galaxies
  Care about AGB Stars III: A Closer Look in Space and Time, ed.
  F.~{Kerschbaum}, R.~F. {Wing}, \& J.~{Hron}, 73

\bibitem[{{De Beck} {et~al.}(2013){De Beck}, {Kami{\'n}ski}, {Patel}, {Young},
  {Gottlieb}, {Menten}, \& {Decin}}]{DeBeck2013A&A...558A.132D}
{De Beck}, E., {Kami{\'n}ski}, T., {Patel}, N.~A., {et~al.} 2013, \aap, 558,
  A132

\bibitem[{{de Vries} {et~al.}(2010){de Vries}, {Min}, {Waters}, {Blommaert}, \&
  {Kemper}}]{deVries2010A&A...516A..86D}
{de Vries}, B.~L., {Min}, M., {Waters}, L.~B.~F.~M., {Blommaert}, J.~A.~D.~L.,
  \& {Kemper}, F. 2010, \aap, 516, A86

\bibitem[{{Decin} {et~al.}(2010{\natexlab{a}}){Decin}, {De Beck},
  {Br{\"u}nken}, {M{\"u}ller}, {Menten}, {Kim}, {Willacy}, {de Koter}, \&
  {Wyrowski}}]{Decin2010A&A...516A..69D}
{Decin}, L., {De Beck}, E., {Br{\"u}nken}, S., {et~al.} 2010{\natexlab{a}},
  \aap, 516, A69

\bibitem[{{Decin} {et~al.}(2010{\natexlab{b}}){Decin}, {Justtanont}, {De Beck},
  {Lombaert}, {de Koter}, {Waters}, {Marston}, {Teyssier}, {Sch{\"o}ier},
  {Bujarrabal}, {Alcolea}, {Cernicharo}, {Dominik}, {Melnick}, {Menten},
  {Neufeld}, {Olofsson}, {Planesas}, {Schmidt}, {Szczerba}, {de Graauw},
  {Helmich}, {Roelfsema}, {Dieleman}, {Morris}, {Gallego},
  {D{\'{\i}}ez-Gonz{\'a}lez}, \& {Caux}}]{Decin2010A&A...521L...4D}
{Decin}, L., {Justtanont}, K., {De Beck}, E., {et~al.} 2010{\natexlab{b}},
  \aap, 521, L4

\bibitem[{{Decin} {et~al.}(2017){Decin}, {Richards}, {Danilovich}, {Van de
  Sande}, {De Beck}, {Waters}, {Wittkowski}, {Homan}, {Zijlstra}, {Paladini},
  {Lombaert}, \& {Cox}}]{Decin2016scan}
{Decin}, L., {Richards}, A.~M.~S., {Danilovich}, T., {et~al.} 2017, \aap, {in
  prep.}

\bibitem[{{Dehaes} {et~al.}(2007){Dehaes}, {Groenewegen}, {Decin}, {Hony},
  {Raskin}, \& {Blommaert}}]{Dehaes2007MNRAS.377..931D}
{Dehaes}, S., {Groenewegen}, M.~A.~T., {Decin}, L., {et~al.} 2007, \mnras, 377,
  931

\bibitem[{{Demyk} {et~al.}(2004){Demyk}, {van Heijnsbergen}, {von Helden}, \&
  {Meijer}}]{Demyk2004A&A...420..547D}
{Demyk}, K., {van Heijnsbergen}, D., {von Helden}, G., \& {Meijer}, G. 2004,
  \aap, 420, 547

\bibitem[{{Dominik} {et~al.}(1989){Dominik}, {Gail}, \&
  {Sedlmayr}}]{Dominik1989A&A...223..227D}
{Dominik}, C., {Gail}, H.-P., \& {Sedlmayr}, E. 1989, \aap, 223, 227

\bibitem[{{Fleischer} {et~al.}(1995){Fleischer}, {Gauger}, \&
  {Sedlmayr}}]{Fleischer1995A&A...297..543F}
{Fleischer}, A.~J., {Gauger}, A., \& {Sedlmayr}, E. 1995, \aap, 297, 543

\bibitem[{{Frisch} {et~al.}(2009){Frisch}, {Trucks}, {Schlegel}, {Scuseria}, \&
  {Robb}}]{Frisch09}
{Frisch}, M.~J., {Trucks}, G.~W., {Schlegel}, H.~B., {Scuseria}, G.~E., \&
  {Robb}, M.~A. e.~a. 2009, {Gaussian 09 {R}evision {D}.01, Gaussian Inc.
  Wallingford CT 2009}

\bibitem[{{Gail} \& {Sedlmayr}(1988)}]{Gail1988A&A...206..153G}
{Gail}, H.-P. \& {Sedlmayr}, E. 1988, \aap, 206, 153

\bibitem[{{Gail} {et~al.}(2013){Gail}, {Wetzel}, {Pucci}, \&
  {Tamanai}}]{Gail2013A&A...555A.119G}
{Gail}, H.-P., {Wetzel}, S., {Pucci}, A., \& {Tamanai}, A. 2013, \aap, 555,
  A119

\bibitem[{{Gobrecht} {et~al.}(2016){Gobrecht}, {Cherchneff}, {Sarangi},
  {Plane}, \& {Bromley}}]{Gobrecht2016A&A...585A...6G}
{Gobrecht}, D., {Cherchneff}, I., {Sarangi}, A., {Plane}, J.~M.~C., \&
  {Bromley}, S.~T. 2016, \aap, 585, A6

\bibitem[{{Goumans} \& {Bromley}(2012)}]{Goumans2012MNRAS.420.3344G}
{Goumans}, T. \& {Bromley}, S.~T. 2012, \mnras, 420, 3344

\bibitem[{{Green} \& {Thaddeus}(1974)}]{Green1974ApJ...191..653G}
{Green}, S. \& {Thaddeus}, P. 1974, \apj, 191, 653

\bibitem[{{Heras} \& {Hony}(2005)}]{Heras2005A&A...439..171H}
{Heras}, A.~M. \& {Hony}, S. 2005, \aap, 439, 171

\bibitem[{{H{\"o}fner}(2008)}]{Hofner2008A&A...491L...1H}
{H{\"o}fner}, S. 2008, \aap, 491, L1

\bibitem[{{H{\"o}fner} \& {Andersen}(2007)}]{Hofner2007A&A...465L..39H}
{H{\"o}fner}, S. \& {Andersen}, A.~C. 2007, \aap, 465, L39

\bibitem[{{H{\"o}fner} {et~al.}(2016){H{\"o}fner}, {Bladh}, {Aringer}, \&
  {Ahuja}}]{Hofner2016arXiv160509730H}
{H{\"o}fner}, S., {Bladh}, S., {Aringer}, B., \& {Ahuja}, R. 2016, \aap, 594,
  A108

\bibitem[{{Jeong} {et~al.}(2003){Jeong}, {Winters}, {Le Bertre}, \&
  {Sedlmayr}}]{Jeong2003A&A...407..191J}
{Jeong}, K.~S., {Winters}, J.~M., {Le Bertre}, T., \& {Sedlmayr}, E. 2003,
  \aap, 407, 191

\bibitem[{{Justtanont} {et~al.}(2004){Justtanont}, {de Jong}, {Tielens},
  {Feuchtgruber}, \& {Waters}}]{Justtanont2004A&A...417..625J}
{Justtanont}, K., {de Jong}, T., {Tielens}, A.~G.~G.~M., {Feuchtgruber}, H., \&
  {Waters}, L.~B.~F.~M. 2004, \aap, 417, 625

\bibitem[{{Justtanont} {et~al.}(1998){Justtanont}, {Feuchtgruber}, {de Jong},
  {Cami}, {Waters}, {Yamamura}, \& {Onaka}}]{Justtanont1998A&A...330L..17J}
{Justtanont}, K., {Feuchtgruber}, H., {de Jong}, T., {et~al.} 1998, \aap, 330,
  L17

\bibitem[{{Kami{\'n}ski} {et~al.}(2016){Kami{\'n}ski}, {Wong}, {Schmidt},
  {M{\"u}ller}, {Gottlieb}, {Cherchneff}, {Menten}, {Keller}, {Br{\"u}nken},
  {Winters}, \& {Patel}}]{Kaminski2016A&A...592A..42K}
{Kami{\'n}ski}, T., {Wong}, K.~T., {Schmidt}, M.~R., {et~al.} 2016, \aap, 592,
  A42

\bibitem[{{Karovicova} {et~al.}(2013){Karovicova}, {Wittkowski}, {Ohnaka},
  {Boboltz}, {Fossat}, \& {Scholz}}]{Karovicova2013arXiv1310.1924K}
{Karovicova}, I., {Wittkowski}, M., {Ohnaka}, K., {et~al.} 2013, ArXiv e-prints

\bibitem[{{Khouri}(2014)}]{KhouriPhD}
{Khouri}, T. 2014, PhD thesis, University of Amsterdam

\bibitem[{{Khouri} {et~al.}(2014{\natexlab{a}}){Khouri}, {de Koter}, {Decin},
  {Waters}, {Lombaert}, {Royer}, {Swinyard}, {Barlow}, {Alcolea}, {Blommaert},
  {Bujarrabal}, {Cernicharo}, {Groenewegen}, {Justtanont}, {Kerschbaum},
  {Maercker}, {Marston}, {Matsuura}, {Melnick}, {Menten}, {Olofsson},
  {Planesas}, {Polehampton}, {Posch}, {Schmidt}, {Szczerba}, {Vandenbussche},
  \& {Yates}}]{Khouri2014A&A...561A...5K}
{Khouri}, T., {de Koter}, A., {Decin}, L., {et~al.} 2014{\natexlab{a}}, \aap,
  561, A5

\bibitem[{{Khouri} {et~al.}(2014{\natexlab{b}}){Khouri}, {de Koter}, {Decin},
  {Waters}, {Maercker}, {Lombaert}, {Alcolea}, {Blommaert}, {Bujarrabal},
  {Groenewegen}, {Justtanont}, {Kerschbaum}, {Matsuura}, {Menten}, {Olofsson},
  {Planesas}, {Royer}, {Schmidt}, {Szczerba}, {Teyssier}, \&
  {Yates}}]{Khouri2014A&A...570A..67K}
{Khouri}, T., {de Koter}, A., {Decin}, L., {et~al.} 2014{\natexlab{b}}, \aap,
  570, A67

\bibitem[{{Khouri} {et~al.}(2016){Khouri}, {Maercker}, {Waters}, {Vlemmings},
  {Kervella}, {de Koter}, {Ginski}, {De Beck}, {Decin}, {Min}, {Dominik},
  {O'Gorman}, {Schmid}, {Lombaert}, \& {Lagadec}}]{Khouri2016A&A...591A..70K}
{Khouri}, T., {Maercker}, M., {Waters}, L.~B.~F.~M., {et~al.} 2016, \aap, 591,
  A70

\bibitem[{{Khouri} {et~al.}(2015){Khouri}, {Waters}, {de Koter}, {Decin},
  {Min}, {de Vries}, {Lombaert}, \& {Cox}}]{Khouri2015A&A...577A.114K}
{Khouri}, T., {Waters}, L.~B.~F.~M., {de Koter}, A., {et~al.} 2015, \aap, 577,
  A114

\bibitem[{{Kim} {et~al.}(2010){Kim}, {Wyrowski}, {Menten}, \&
  {Decin}}]{Kim2010A&A...516A..68K}
{Kim}, H., {Wyrowski}, F., {Menten}, K.~M., \& {Decin}, L. 2010, \aap, 516, A68

\bibitem[{{Knapp} {et~al.}(2003){Knapp}, {Pourbaix}, {Platais}, \&
  {Jorissen}}]{Knapp2003A&A...403..993K}
{Knapp}, G.~R., {Pourbaix}, D., {Platais}, I., \& {Jorissen}, A. 2003, \aap,
  403, 993

\bibitem[{{Knauer}(1987)}]{Knauer1987}
{Knauer}, W. 1987, J. Appl. Phys., 62, 841

\bibitem[{{Koike} {et~al.}(1995){Koike}, {Kaito}, {Yamamoto}, {Shibai},
  {Kimura}, \& {Suto}}]{Koike1995Icar..114..203K}
{Koike}, C., {Kaito}, C., {Yamamoto}, T., {et~al.} 1995, \icarus, 114, 203

\bibitem[{{Koput} \& {Gertych}(2004)}]{Koput2004JChPh.121..130K}
{Koput}, J. \& {Gertych}, A. 2004, \jcp, 121, 130

\bibitem[{{Lee} {et~al.}(1988){Lee}, {Yang}, \&
  {Parr}}]{Lee1988PhRvB..37..785L}
{Lee}, C., {Yang}, W., \& {Parr}, R.~G. 1988, \prb, 37, 785

\bibitem[{{Levin} \& {Brandon}(2005)}]{Levin2005}
{Levin}, I. \& {Brandon}, D.~G. 2005, J.\ Am.\ Ceram.\ Soc., 81, 1995

\bibitem[{{Levin} {et~al.}(1998){Levin}, {Gemming}, \&
  {Brandon}}]{Levin1998PSSAR.166..197L}
{Levin}, I., {Gemming}, T., \& {Brandon}, D.~G. 1998, Physica Status Solidi
  Applied Research, 166, 197

\bibitem[{{Li} \& {Cheng}(2012)}]{Li2012}
{Li}, R. \& {Cheng}, L. 2012, {Comp. and Theor. Chem.}, 996, 125

\bibitem[{{Maercker} {et~al.}(2016){Maercker}, {Danilovich}, {Olofsson}, {De
  Beck}, {Justtanont}, {Lombaert}, \& {Royer}}]{Maercker2016A&A...591A..44M}
{Maercker}, M., {Danilovich}, T., {Olofsson}, H., {et~al.} 2016, \aap, 591, A44

\bibitem[{{Maercker} {et~al.}(2008){Maercker}, {Sch{\"o}ier}, {Olofsson},
  {Bergman}, \& {Ramstedt}}]{Maercker2008A&A...479..779M}
{Maercker}, M., {Sch{\"o}ier}, F.~L., {Olofsson}, H., {Bergman}, P., \&
  {Ramstedt}, S. 2008, \aap, 479, 779

\bibitem[{{Mathis} {et~al.}(1977){Mathis}, {Rumpl}, \&
  {Nordsieck}}]{Mathis1977ApJ...217..425M}
{Mathis}, J.~S., {Rumpl}, W., \& {Nordsieck}, K.~H. 1977, \apj, 217, 425

\bibitem[{{Matsumoto} {et~al.}(2008){Matsumoto}, {Omodaka}, {Imai}, {Shimizu},
  {Bushimata}, {Choi}, {Hirota}, {Honma}, {Inomata}, {Iwadate}, {Jike},
  {Kameno}, {Kameya}, {Kamohara}, {Kan-Ya}, {Kawaguchi}, {Kobayashi}, {Kuji},
  {Kurayama}, {Maeda}, {Manabe}, {Miyaji}, {Nakagawa}, {Nagayama}, {Nakashima},
  {Oh}, {Oyama}, {Sakai}, {Sakakibara}, {Sasao}, {Sato}, {Shibata}, {Shintani},
  {Sofue}, {Sora}, {Suda}, {Tamura}, {Tsushima}, \&
  {Yamashita}}]{Matsumoto2008PASJ...60.1039M}
{Matsumoto}, N., {Omodaka}, T., {Imai}, H., {et~al.} 2008, \pasj, 60, 1039

\bibitem[{{Miehlich} {et~al.}(1989){Miehlich}, {Savin}, {Stoll}, \&
  {Preuss}}]{Miehlich1989CPL...157..200M}
{Miehlich}, B., {Savin}, A., {Stoll}, H., \& {Preuss}, H. 1989, Chemical
  Physics Letters, 157, 200

\bibitem[{{Milam} {et~al.}(2007){Milam}, {Apponi}, {Woolf}, \&
  {Ziurys}}]{Milam2007ApJ...668L.131M}
{Milam}, S.~N., {Apponi}, A.~J., {Woolf}, N.~J., \& {Ziurys}, L.~M. 2007,
  \apjl, 668, L131

\bibitem[{{M{\"u}ller} {et~al.}(2005){M{\"u}ller}, {Schl{\"o}der}, {Stutzki},
  \& {Winnewisser}}]{Muller2005JMoSt.742..215M}
{M{\"u}ller}, H.~S.~P., {Schl{\"o}der}, F., {Stutzki}, J., \& {Winnewisser}, G.
  2005, Journal of Molecular Structure, 742, 215

\bibitem[{{M{\"u}ller} {et~al.}(2001){M{\"u}ller}, {Thorwirth}, {Roth}, \&
  {Winnewisser}}]{Muller2001A&A...370L..49M}
{M{\"u}ller}, H.~S.~P., {Thorwirth}, S., {Roth}, D.~A., \& {Winnewisser}, G.
  2001, \aap, 370, L49

\bibitem[{{Nittler} {et~al.}(2008){Nittler}, {Alexander}, {Gallino}, {Hoppe},
  {Nguyen}, {Stadermann}, \& {Zinner}}]{Nittler2008ApJ...682.1450N}
{Nittler}, L.~R., {Alexander}, C.~M.~O., {Gallino}, R., {et~al.} 2008, \apj,
  682, 1450

\bibitem[{{Nittler} {et~al.}(1997){Nittler}, {Alexander}, {Gao}, {Walker}, \&
  {Zinner}}]{Nittler1997ApJ...483..475N}
{Nittler}, L.~R., {Alexander}, O., {Gao}, X., {Walker}, R.~M., \& {Zinner}, E.
  1997, \apj, 483, 475

\bibitem[{{Norris} {et~al.}(2012){Norris}, {Tuthill}, {Ireland}, {Lacour},
  {Zijlstra}, {Lykou}, {Evans}, {Stewart}, \&
  {Bedding}}]{Norris2012Natur.484..220N}
{Norris}, B.~R.~M., {Tuthill}, P.~G., {Ireland}, M.~J., {et~al.} 2012, \nat,
  484, 220

\bibitem[{{Nuth} \& {Ferguson}(2006)}]{Nuth2006ApJ...649.1178N}
{Nuth}, III, J.~A. \& {Ferguson}, F.~T. 2006, \apj, 649, 1178

\bibitem[{{Ohnaka} {et~al.}(2016){Ohnaka}, {Weigelt}, \&
  {Hofmann}}]{Ohnaka2016A&A...589A..91O}
{Ohnaka}, K., {Weigelt}, G., \& {Hofmann}, K.-H. 2016, \aap, 589, A91

\bibitem[{{Raskin} {et~al.}(2011){Raskin}, {van Winckel}, {Hensberge},
  {Jorissen}, {Lehmann}, {Waelkens}, {Avila}, {de Cuyper}, {Degroote},
  {Dubosson}, {Dumortier}, {Fr{\'e}mat}, {Laux}, {Michaud}, {Morren}, {Perez
  Padilla}, {Pessemier}, {Prins}, {Smolders}, {van Eck}, \&
  {Winkler}}]{Raskin2011A&A...526A..69R}
{Raskin}, G., {van Winckel}, H., {Hensberge}, H., {et~al.} 2011, \aap, 526, A69

\bibitem[{{Richards} {et~al.}(2012){Richards}, {Etoka}, {Gray}, {Lekht},
  {Mendoza-Torres}, {Murakawa}, {Rudnitskij}, \&
  {Yates}}]{Richards2012A&A...546A..16R}
{Richards}, A.~M.~S., {Etoka}, S., {Gray}, M.~D., {et~al.} 2012, \aap, 546, A16

\bibitem[{{Ryde} {et~al.}(1999){Ryde}, {Eriksson}, \&
  {Gustafsson}}]{Ryde1999A&A...341..579R}
{Ryde}, N., {Eriksson}, K., \& {Gustafsson}, B. 1999, \aap, 341, 579

\bibitem[{{Sacuto} {et~al.}(2013){Sacuto}, {Ramstedt}, {H{\"o}fner},
  {Olofsson}, {Bladh}, {Eriksson}, {Aringer}, {Klotz}, \&
  {Maercker}}]{Sacuto2013A&A...551A..72S}
{Sacuto}, S., {Ramstedt}, S., {H{\"o}fner}, S., {et~al.} 2013, \aap, 551, A72

\bibitem[{{Schick}(1960)}]{Schick1960}
{Schick}, H.~L. 1960, {Chem. Rev.}, 60, 331

\bibitem[{{Sch{\"o}ier} {et~al.}(2004){Sch{\"o}ier}, {Olofsson}, {Wong},
  {Lindqvist}, \& {Kerschbaum}}]{Schoier2004A&A...422..651S}
{Sch{\"o}ier}, F.~L., {Olofsson}, H., {Wong}, T., {Lindqvist}, M., \&
  {Kerschbaum}, F. 2004, \aap, 422, 651

\bibitem[{{Sch{\"o}ier} {et~al.}(2005){Sch{\"o}ier}, {van der Tak}, {van
  Dishoeck}, \& {Black}}]{Schoier2005A&A...432..369S}
{Sch{\"o}ier}, F.~L., {van der Tak}, F.~F.~S., {van Dishoeck}, E.~F., \&
  {Black}, J.~H. 2005, \aap, 432, 369

\bibitem[{{Seinfeld} {et~al.}(1998){Seinfeld}, {Pandis}, \&
  {Noone}}]{Seinfeld1998PhT....51j..88S}
{Seinfeld}, J.~H., {Pandis}, S.~N., \& {Noone}, K. 1998, Physics Today, 51, 88

\bibitem[{{Sierka} {et~al.}(2007){Sierka}, {D\"obler}, {Sauer}, {Santambrogio},
  {Br\"ummer}, {W\"oste}, {Janssens}, {Meijer}, \& {Asmis}}]{Sierka2007}
{Sierka}, M., {D\"obler}, J., {Sauer}, J., {et~al.} 2007, {Angew.\ Chem.\ Int.\
  Ed.}, 46, 3372

\bibitem[{{Stranz}(1980)}]{Stranz1980PhDT........80S}
{Stranz}, D.~D. 1980, PhD thesis, Maryland Univ., College Park

\bibitem[{{Stroud} {et~al.}(2004){Stroud}, {Nittler}, \&
  {Alexander}}]{Stroud2004Sci...305.1455S}
{Stroud}, R.~M., {Nittler}, L.~R., \& {Alexander}, C.~M.~O. 2004, Science, 305,
  1455

\bibitem[{{Takigawa} {et~al.}(2016){Takigawa}, {Kamizuka}, {Tachibana}, \&
  {Yamamura}}]{Takigawa2016LPICo1921.6543T}
{Takigawa}, A., {Kamizuka}, T., {Tachibana}, S., \& {Yamamura}, I. 2016, LPI
  Contributions, 1921, 6543

\bibitem[{{Tenenbaum} \& {Ziurys}(2009)}]{Tenenbaum2009ApJ...694L..59T}
{Tenenbaum}, E.~D. \& {Ziurys}, L.~M. 2009, \apjl, 694, L59

\bibitem[{{Tenenbaum} \& {Ziurys}(2010)}]{Tenenbaum2010ApJ...712L..93T}
{Tenenbaum}, E.~D. \& {Ziurys}, L.~M. 2010, \apjl, 712, L93

\bibitem[{{Tielens}(1990)}]{Tielens1990fmpn.coll..186T}
{Tielens}, A.~G.~G.~M. 1990, in From Miras to Planetary Nebulae: Which Path for
  Stellar Evolution?, ed. M.~O. {Mennessier} \& A.~{Omont}, 186--200

\bibitem[{{Turner} {et~al.}(1992){Turner}, {Chan}, {Green}, \&
  {Lubowich}}]{Turner1992ApJ...399..114T}
{Turner}, B.~E., {Chan}, K.-W., {Green}, S., \& {Lubowich}, D.~A. 1992, \apj,
  399, 114

\bibitem[{{Uttenthaler} {et~al.}(2011){Uttenthaler}, {van Stiphout}, {Voet},
  {van Winckel}, {van Eck}, {Jorissen}, {Kerschbaum}, {Raskin}, {Prins},
  {Pessemier}, {Waelkens}, {Fr{\'e}mat}, {Hensberge}, {Dumortier}, \&
  {Lehmann}}]{Uttenthaler2011A&A...531A..88U}
{Uttenthaler}, S., {van Stiphout}, K., {Voet}, K., {et~al.} 2011, \aap, 531,
  A88

\bibitem[{{Van de Sande} {et~al.}(2017){Van de Sande}, {Decin}, {Lombaert},
  {Khouri}, {de Koter}, {Wyrowski}, {De Nutte}, \& {Homan}}]{VandeSande2017}
{Van de Sande}, M., {Decin}, L., {Lombaert}, R., {et~al.} 2017, \aap, {subm.}

\bibitem[{{van Heijnsbergen} {et~al.}(2003){van Heijnsbergen}, {Demyk},
  {Duncan}, {Meijer}, \& {von Helden}}]{vanHeijnsbergen2003PCCP....5.2515V}
{van Heijnsbergen}, D., {Demyk}, K., {Duncan}, M.~A., {Meijer}, G., \& {von
  Helden}, G. 2003, Physical Chemistry Chemical Physics (Incorporating Faraday
  Transactions), 5, 2515

\bibitem[{{Velilla Prieto} {et~al.}(2016){Velilla Prieto}, {S{\'a}nchez
  Contreras}, {Cernicharo}, {Ag{\'u}ndez}, {Quintana-Lacaci}, {Bujarrabal},
  {Alcolea}, {Balan{\c c}a}, {Herpin}, {Menten}, \&
  {Wyrowski}}]{Prieto2016arXiv160901904V}
{Velilla Prieto}, L., {S{\'a}nchez Contreras}, C., {Cernicharo}, J., {et~al.}
  2016, ArXiv e-prints

\bibitem[{{Vollmer} {et~al.}(2006){Vollmer}, {Hoppe}, {Brenker}, \&
  {Palme}}]{Vollmer2006LPI....37.1284V}
{Vollmer}, C., {Hoppe}, P., {Brenker}, F., \& {Palme}, H. 2006, in Lunar and
  Planetary Science Conference, Vol.~37, 37th Annual Lunar and Planetary
  Science Conference, ed. S.~{Mackwell} \& E.~{Stansbery}

\bibitem[{{Wing} \& {Lockwood}(1973)}]{Wing1973ApJ...184..873W}
{Wing}, R.~F. \& {Lockwood}, G.~W. 1973, \apj, 184, 873

\bibitem[{{Woitke}(2006)}]{Woitke2006A&A...460L...9W}
{Woitke}, P. 2006, \aap, 460, L9

\bibitem[{{Yamada} {et~al.}(1987){Yamada}, {Usui}, \& {Takagi}}]{Yamada1987}
{Yamada}, I., {Usui}, H., \& {Takagi}, T. 1987, J. Phys. Chem., 91, 2463

\bibitem[{{Zhao-Geisler} {et~al.}(2015){Zhao-Geisler}, {K{\"o}hler}, {Kemper},
  {Kerschbaum}, {Mayer}, {Quirrenbach}, \& {Lopez}}]{Zhao2015PASP..127..732Z}
{Zhao-Geisler}, R., {K{\"o}hler}, R., {Kemper}, F., {et~al.} 2015, \pasp, 127,
  732

\bibitem[{{Zhao-Geisler} {et~al.}(2012){Zhao-Geisler}, {Quirrenbach},
  {K{\"o}hler}, \& {Lopez}}]{Zhao2012A&A...545A..56Z}
{Zhao-Geisler}, R., {Quirrenbach}, A., {K{\"o}hler}, R., \& {Lopez}, B. 2012,
  \aap, 545, A56

\bibitem[{{Zhao-Geisler} {et~al.}(2011){Zhao-Geisler}, {Quirrenbach},
  {K{\"o}hler}, {Lopez}, \& {Leinert}}]{Zhao2011A&A...530A.120Z}
{Zhao-Geisler}, R., {Quirrenbach}, A., {K{\"o}hler}, R., {Lopez}, B., \&
  {Leinert}, C. 2011, \aap, 530, A120

\end{thebibliography}

 \newpage
 \begin{appendix}
 
 \end{appendix}
 \end{document}